\documentclass[preprint,authoryear,12pt]{elsarticle}

\usepackage{amssymb}
\usepackage{amsmath}
\usepackage{fancyhdr}
\usepackage{graphicx}
\usepackage{epsfig}
\usepackage{natbib}
\usepackage{multicol}
\usepackage{pstricks,pst-grad}
\usepackage{aas_macros}
\usepackage{multirow}
\usepackage{amssymb}
\usepackage{graphics}
\usepackage{xspace}
\usepackage{lineno}
\usepackage{color}
\newcommand{\unitSI}[2][{}]{\ensuremath{\ifx\\#1\else#1\,\fi\mathrm{#2}}}
\newcommand{\unit}[2][{}]{\ensuremath{\ifx\\#1\else#1\,\fi{#2}}}%\mathit

\newcommand{\pdiff}[2][{}]{\ensuremath{\partial_{#2}\ifx\\#1\else^{#1}\fi}}

\newcommand{\Eq}[1]{Eq.~({#1})}
\newcommand{\Eqs}[1]{Eqs.~({#1})}
\newcommand{\Fig}[1]{Fig.~{#1}}
\newcommand{\Figs}[1]{Figs.~{#1}}
\newcommand{\Table}[1]{Table~{#1}}

\newlength{\figwidth}
\newcommand{\Rev}[1]{{\textcolor{black}{#1}}} % Used to annotate document to highlight changes made.
\newcommand{\Revnew}[1]{{\textcolor{black}{#1}}} % Used to annotate document to highlight changes made.
\newcommand{\FAC}{\renewcommand{\FAC}{FAC\xspace}field-aligned current
(FAC)\xspace}
\newcommand{\MI}{\renewcommand{\MI}{M-I\xspace}magnetosphere-ionosphere
(M-I)\xspace}
\newcommand{\GCM}{\renewcommand{\GCM}{GCM\xspace}Global Circulation Model
(GCM)\xspace}

\journal{Planetary and Space Science}

\begin{document}

\begin{frontmatter}



\title{Response of the Jovian thermosphere to a transient `pulse' in solar wind pressure}


%\author{J. N. Yates, N. Achilleos and P. Guio}

%\address{}
\author[ad1,ad2,ad3]{J. N. Yates\corref{cor1}}
\ead{japheth.yates@imperial.ac.uk}

\author[ad1,ad2]{N. Achilleos}
%\ead{nick@star.ucl.ac.uk} 

\author[ad1,ad2]{P. Guio}
%\ead{patrick@star.ucl.ac.uk}

\address[ad1]{Department of Physics and Astronomy, University College London, Gower Street, London, UK}
\address[ad2]{Centre for Planetary Sciences at UCL / Birkbeck, University College London, Gower Street, London, UK}
\address[ad3]{Now at Department of Physics, Space and Atmospheric Physics, Imperial College London, London, UK}
\cortext[cor1]{Corresponding author: Tel.: +44 (0)20 7594 1155; fax: +44 (0)20 7594 7772.}

\begin{abstract}
  
The importance of the Jovian thermosphere with regard to magnetosphere-ionosphere coupling 
is often neglected in magnetospheric physics. We present the first study to investigate the 
response of the Jovian thermosphere to transient variations in solar wind dynamic pressure, 
using an azimuthally symmetric global circulation model coupled to a simple magnetosphere and fixed 
auroral conductivity model. In our simulations, the Jovian magnetosphere encounters a solar wind shock or rarefaction 
region and is subsequently compressed or expanded. We present the ensuing response of the 
coupling currents, thermospheric flows, heating and cooling terms, and the aurora to these 
transient events. Transient compressions cause the reversal\Rev{, with respect to steady state,} of 
magnetosphere-ionosphere coupling currents 
and momentum transfer between the thermosphere and magnetosphere. They also cause at least a factor of two 
increase in the Joule heating rate. Ion drag significantly changes the kinetic energy of the 
thermospheric neutrals depending on whether the magnetosphere is compressed or expanded. \Rev{Local temperature 
variations appear between ${\sim}\unitSI[{-}45\mbox{ and }175]{K}$ for the compression scenario and 
${\sim}\unitSI[{-}20\mbox{ and }50]{K}$ 
for the expansion case. Extended regions of equatorward flow develop in the wake of compression events - we 
discuss the implications of this behaviour for global energy transport.} Both compressions and expansions lead 
to a ${\sim}\unitSI[2000]{TW}$ 
increase in the total power dissipated or deposited in the thermosphere. In terms of auroral processes, 
transient compressions increase main oval UV emission by a factor of ${\sim}\unitSI[4.5]{}$ whilst transient 
expansions increase this main emission by a more modest $\unitSI[37]{\%}$. Both types of transient event cause 
shifts in the position of the main oval, of up to \unitSI[1]{^{\circ}} latitude.

\end{abstract}

\begin{keyword}

  Jupiter \sep magnetosphere \sep thermosphere \sep angular momentum \sep transient \sep time-dependent 

\end{keyword}

\end{frontmatter}

%\linenumbers

\section{Introduction} 
\label{sec:intro}

\subsection{Jovian magnetosphere-ionosphere coupling}\label{sec:introcoupling}

 The interaction between the Jovian magnetosphere and ionosphere is complex. The current 
 systems which connect the planet's ionosphere and magnetosphere are controlled by a feedback 
 mechanism involving the rotation of magnetospheric plasma, the conductance of the ionosphere and the wind 
 system prevailing in the thermosphere (upper atmosphere). Several studies, however, have made 
 substantial progress in modelling this interaction \citep{hill79,pontius97,hill2001,cowbun2001,cowbun2003a,
 cowbun2003b,nichols04,cowley05,bougher2005,cowley07,majeed2009,tao09,ray2010,nichols2011,ray2012}. The models of 
 \citet{cowbun2003a,cowbun2003b,nichols04} were primarily used to study the interaction of the inner and 
 middle magnetosphere and how these regions couple with the Jovian ionosphere; \citet{cowley05} and 
 \citet{cowley07} expanded on the former studies by incorporating simplified models for the outer 
 magnetosphere and polar cap region, and thus coupling the `entire' magnetosphere to the ionosphere. 
 \Rev{\citet{nichols2011} considered how a whole magnetosphere self-consistently interacted with the 
 magnetosphere-ionosphere system. The force balance formalism of \citet{caudal1986} was used in the 
 \citet{nichols2011}.}\\
 
 In addition, \citet{cowbun2003a,cowbun2003b} and \citet{cowley07} investigated how the coupled 
 \MI system interacts with the solar wind - specifically, transient variations in the solar wind 
 dynamic pressure which cause compressions and expansions of the magnetosphere. These models have 
 made realistic predictions regarding the corresponding response of magnetospheric and ionospheric 
 currents, plasma angular velocity profiles and auroral emission (both in terms of the intensity of emission 
 and its location in the ionosphere). Many of these model predictions are supported by observations 
 and complementary theoretical studies such as \citet{nichols2009,clarke2009} and \citet{southkiv2001}.
 None of these aforementioned studies, however, have \Rev{self-consistently} accounted for the dynamics of the Jovian 
 thermosphere. In these studies the thermosphere is assumed to have an angular velocity \unit{\Omega_T}, 
 independent of altitude, which is derived from a constant `slippage factor', \unit{K}, given by 
 
 \begin{align}
 	\unit{K} &= \unit{ \frac{\left( \Omega_J - \Omega_T \right)}{\left( \Omega_J - \Omega_M \right)} }. \label{eq:slippage}
 \end{align}
 
 \noindent In this expression \unit{\Omega_J} (\unitSI[1.76{\times}10^{-4}]{rad\,s^{-1}}) is the 
 angular velocity of the planet and \unit{\Omega_M} is the angular velocity of the magnetospheric 
 region conjugate to the thermosphere. This ensures the ordering \unit{\Omega_J > \Omega_T > \Omega_M}, 
 for a steady state, where angular momentum is transferred from ionosphere to magnetosphere. \\
 
 \citet{smith09} expanded further on the current body of \MI models by coupling a simplified magnetosphere 
 model with an azimuthally symmetric \GCM of Jupiter. Their approach allowed for the self-consistent 
 calculation of the Jovian thermospheric angular velocity, in a coupled \MI system which had reached a steady state. \\
 
 The study by \citet{smith09} produced some notable results such as: 
 
 \emph{i) Angular momentum transfer:} meridional advection of momentum, rather than vertical viscous 
 transport, is the main mechanism for transferring angular momentum in the high latitude thermosphere.
 
 \emph{ii) Thermospheric super-corotation:} \Rev{largely due to \emph{(i)},} the thermosphere super-corotates 
 ($\unit{\Omega_T}{=}\unitSI[1.05]{\Omega_J}$) throughout those latitudes 
 (${\sim}\unitSI[65{-}73]{^{\circ}}$) where it magnetically maps to the middle magnetosphere (${\sim}\unitSI[6{-}25]{R_J}$).
 
 \emph{iii) Distribution of heat:} the simulated thermospheric winds develop two main cells of meridional flow, which 
 cool lower latitudes (${\lesssim}\unitSI[75]{^{\circ}}$) whilst heating the polar regions 
 (${\gtrsim}\unitSI[80]{^{\circ}}$). \\

 \citet{yates2012} used the \Rev{model of} \citet{smith09} to study the influence \Rev{of the} solar wind on 
 steady-state thermospheric flows of Jupiter. They found that ionospheric and magnetospheric currents, 
 thermospheric powers, temperature and auroral emission (by proxy of \FAC) all exhibit increases with 
 decreasing solar wind dynamic pressure (from \unitSI[0.213]{nPa} to \unitSI[0.021]{nPa} \citep{joy2002}).\\
 
 \citet{southkiv2001} suggested that a magnetospheric compression would \Rev{cause} an increase 
 in the degree of magnetospheric plasma corotation (i.e. the quantity $(\Omega_J - \Omega_M)$ would 
 decrease), and this would consequently lead to a sizeable decrease in \MI coupling currents and 
 auroral emission. They also argued that the reverse would be true for a magnetospheric expansion. 
 \Rev{Simulations} by \citet{cowley07} and \citet{yates2012} confirmed these predictions, provided \Rev{that} 
 the system is given enough time to achieve steady-state \Rev{(${\geq}\unitSI[50]{}$ rotations)}. On the other hand, 
 the studies of 
 \citet{cowbun2003a,cowbun2003b} and more recently, of \citet{cowley07} 
 simulated the \Rev{`transient'} (short-term) response of the system to rapid \Rev{(${\sim}\unitSI[2-3]{hours}$)} 
 magnetospheric compressions and expansions. This short-term behaviour was found \Rev{to differ from} the 
 steady state \Rev{case}. For rapid compressions (${\lesssim}\unitSI[3]{hours}$), the conservation of plasma 
 angular momentum causes the magnetosphere to super-corotate compared to the planet and thermosphere. The 
 flow shear between the thermosphere and magnetosphere, represented by  $(\Omega_T - \Omega_M)$, is now 
 negative and leads to current reversals at \Rev{magnetic} co-latitudes that are \Rev{conjugate} to the middle and 
 outer magnetospheres (${\sim}\unitSI[10{-}17]{^{\circ}}$). Negative flow shear also causes energy to 
 be transferred from the magnetosphere to thermosphere; in contrast to the steady-state, where energy 
 is transferred from the thermosphere to the magnetosphere, in order to accelerate outflowing, magnetospheric plasma 
 towards corotation. For transient expansions, \citet{cowley07} showed that \unit{\Omega_M} decreases but 
 the flow shear increases, leading to a ${\sim}\unitSI[500]{\%}$ increase in the intensity of \MI currents 
 (for an expansion from a dayside magnetopause radius of \unitSI[45]{R_J} to \unitSI[85]{R_J}, 
 $\unit{R_J}{=}\unitSI[71492]{km}$). \\
 
 For these transient events, where the magnetopause is displaced by ${\sim}\unitSI[40]{R_J}$, 
 \citet{cowley07} predict differing auroral responses dependent on the nature of the event \Rev{(compression 
 or expansion)}. For compressions, electron energy flux (${\sim}\unitSI[10]{\%}$ of which is 
 used to produce ultraviolet (UV) aurora) at the open-closed field line 
 boundary (polar emission) increases by two orders of magnitude, whilst the main emission 
 is halved. In the expansion case, there is a 30-fold increase in main 
 emission mapping to the middle magnetosphere, whilst polar emission decreases to ${\sim}\unitSI[2]{\%}$ of its 
 steady-state value. 
 Recent observations of auroral emission by \citet{clarke2009} show a factor of two 
 increase in total ultraviolet (UV) auroral power, near the arrival of a solar wind compression region, 
 typically corresponding to an increase in solar wind dynamic pressure of ${\sim}\unitSI[0.01{-}0.3]{nPa}$. 
 Furthermore, \citet{nichols2009} showed, using the same \Rev{Hubble Space Telescope (HST)} images as 
 \citet{clarke2009}, that this 
 increase in auroral emission consists of approximately even contributions from the so-called `main oval' 
 and the high-latitude polar emission. \citet{nichols2009} also showed that the location of the `main oval' 
 shifted polewards by ${\sim}\unitSI[1]{^{\circ}}$ in response to solar wind pressure increase of 
 an order of magnitude. For a rarefaction region in the solar wind, an order of magnitude decrease in solar 
 wind pressure, \citet{clarke2009} observed little, if any, change in auroral emission. 
 
\subsection{Jovian atmospheric heating}\label{sec:introheat}

 The Jovian upper atmospheric temperature is up to \unitSI[700]{K} higher than that predicted by solar heating 
 alone \citep{strobel1973,yellemiller2004}. This `energy crisis' at Jupiter and the other giant planets has 
 puzzled scientists for over \unitSI[40]{} years. Different theories have been put forward to 
 explain Jovian upper atmospheric heating: gravity waves \citep{young1997}, auroral particle precipitation 
 \citep{waite1983,grodent01}, Joule heating \citep{waite1983,eviatar1984} and ion drag 
 \citep{miller2000,smith05,millward2005}. None of the aforementioned studies have been able to fully account for the 
 observations. \\

 \MI coupling models by \citet{achilleos98,bougher2005,smith09,tao09,yates2012} have all discussed steady-state heating and 
 cooling terms in the Jovian thermosphere. \citet{yates2012}, whilst investigating the influence of solar wind on 
 the steady-state thermospheric flows of Jupiter, found that \Revnew{ion drag energy} and Joule heating increased 
 by ${\lesssim}\unitSI[200]{\%}$ (from a compressed to expanded magnetospheric configuration) resulting in a 
 thermospheric temperature increase of ${\sim}\unitSI[135]{K}$. \citet{cowley07} discussed `transient' heating and 
 dynamics in terms of power dissipated in the thermosphere via Joule heating and \Revnew{ion drag energy}, as well as 
 power used to accelerate 
 magnetospheric plasma. \citet{cowley07} considered displacements of the Jovian magnetopause by ${\sim}\unitSI[40]{R_J}$. 
 They found that, for compressions, there was a net transfer 
 of power from magnetosphere to planet of ${\sim}\unitSI[325]{TW}$, due to the expected super-corotation of magnetospheric 
 plasma. For expansions, \citet{cowley07} found that the power dissipated in the thermosphere (and used to accelerate 
 magnetospheric plasma) increased by a factor of ${\sim}\unitSI[2.5]{}$ resulting from a large increase in azimuthal 
 flow shear between the expanded magnetosphere and the thermosphere. \\
 
 \citet{melin2006} analysed infrared data from an auroral heating event observed by \citet{stallard2001,stallard2002} 
 (from September 8-11, 1998) and found that particle precipitation could not account for the observed  
 increase in ionospheric temperature (\unitSI[940{-}1065]{K}). The combined estimate of ion drag and Joule heating rates 
 increased from \unitSI[67]{mW\,m^{-2}} (on September 8) to \unitSI[277]{mW\,m^{-2}} (on September 11) resulting from 
 a doubling of the ionospheric electric field (inferred from spectroscopic observations); this increase in heating was 
 able to account for the observed rise in temperature. Cooling \Rev{rates} (by Hydrocarbons and \unit{H_{3}^{+}} emission) 
 also increased during the event but only by ${\sim}\unitSI[20]{\%}$ of the total inferred \Rev{heating rate}. Thus\Rev{,} 
 a net increase in ionospheric temperature resulted. More detailed analysis showed that these cooling mechanisms 
 would be unlikely to return the thermosphere to its initial temperature before the onset of subsequent heating events. 
 \citet{melin2006} thus concluded that the temperature increases could plausibly lead to an increase in equatorward 
 winds, which transport thermal energy to lower latitudes \citep{waite1983}.\\
 
 In this study, we use the \citet{yates2012} model, `JASMIN' \Rev{(Jovian Axisymmetric Simulator, with 
 Magnetosphere, Ionosphere and Neutrals)}, to estimate the response of Jovian thermospheric dynamics, 
 heating and aurora to transient changes in the solar wind dynamic pressure and, consequently, 
 magnetospheric size. By transient, we mean changes on time scales ${\lesssim}\unitSI[3]{hours}$, where the 
 angular momentum of the magnetospheric plasma is approximately conserved \citep{cowley07} as the time scales 
 required for changes in the \MI currents to affect \unit{\Omega_M} are much longer, ${\sim}\unitSI[10{-}20]{hours}$. 
 Our coupled model responds to time-dependent profiles of plasma angular velocity in the magnetosphere. We employ 
 different \unit{\Omega_M(\rho_e,t)} profiles (\unit{\rho_e} represents equatorial radial distance; \unit{t} denotes 
 time) to represent compressions and expansions of the middle magnetosphere. This is the first study to investigate how 
 \Rev{time-dependent} variations in solar wind pressure influence both magnetospheric and thermospheric properties of the 
 Jovian system, and to use a realistic \GCM to represent the thermosphere.\\
 
 In section~\ref{sec:time} we summarise the time scales involved in the Jovian system and section~\ref{sec:background} 
 describes the model used in this study. In sections~\ref{sec:cmp} and \ref{sec:exp} we present our findings for the 
 transient compression and expansion scenarios respectively. We discuss our findings and \Rev{and their limitations} 
 in section~\ref{sec:discussion} and conclude in section~\ref{sec:conclusion}.
 
%%%%%%%%%%%%%%%%%%%%%%%%%%%%%%%%%%%%%%%%%%%%%%%%%%%%%%%%%%%%%%%%%%%%%% 

\section{Time-dependence of the Jovian system}
\label{sec:time}
 
 Variations in magnetic field, plasma angular velocity and thermospheric flow patterns due to solar wind 
 pressure changes present challenges for modelling the Jovian system. Various time-scales, such as those 
 associated with \MI coupling, compression or expansion of the magnetosphere and thermospheric response, 
 need to be considered. The studies by \citet{cowbun2003a,cowbun2003b} and \citet{cowley07} are among the 
 few to have addressed these issues, using the simplifying approximations discussed hereafter.\\
 
 \emph{i) \MI coupling time scale:} 
    The neutral atmosphere transfers angular momentum to the magnetosphere along magnetic field lines 
    in order to accelerate the radially outflowing magnetospheric plasma towards corotation. The 
    time-scale on which this angular momentum is transferred has been estimated by \citet{cowbun2003a} to 
    be ${\sim}\unitSI[5{-}20]{}$ hours\Rev{, similarly to that found by \citet{vasyliunas1994}.} \\
	
 \emph{ii) Compression (and expansion) of the magnetosphere:} 
    Large changes in magnetospheric size (${\sim}\unitSI[40]{R_J}$) can occur when the Jovian 
    magnetosphere encounters a sudden change in solar wind dynamic pressure, such as would be 
    caused by a Coronal Mass Ejection (CME) or Corotating Interaction Regions (CIR).
    \citet{cowbun2003a} and \citet{cowley07} considered compressions (and expansions) occurring over 
    ${\sim}\unitSI[2{-}3]{hours}$, and were thus able to assume conservation of plasma angular momentum 
    when calculating the response of the \MI system\Rev{, since the coupling time scale discussed in i) is 
    large, by comparison}. \\
 
 \emph{iii) Thermospheric response time:} 
	The thermosphere and magnetosphere are coupled together via ion-neutral collisions in the 
	ionosphere; therefore a change in plasma angular velocity would cause a corresponding change 
	in the thermosphere's effective angular velocity. 
	Recent models for the thermospheric response are generally divided in two scenarios: (i) A system 
	where the thermosphere responds promptly, on the order of a few tens of minutes \Rev{as found by} 
	\citep{millward2005}, 
	and (ii) a system where the thermosphere responds on the order of two days and, as such, is essentially 
	unresponsive to transient events \citep{gong2005phd}. However, in this study, we do not make a 
	distinction between thermospheric response models. We simply allow the \GCM to \Rev{respond self-consistently} 
	to the imposed changes in plasma angular velocity assumed for the transient compressions and expansions, 
	thus allowing a realistic, thermospheric response to these changes.
 
%%%%%%%%%%%%%%%%%%%%%%%%%%%%%%%%%%%%%%%%%%%%%%%%%%%%%%%%%%%%%%%%%%%%%%
\section{Model description}
\label{sec:background}

 \subsection{Thermosphere model}
 \label{sec:therm_model}

  The thermosphere model used in this study is a \GCM which solves the non-linear Navier-Stokes equations 
  of energy, momentum and continuity, using explicit time integration \citep{mullerwodarg2006}. The 
  \citet{mullerwodarg2006} three-dimensional (3-D) \GCM was created for Saturn's thermosphere, and later 
  modified for Saturn and Jupiter respectively in \citet{smith08saturn} and \citet{smith09}. It is 
  the \citet{smith09} modified \GCM that we use in this study. The model assumes azimuthal symmetry, and is 
  thus two-dimensional (pressure/altitude and latitude) \Rev{whilst still solving the 3-D equations}. The 
  Navier-Stokes equations are solved in the pressure coordinate system, providing time dependent 
  distributions of thermospheric wind, temperature and energy. The zonal and meridional momentum equations, 
  and the energy equation forming the basis of this particular \GCM can be found in 
  \citet{achilleos98,mullerwodarg2006} or \citet{tao09}, should the reader be interested. Our model is resolved on 
  a \unitSI[0.2]{^{\circ}} latitude and \unitSI[0.4]{} pressure scale height grid, with a lower boundary 
  at \unitSI[2]{\mu bar} (\unitSI[300]{km} above the \unitSI[1]{bar(B)} level) and upper boundary at 
  \unitSI[0.02]{nbar}.   

%%%%%%%%%%%%%%%%%%%%%%%%%%%%%%%%%%%%%%%%%%%%%%%%%%%%%%%%%%%%%%%%%%%%%% 

 \subsection{Ionosphere model}
 \label{sec:ion_model}
 
  \Rev{ We employ a simplified model of the ionosphere used in \citet{smith09} and \citet{yates2012}, who separate the 
  model into two components: i) a vertical part describing the relative change of conductivity with altitude, as defined 
  by the 1D model of \citet{grodent01}. ii) \Revnew{A horizontal part, which linearly scales} the 
  \citet{grodent01} model at all altitudes so that the height-integrated Pedersen conductance \unit{\Sigma_P} matches a 
  global pattern prescribed by the user. The sole difference between the model used in \citet{yates2012} and that 
  presented here lies in \Revnew{the horizontal part. Here we} employ a 
  fixed value of \unit{\Sigma_P} between latitudes \unitSI[60]{^{\circ}} and \unitSI[74]{^{\circ}} whilst in the above 
  studies, conductances in this region may be enhanced, above background levels, by \FAC{}s \citep{nichols04}. 
  \Table{\ref{tb:model:tspedcond}} shows the three different height-integrated Pedersen conductance regions employed in 
  our model, their corresponding ionospheric latitudes and their assigned values of \unit{\Sigma_P}. 
  Section~\ref{sec:mod_lim} discusses the limitations in using such assumptions.}
  
% table 1 here
 
%%%%%%%%%%%%%%%%%%%%%%%%%%%%%%%%%%%%%%%%%%%%%%%%%%%%%%%%%%%%%%%%%%%%%% 

\subsection{Magnetosphere model}
 \label{sec:mag_model}

 \Rev{The axisymmetric Jovian magnetosphere model employed in this study is the same as that described 
 in \citet{yates2012}. It combines a detailed model of the inner and middle magnetosphere \citep{nichols04} 
 with a simplified model of the outer magnetosphere and region of open field lines \citep{cowley05}. The ability to 
 reconfigure the magnetosphere depending on its size is also included by assuming that magnetic flux is 
 conserved \citep{cowley07}. Surfaces of constant flux function define shells of magnetic field lines with common 
 equatorial radii \unit{\rho_e} and ionospheric co-latitude \unit{\theta_i}. These surfaces also allow for 
 the magnetic mapping from the equatorial plane of the magnetodisc (middle magnetosphere) to the ionosphere. 
 This mapping requires an ionospheric flux function \unit{F_i (\theta_i)}, a magnetospheric 
 counterpart \unit{F_e (\rho_e)} (representing the magnetic flux integrated between a given equatorial radial 
 distance \unit{\rho_e} and infinity) and the equality $\unit{F_i (\theta_i)}{=}\unit{F_e (\rho_e)}$, which 
 represents the mapping between \unit{\theta_i} and \unit{\rho_e} to which the corresponding magnetic field 
 line extends \citep{nichols04}. The ionospheric flux function is given by:}

  \begin{align}
      \unit{F_i} = \unit{B_J \rho_{i}^{2}} = \unit{B_J R_{i}^{2} \sin^2\theta_i},\label{eq:ionflux}
  \end{align}

 \noindent where \unit{B_J} is the equatorial magnetic field strength at the planet's surface and 
 \unit{\rho_i} is the perpendicular distance to the planet's magnetic/rotation axis 
 ($\unit{\rho_i}{=}\unit{ R_i \sin\theta_i}$, where \unit{R_i} is the ionospheric radius. We adopt 
 $\unit{B_J}{=}\unitSI[426 400]{nT}$ \citep{connerney1998}, and $\unit{R_i}{=}\unitSI[67 350]{km}$ 
 \citep{cowley07}. Note $\unit{R_i}{<}\unit{R_J}$ due to polar flattening at Jupiter. For further 
 details on the magnetosphere model employed here the reader is referred to \citet{yates2012} and 
 the references therein. \Rev{A discussion on the currents which couple our model can be found in 
 \ref{sec:mi_coup}.}

%%%%%%%%%%%%%%%%%%%%%%%%%%%%%%%%%%%%%%%%%%%%%%%%%%%%%%%%%%%%%%%%%%%%%% 

 \subsection{Obtaining the transient plasma angular velocity}
 \label{sec:tran_plas}
 
 \Rev{In steady state, plasma angular velocity profiles are obtained in a similar manner to that discussed 
 in \citet{smith09} and \citet{yates2012}; by solving the Hill-Pontius equation in the inner and middle magnetosphere, 
 but assuming a constant Pedersen conductance.} \\
 
 We now discuss the calculation of plasma angular velocity once the model has entered the transient regime 
 i.e. once our initial, steady-state system begins to undergo a transient compression/expansion of the 
 magnetosphere. Our method of calculating transient plasma angular velocities follows that of \citet{cowley07}. 
 Prior to the rapid compression or expansion, the system exists in a steady state, with plasma angular velocity 
 \unit{\Omega_{M} (\theta_i, t{=}0)} as a function of co-latitude \unit{\theta_i} and time \unit{t}. Using the 
 magnetic mapping method discussed in section~\ref{sec:mag_model}, the equatorial radial distance 
 \unit{\rho_{e} (\theta_i, t{=}0)} of the local field line can be found. The arrival of the solar wind pulse or 
 rarefaction causes the magnetosphere to compress or expand by several tens of \unit{R_J} (typical choice for the 
 simulations) and the model enters the transient (time-dependent) regime. Thus, a given co-latitude \unit{\theta_i} 
 now maps to a new radial distance \unit{\rho_{e} (\theta_i, t)}. If, as discussed in section~\ref{sec:time}, the 
 solar wind pulse causes perturbations that occur on sufficiently small time scales (${\sim}\unitSI[2{-}3]{hours}$), 
 we can assume that plasma angular momentum is approximately conserved. The plasma angular velocity profile 
 throughout the `pulse' in solar wind pressure is then given by:
 
 \begin{align}
 	\unit{\Omega_{M} (\theta_i, t)} =& \unit{\Omega_{M} (\theta_i, t{=}0) \left( \frac{\rho_{e} (\theta_i, t{=}0)}{\rho_{e} (\theta_i, t)}\right)^2}, 
 	\label{eq:omega_tran}
 \end{align}
 
 \noindent where the notation \unit{t{=}0} and \unit{t} denote the initial (steady-state) and transient state 
 (at each time-step throughout the event) respectively.  \\
 
\begin{figure}
      \centering
      \includegraphics[width= 0.99\figwidth]{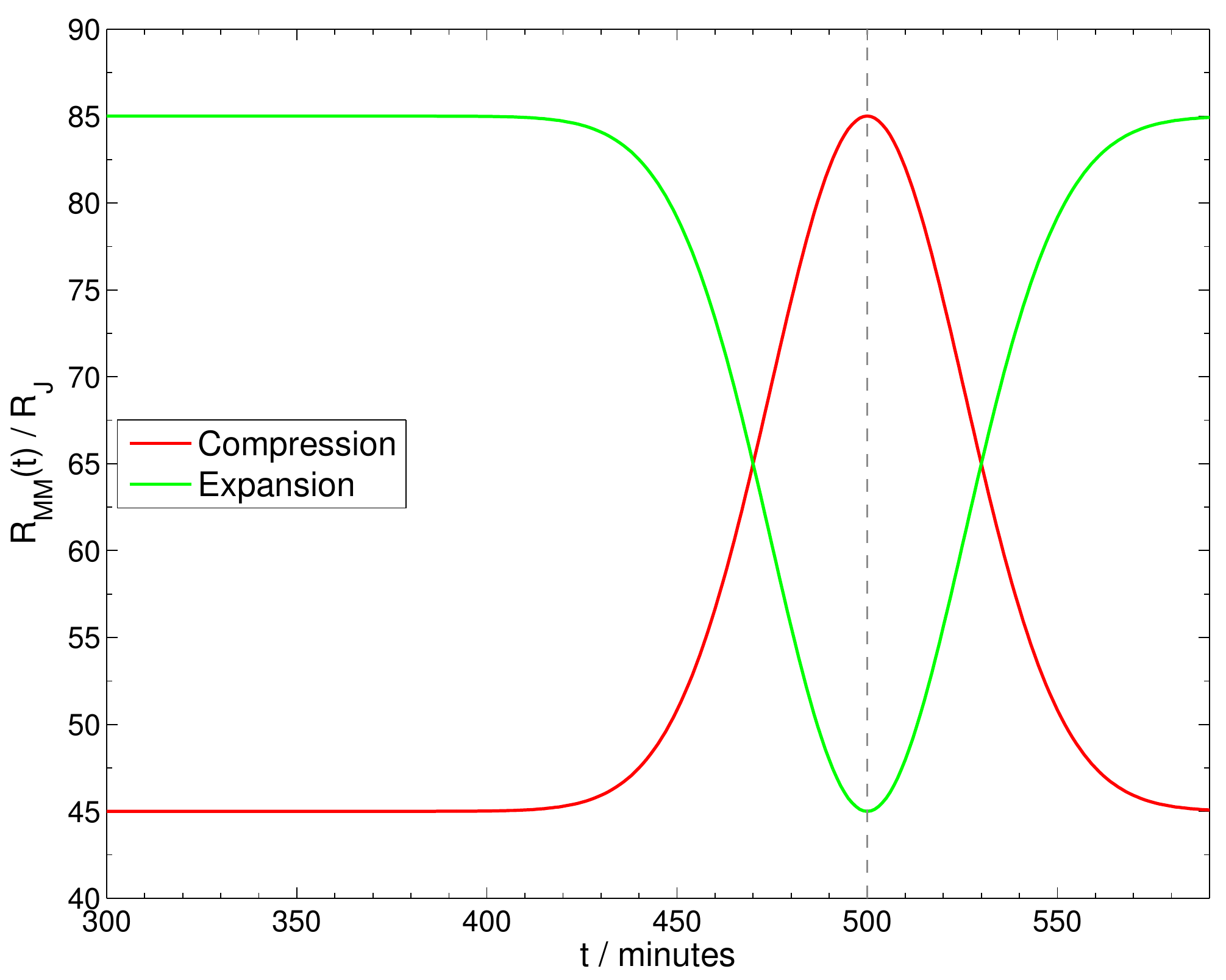}
      
      \caption{ The variation of magnetodisc radius \unit{R_{MM}(t)} with time during a pulse in the solar wind. 
      The \Rev{red} and \Rev{green} lines represent a compression and expansion respectively throughout the entire pulse. 
      The grey dashed line indicates the point of maximum variation.}

      \label{fig:rmmt}
  \end{figure}

 For this study, the time evolution of solar wind dynamic pressure, and thus magnetodisc size, is represented by a 
 Gaussian function. \unit{R_{MM}(t)} represents the magnetodisc radius as a function of time and is given by
	
 \begin{align}
	\unit{R_{MM}(t)} = \unit{ A e^{-\frac{\left( t - t_o\right)^2}{2\Delta t^2} } } + \unit{R_{MMO}}, \label{eq:rmmt}
 \end{align}
	
 \noindent where $\unit{A}{=}\unit{R_{MM}(t_o) - R_{MMO}}$ and is the amplitude of the corresponding curve, 
 \unit{R_{MMO}} is the initial magnetodisc radius, \unit{R_{MM}(t_o)} is the maximum or minimum radius, 
 \unit{t_o} is the time at which 
 $\unit{R_{MM}(t)}{=}\unit{R_{MM}(t_o)}$ (\unitSI[90]{minutes} after pulse start time \unit{t_s}), and 
 \unit{\Delta t} controls the width of the `bell' (obtained using 
 $\unit{(2/3)(t_o - t_s)} = \unit{2\sqrt{2\ln2}\,\,\Delta t}$ ). After achieving steady-state, we run the model 
 for a single Jovian day, transient mode is then initialised \unitSI[3]{hours} prior to the end of the Jovian day (and 
 model runtime). Profiles of \unit{R_{MM}(t)} for compressions and expansions are shown in \Fig{\ref{fig:rmmt}}.\\
 
% FIG 1 here
 
 As indicated in \Fig{\ref{fig:rmmt}}, the simulated pulse lasts for a total of three hours, after which the 
 magnetodisc returns to its initial size. This is represented by the \Rev{red} (compression) and \Rev{green} 
 (expansion) lines. The black dashed line indicates the point of maximum compression/expansion (at $t{=}\unit{t_o}$) 
 where we take a `snapshot' of model outputs in order to investigate the thermospheric response midway through the 
 transient pulse (henceforth, this phase of the event is referred to as `half-pulse'). \\
 
% FIG 2 here
 
 As in \citet{yates2012}, we divided the magnetosphere into four regions: region I, representing 
 open field lines of the polar cap; region II containing the closed field lines of the outer magnetosphere; 
 region III (shaded in figures) is the middle magnetosphere (magnetodisc) where we assume the Hill-Pontius 
 equation is valid for steady-state conditions. Region IV is the inner magnetosphere (which is assumed to be 
 fully corotating in steady state). Region III is our main region of interest throughout this study since it 
 plays a central role in determining the morphology of auroral currents.\\
 
% table 2 here
  
\begin{figure}
      \centering
      \includegraphics[width= 0.99\figwidth]{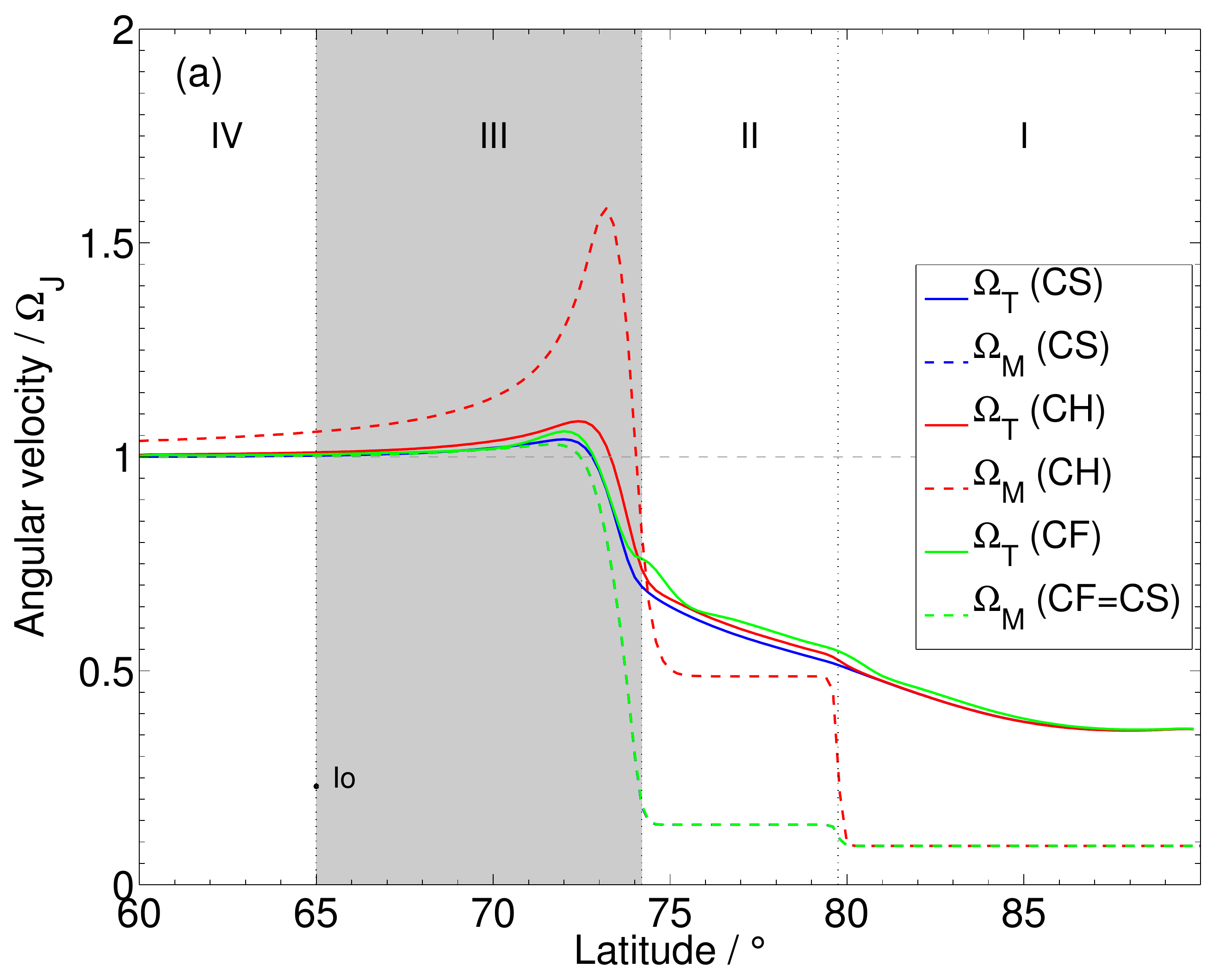}
      \includegraphics[width= 0.99\figwidth]{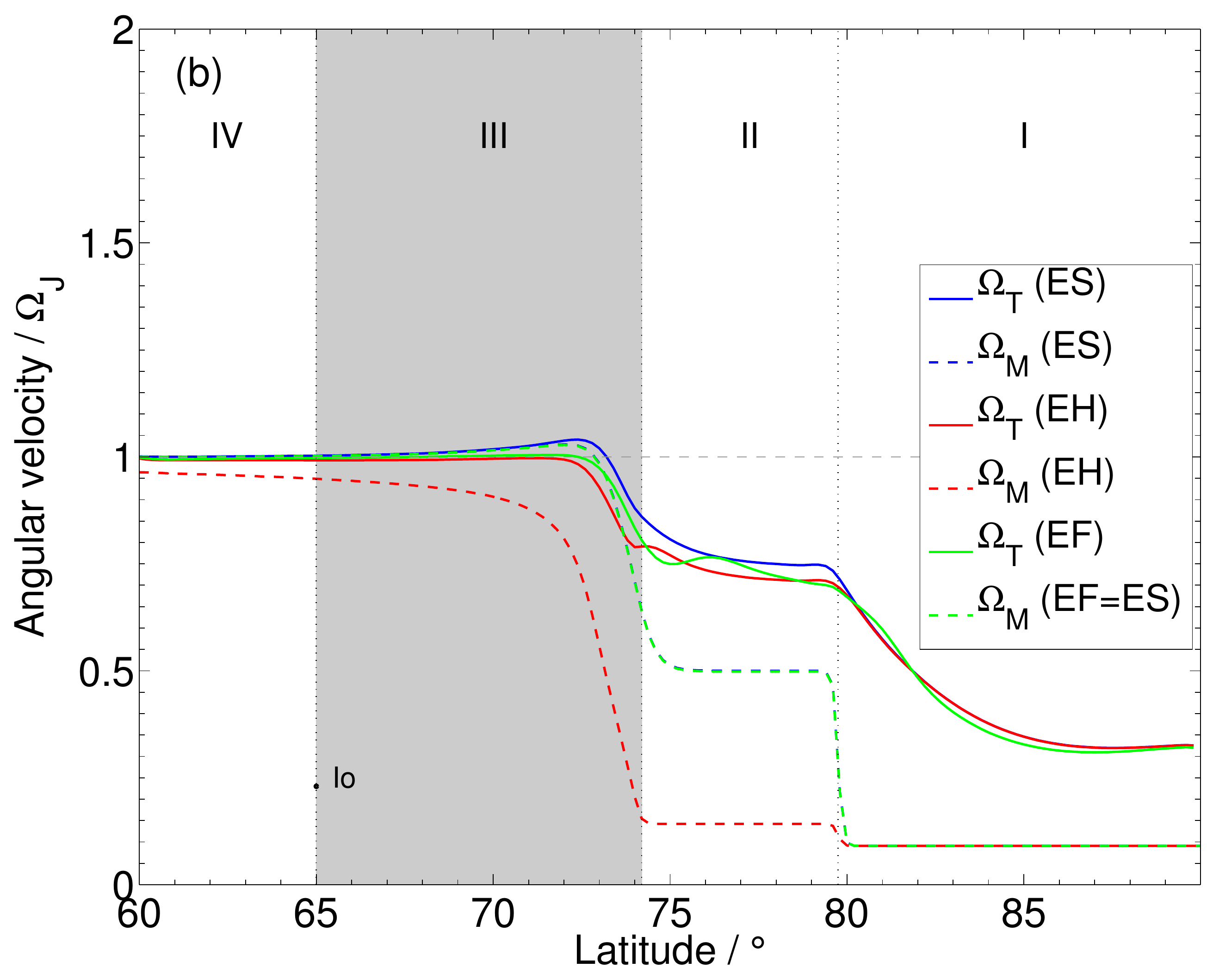}

      \caption{ (a) Thermospheric and plasma angular velocity profiles for the transient compression cases
               as a function of ionospheric latitude. \Rev{Solid} lines represent thermospheric profiles whilst 
               \Rev{dashed} lines represent plasma profiles. The \Rev{blue} lines represent case CS (steady state 
               before compression) whilst the \Rev{red} and \Rev{green} lines indicate cases CH (system at minimum 
               disc radius) and CF (system just returned to initial disc radius) respectively. The magnetospheric 
               regions (region III shaded) are labelled and separated by the black dotted lines. The magnetically 
               mapped location of Io on the ionosphere is marked and labelled. (b) Thermosphere and plasma angular 
               velocity profiles for the transient expansion cases as a function of ionospheric latitude. The line 
               styles are the same as (a) but the cases are now ES, EH and EF respectively, where `E' denotes expansion, 
               and the `S', `H' and `F' symbols represent the same phases of the event as for \Fig{\ref{fig:angvel}a}. 
%               \Rev{(c) and (d) respectively shows the perpendicular electric field as a function of ionospheric latitude 
%               for our compression and expansion scenarios. The colour code is as for (a), where blue lines represent 
%               case CS (or ES), red lines indicate case CH (or EH) and green lines represent case CF (or EF). 
%               Note that this model is axisymmetric.}}
			   }
      \label{fig:angvel}
  \end{figure}

 Plasma velocities are shown in \Fig{\ref{fig:angvel}a-b} (dashed lines) along with their 
 corresponding thermospheric angular velocities (solid lines). \Fig{\ref{fig:angvel}a} shows angular velocity 
 profiles pertaining to the transient compression scenario. The starting configuration (steady-state) 
 is indicated by \Rev{blue} lines and is henceforth, referred to as `case CS' (pre-Compression Steady-state). Halfway 
 through the pulse, when the magnetodisc radius is a minimum, angular velocity profiles are represented by \Rev{red} 
 lines and will be referred to as case CH (Compression Half-pulse). Case CF (Compression Full-pulse) profiles are 
 indicated by \Rev{green} lines and represent the state of the system after the pulse subsides 
 \Rev{(see \Table{\ref{tb:trancases}} for description of different cases)}. In \Fig{\ref{fig:angvel}a}, there is 
 significant super-corotation of the magnetodisc plasma throughout most of regions IV and III. Plasma rotating faster 
 than both the thermosphere and deep planet	creates a reversal of currents and angular momentum transfer between the 
 ionosphere and magnetosphere \citep{cowley07}. Thus angular momentum is transported from the magnetosphere to the 
 thermosphere, where rotation rate increases from its initial state. We see an average of ${\sim}\unitSI[3]{\%}$ 
 increase in peak \unit{\Omega_T} in response to the transient compression event. This is small compared to the 
 factor of two increase in peak \unit{\Omega_M} (for case CH). The significant difference in response between 
 the thermosphere and magnetosphere is due to the larger mass of the neutral thermosphere and thus, its greater 
 resistance to change (inertia). After the subsidence of the pulse, the magnetosphere returns 
 to its initial size and, thus, the \unit{\Omega_M} profile for case CF is equal to that of CS at all latitudes. The 
 same cannot be said for the thermospheric angular velocities; the CF thermosphere rotates slightly faster 
 (${\sim}\unitSI[2]{\%}$ at maximum \unit{\Omega_T}) for parts of regions III and I and all of region II. This comparison 
 highlights the difference in response between the thermosphere and magnetosphere to the prescribed changes in solar 
 wind pressure.\\
 
 \Fig{\ref{fig:angvel}b} shows angular velocity profiles corresponding to the transient expansion scenario. Like the 
 compression scenario, we have cases ES (pre-Expansion Steady-state (initial value of $\unit{R_{MM}{=}\unit[45]{R_J}}$)), 
 EH (Expansion Half-pulse $\unit{R_{MM}{=}\unit[85]{R_J}}$) and EF (Expansion Full-pulse) indicated by \Rev{blue, red 
 and green} lines respectively. 
 The behaviour is very different from the compression: 
 midway through the event (case EH), the magnetodisc plasma sub-corotates to an even greater degree in regions IV 
 and III compared to the initial steady-state case, ES. The thermosphere also sub-corotates to a greater degree, 
 but \Rev{maintains a} higher angular velocity than the disc plasma, meaning that current reversal does not occur. 
 Thermospheric angular velocities for cases ES and EF differ slightly, as in the compression scenario i.e due to 
 the greater lag in the thermospheric response time.\\
 
 \Fig{\ref{fig:angvel}} theoretically demonstrates the effect that transient shocks and rarefactions in the solar 
 wind have on both plasma and thermospheric angular velocities. Sections~\ref{sec:cmp} and \ref{sec:exp} 
 will discuss the effects on the \MI coupling currents and the global thermospheric dynamics.

%%%%%%%%%%%%%%%%%%%%%%%%%%%%%%%%%%%%%%%%%%%%%%%%%%%%%%%%%%%%%%%%%%%%%%
\section{Magnetospheric Compressions}
\label{sec:cmp}
  
  In this section we present findings for our transient magnetospheric compression scenario which 
  lasts for a total of three hours.
  
 \subsection{Auroral currents}
 \label{sec:current_cmp}

% FIG 3 here
 
\begin{figure}
      \centering
      \includegraphics[width= 0.99\figwidth]{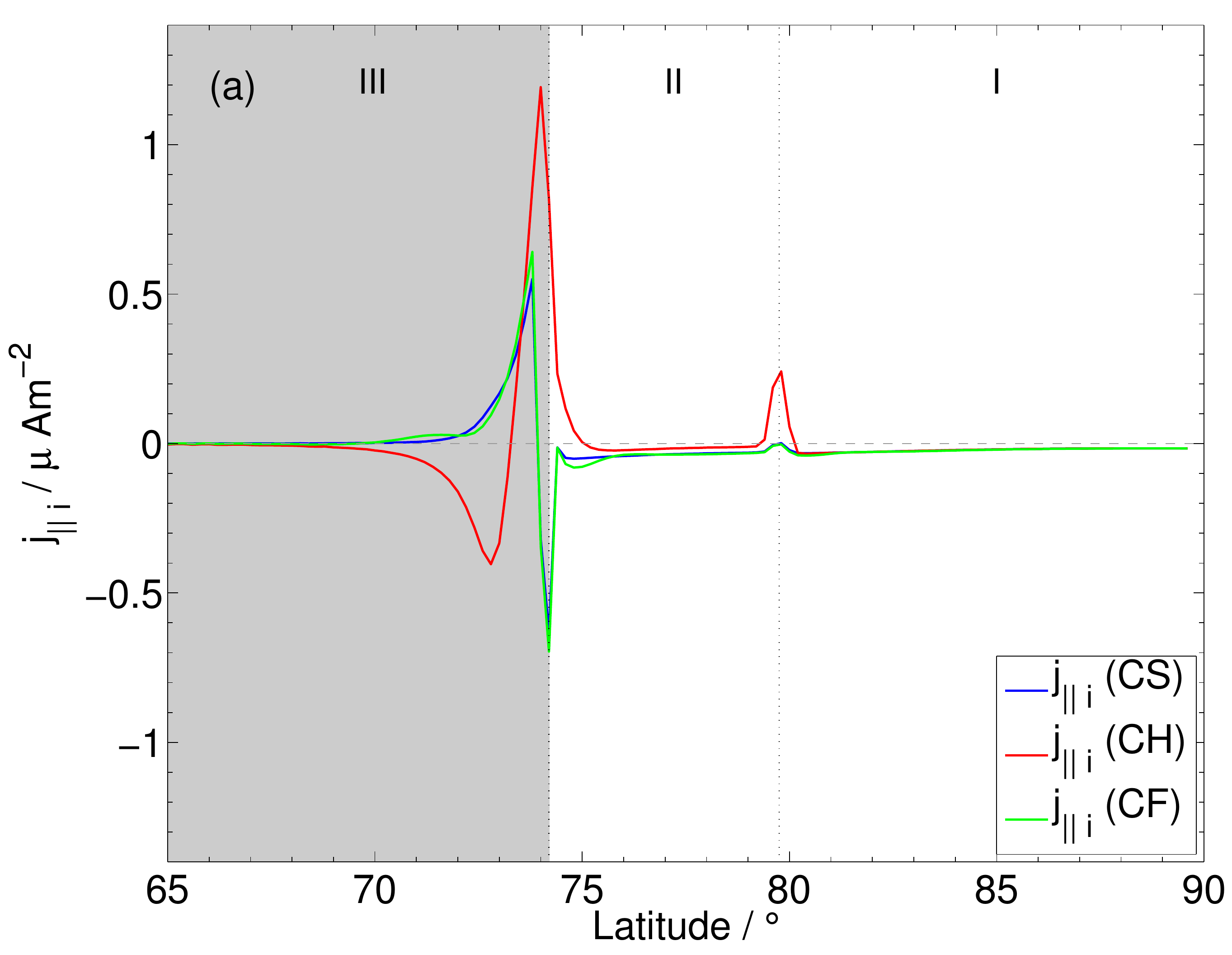}
      \includegraphics[width= 0.99\figwidth]{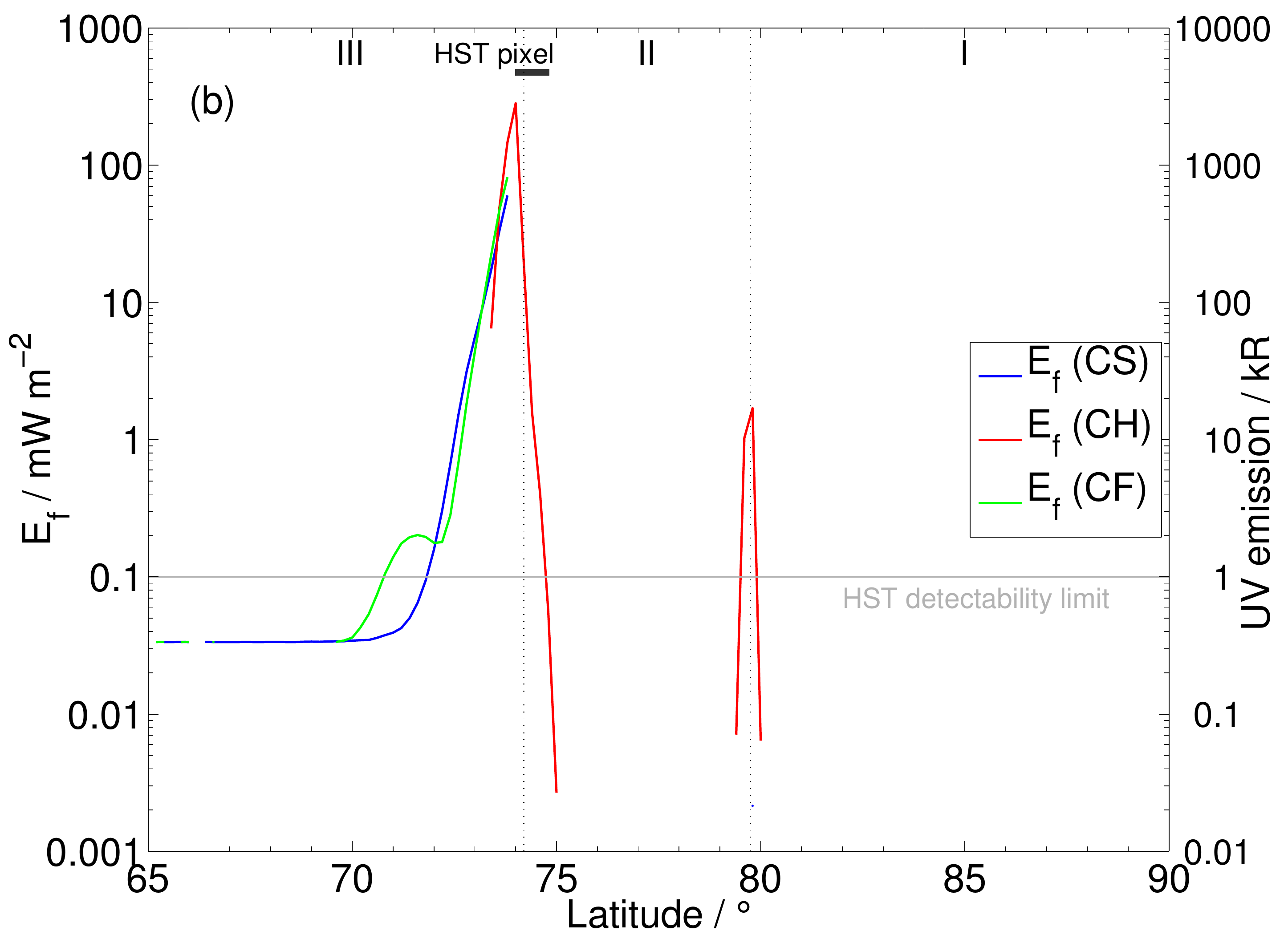}
      
      \caption{ (a) \FAC densities in the high latitude ionospheric region for our transient compression cases. 
      			The \Rev{blue} line represents case CS whilst the \Rev{red} and \Rev{green} lines indicate cases CH and 
      			CF respectively. The conjugate magnetospheric regions (region III is shaded) are separated by 
      			dotted black lines and labelled. 
                (b) Shows the latitudinal variation of the precipitating electron flux \Rev{(on the left axis) and 
                the corresponding UV auroral emission (on the right axis)} for the transient compression cases. 
                The colour codes and in plot labels are the same as (a). The latitudinal size of an ACS-SBC HST pixel 
                located near the main auroral emission is represented by the dark grey box. The solid grey line indicates 
                the limit of detectability of the HST \citep{cowley07}. 
      			}
      \label{fig:jpar_cmp}
 \end{figure}

% jpar cmp 
 \Fig{\ref{fig:jpar_cmp}\Rev{a}} shows \FAC density as a function of latitude (computed from the horizontal 
 divergence of \unit{I_P} (ionospheric Pedersen current density)\Rev{; see \Eq{\ref{eq:AIpedcur}}}) for 
 cases CS, CH and CF. \Rev{The blue line represents 
 case CS, whilst the red and green lines respectively show cases CH and CF}. \Rev{Both cases 
 CS and CF possess upward (positive) \FAC density (indicating downward moving electrons) peaking at 
 ${\sim}\unitSI[74]{^{\circ}}$, corresponding to the `main auroral oval'. Strong downward (negative) \FAC densities 
 are located at the region III/II boundary, indicating that electrons in this region are moving upwards along the 
 magnetic field lines. In regions II and I, \FAC density profiles remain slightly negative.} Peaks in upward \FAC arise 
 from strong spatial gradients in \unit{\Omega_M} (\unit{\Omega_M} decreases by ${\sim}\unitSI[78]{\%}$ across 
 ${\sim}\unitSI[2]{^{\circ}}$ \Rev{caused by the breakdown in corotation of magnetodisc plasma}), and consequently, 
 flow shears located at or near magnetospheric region boundaries. 
 Downward \FAC{}s at the region III boundary are also caused by large spatial gradients in \unit{\Omega_M} and to a 
 lesser extent the change in \unit{\Sigma_P} encountered as we traverse this boundary \citep{yates2012}. \Rev{The minor 
 differences between these two cases are attributed to the response of the thermosphere to the transient pulse. 
 At full-pulse, $\unit{\Omega_M}(CF){=}\unit{\Omega_M}(CS)$ but $\unit{\Omega_T}(CF){\neq}\unit{\Omega_T}(CS)$ as the 
 thermosphere has not had sufficient time to settle back to a steady-state (due to its large inertia, as discussed in 
 section~\ref{sec:tran_plas}). Although this is a subtle example of the atmospheric modulation of auroral currents, 
 future simulations will aim at further exploring how this effect changes within the parameter space of the pulse 
 duration and its change in solar wind pressure.}\\

 Case CH shows the largest deviation from steady state. Its \FAC density profile is directed downwards at 
 latitudes up to ${\sim}\unitSI[73]{^{\circ}}$. \Rev{This 
 current reversal (compared to case CS) is due to a negative flow shear (\unit{\Omega_T - \Omega_M}) caused by the 
 significant super-corotation of the magnetosphere compared to the thermosphere (see section~\ref{sec:tran_plas} 
 and \citet{cowley07}). Poleward of the main downward current region, the \FAC density remains negative except for two 
 locations:} \\
 
 \Rev{\emph{i)} the 'main auroral oval': where a peak upward \FAC density of ${\sim}\unitSI[1.2]{\mu\,A\,m^{-2}}$ (a 
 factor-of-two increase compared to case CS) is due to magnetodisc plasma transitioning from a super-corotational 
 state to a significantly sub-corotational state.}\\
 
 \Rev{\emph{ii)} the region II/I (open-closed field line) boundary: with an upward \FAC density peak of 
 ${\sim}\unitSI[0.2]{\mu\,A\,m^{-2}}$ caused by the differing \unit{\Omega_M} in these two regions. In region II 
 \unit{\Omega_M} is fixed at a value depending on magnetodisc size \citep{cowley05}. In region I, we set 
 $\unit{\Omega_M}=\unitSI[0.10]{\Omega_J}$, for all cases, in accordance with the formula of \citet{isbell1984}.}\\

 We briefly compare \FAC densities from case CH with transient results from \citet{cowley07} (compression from 
 \unitSI[85{-}45]{R_J}). Despite a resemblance in \FAC profiles, upward FACs in the magnetodisc (region III) are 
 ${\sim}\unitSI[2.5]{}$ times larger in case CH than the equivalent case (with a responsive thermosphere) in 
 \citet{cowley07}. FACs in case CH are actually closer to those in \citet{cowley07}'s 
 non-responsive thermosphere compression case. This suggests that, the thermosphere (represented by a \GCM) in 
 our study lies somewhere in between a responsive and non-responsive thermosphere (although closer to the 
 latter, for the pulse parameters assumed).\\ 
 
% e_f cmp
 Corresponding precipitating electron energy fluxes are shown in \Fig{\ref{fig:jpar_cmp}b}. These fluxes are plotted 
 as a function of latitude and obtained using \Eq{\ref{eq:eflux}} and \Table{\ref{tb:aur_parameters}}, which uses only 
 the upward (positive) \FAC densities presented in \Fig{\ref{fig:jpar_cmp}a}. The line style code and labels are the 
 same as in \Fig{\ref{fig:jpar_cmp}a}. \Rev{The 
 latitudinal size of a Hubble Space Telescope (HST) ACS-SBC pixel (\unitSI[0.03\mbox{x}0.03]{arc\,sec}) is represented 
 by the dark grey rectangle (assuming that the magnetic axis of the Jovian dipole is perpendicular to the observer's line 
 of sight) and the grey solid line indicates the limit of present detectability with HST instrumentation 
 (${\sim}\unitSI[1]{kR}$; \citet{cowley07})}. We initially compare electron energy fluxes for 
 cases CS and CF. These profiles are \Rev{non-existent} poleward of ${\sim}\unitSI[74]{^{\circ}}$; equatorward of this 
 location, case CF shows little deviation from CS, except \Rev{that caused by} the thermospheric lag discussed \Rev{above}. 
 In region III, we find that the peak energy flux for case CF is ${\sim}\unitSI[35]{\%}$ larger than that in case CS and 
 the location of these peaks coincide with the location of the main auroral oval (${\sim}\unitSI[74]{^{\circ}}$). The 
 slight increase in peak energy flux is due to a relative increase in flow shear as seen in \Fig{\ref{fig:angvel}a}. 
 Case CF would therefore produce main oval emission approximately ${\sim}\unitSI[200]{kR}$ brighter than that of CS 
 as indicated by the right axis in the Figure (assuming that \unitSI[1]{mW\,m^{-2}} of precipitation creates 
 ${\sim}\unitSI[10]{kR}$ of UV output \citep{cowley07}). \\
 
 The \unit{E_f} profile for case CH is different from those of both cases CS and CF. There are 
 three main changes in CH compared to CS: \emph{i)} peak energy flux in region III is ${\sim}\unitSI[280]{mW\,m^{-2}}$, 
 almost a factor of five larger, \emph{ii)} location of peak energy flux has shifted polewards by 
 ${\sim}\unitSI[0.2]{^{\circ}}$ and, \emph{iii)} presence of a second peak with an energy flux of \unitSI[1.7]{mW\,m^{-2}} 
 at the region II/I boundary. The large increase in electron energy flux is caused by a substantial increase in flow 
 shear between the thermosphere and magnetosphere, resulting from the super-corotation of the magnetodisc plasma (see 
 \Fig{\ref{fig:angvel}}). The 
 presence of a second upward \FAC region at the region II/I boundary is also due to flow 
 shear increase across the boundary, as the magnetosphere in region II corotates at a larger fraction of \unit{\Omega_J} 
 compared to case CS. The result for this higher-latitude boundary should be regarded as preliminary, since it is sensitive 
 to the values of \unit{\Omega_M} we assume in the outer magnetospheric region and polar cap. Flow velocities in these regions 
 are poorly constrained, with few observations \citep{stallard2003}.  \Rev{The increase in \unit{E_f} for case CH would lead to 
 corresponding increases in auroral emission. As such, we would expect `main oval' emission for case CH to shift polewards 
 by ${\sim}\unitSI[0.2]{^{\circ}}$ and be ${\sim}\unitSI[4.7]{{\times}}$ larger than emission in case CS i.e. 
 ${\sim}\unitSI[2800]{kR}$ compared to ${\sim}\unitSI[600]{kR}$}.\\
 
 Comparing the energy flux profile of case CH with the equivalent case in \citet{cowley07}, we see that in the closed field 
 regions (III and II), peak energy fluxes are two orders of magnitude larger in case CH. This demonstrates the differences 
 between using a \GCM to represent the thermosphere and using a simple `slippage' relation between thermospheric and 
 magnetospheric angular velocities. At the open-closed field line boundary (II/I boundary) our peak flux is an order of 
 magnitude smaller than that in \citet{cowley07}; this difference arises from the different models used to represent the 
 outer magnetosphere. The outer magnetosphere (region II) and open field line region (region I) in this study is modelled 
 using plasma angular velocities from \citet{cowley05}.
 
 \subsection{Thermospheric dynamics}
 \label{sec:dyn_cmp}

 In this section, we discuss the thermospheric response to the simulated transient magnetospheric 
 compressions. \\
 
% FIG 4 here
 
\begin{figure}
      \centering
      \includegraphics[width= 0.65\figwidth]{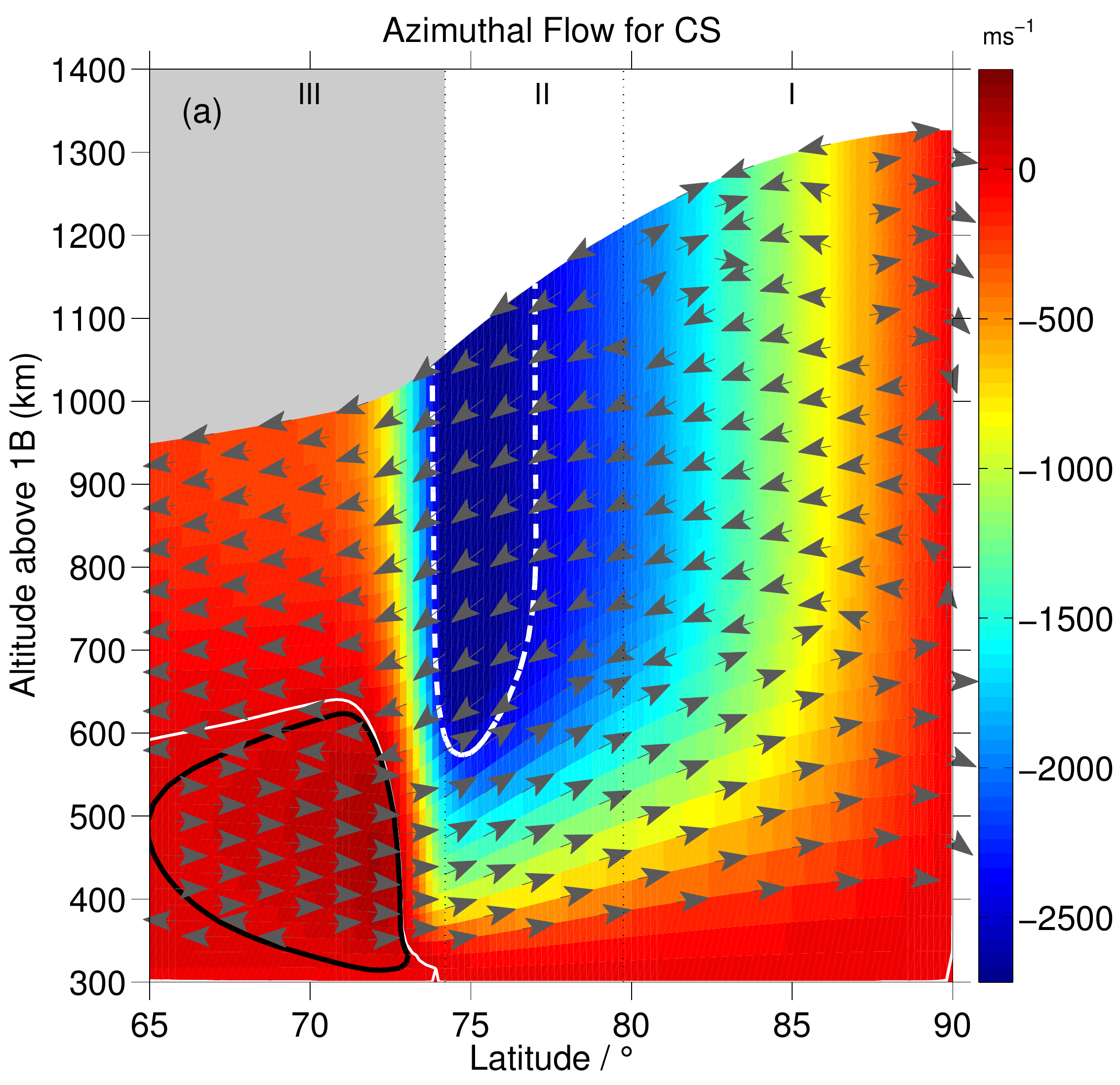}
      \includegraphics[width= 0.65\figwidth]{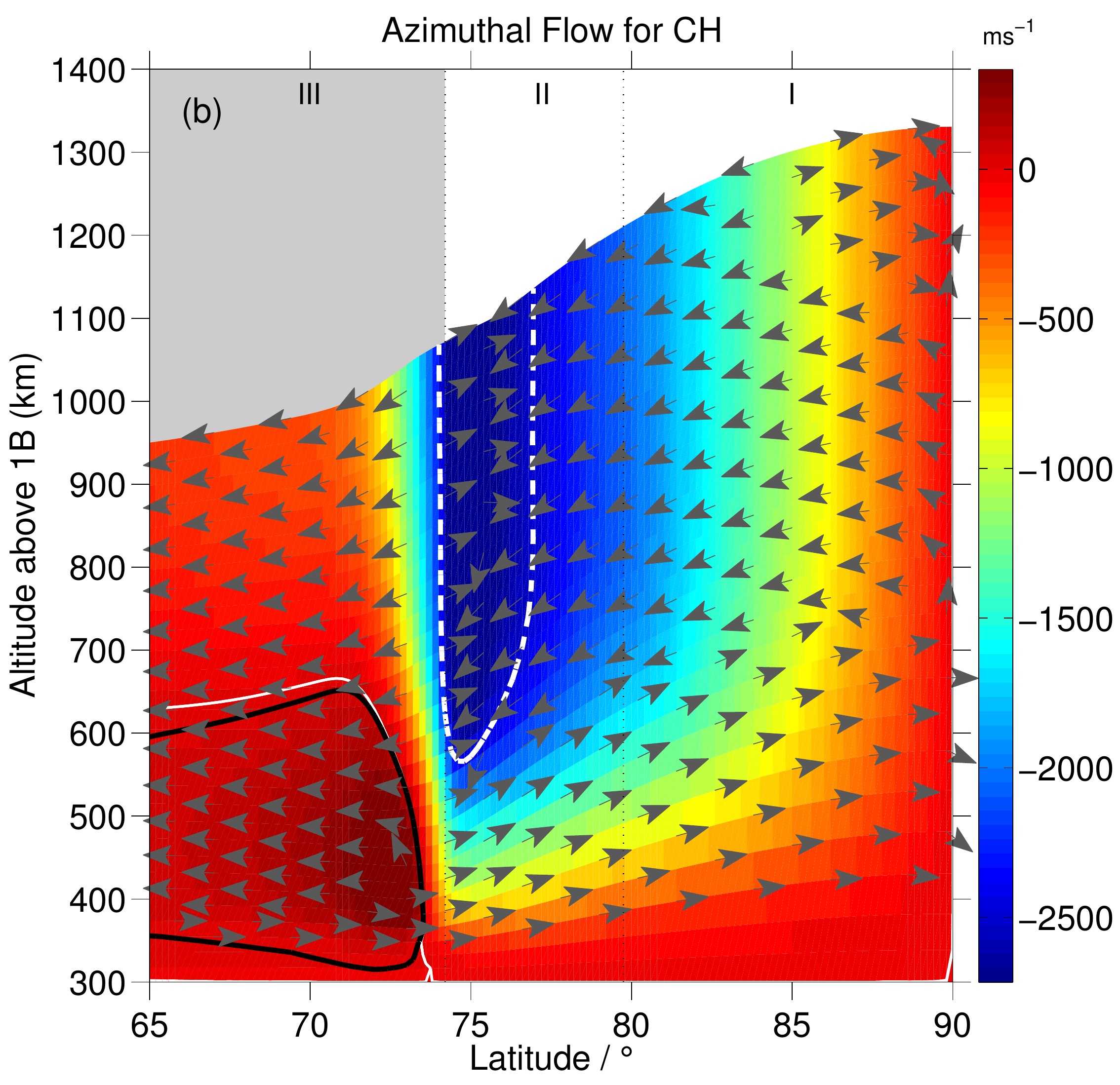}
      \includegraphics[width= 0.65\figwidth]{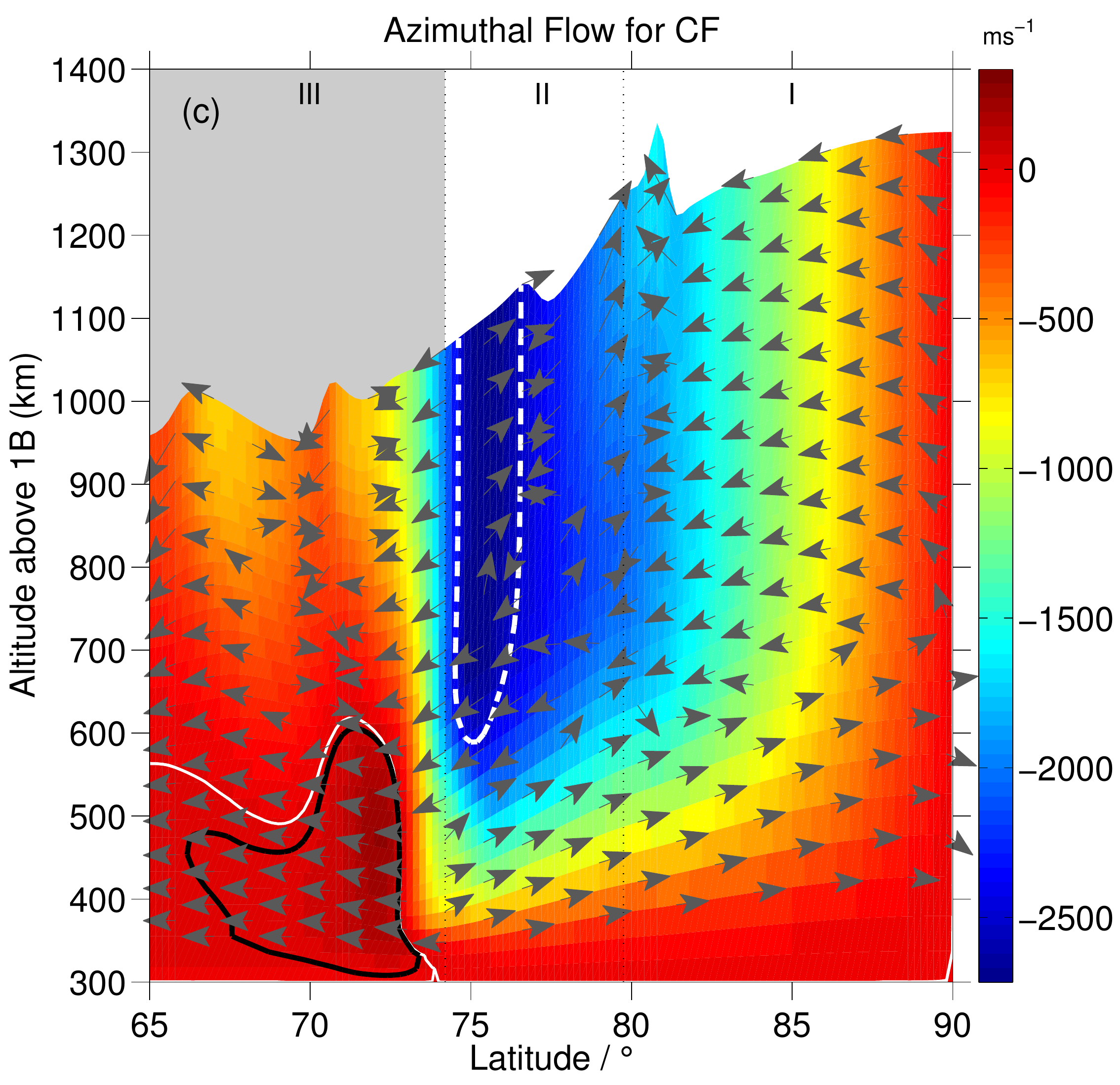} \\
      
      \includegraphics[width= 0.65\figwidth]{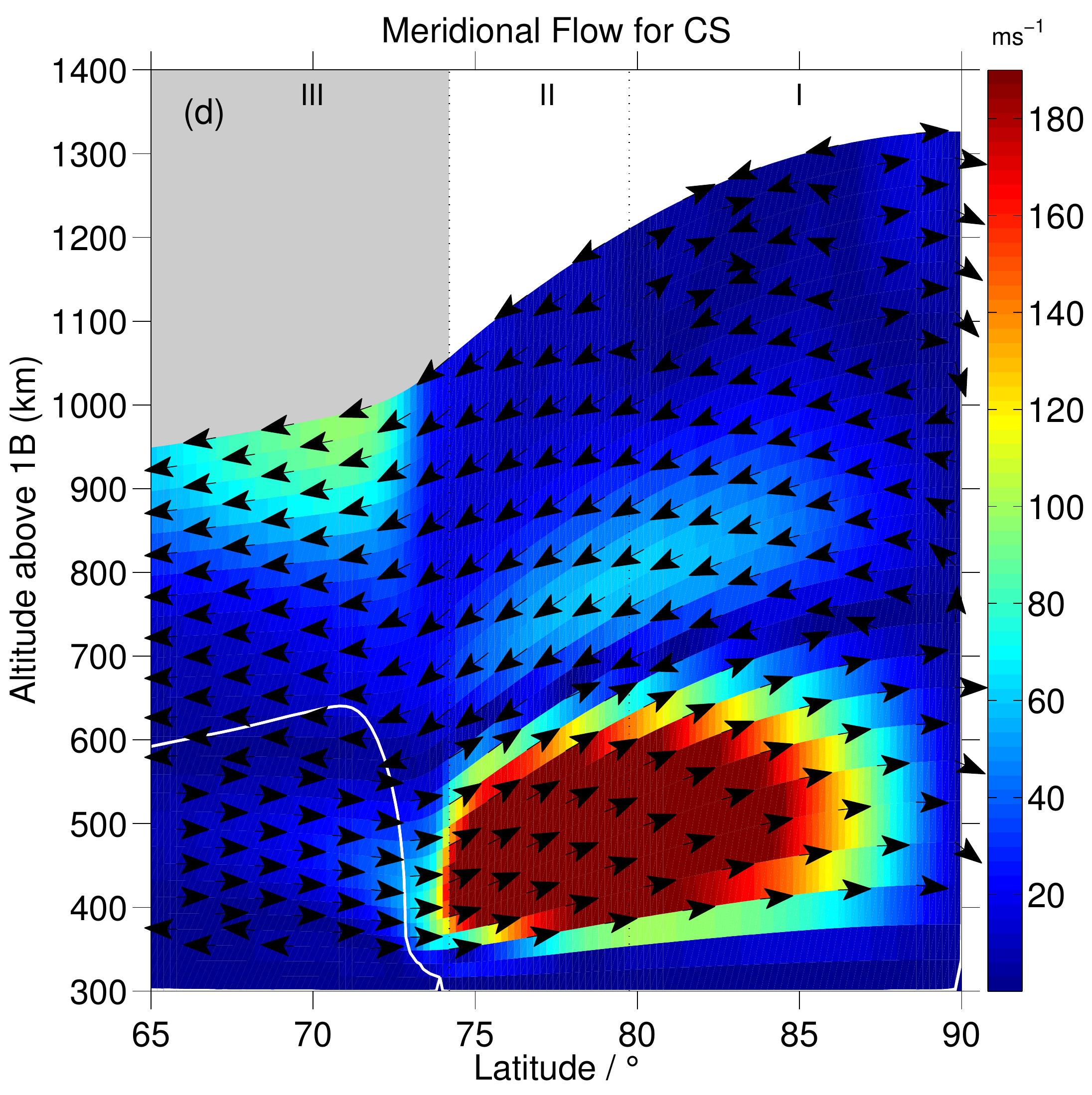}
      \includegraphics[width= 0.65\figwidth]{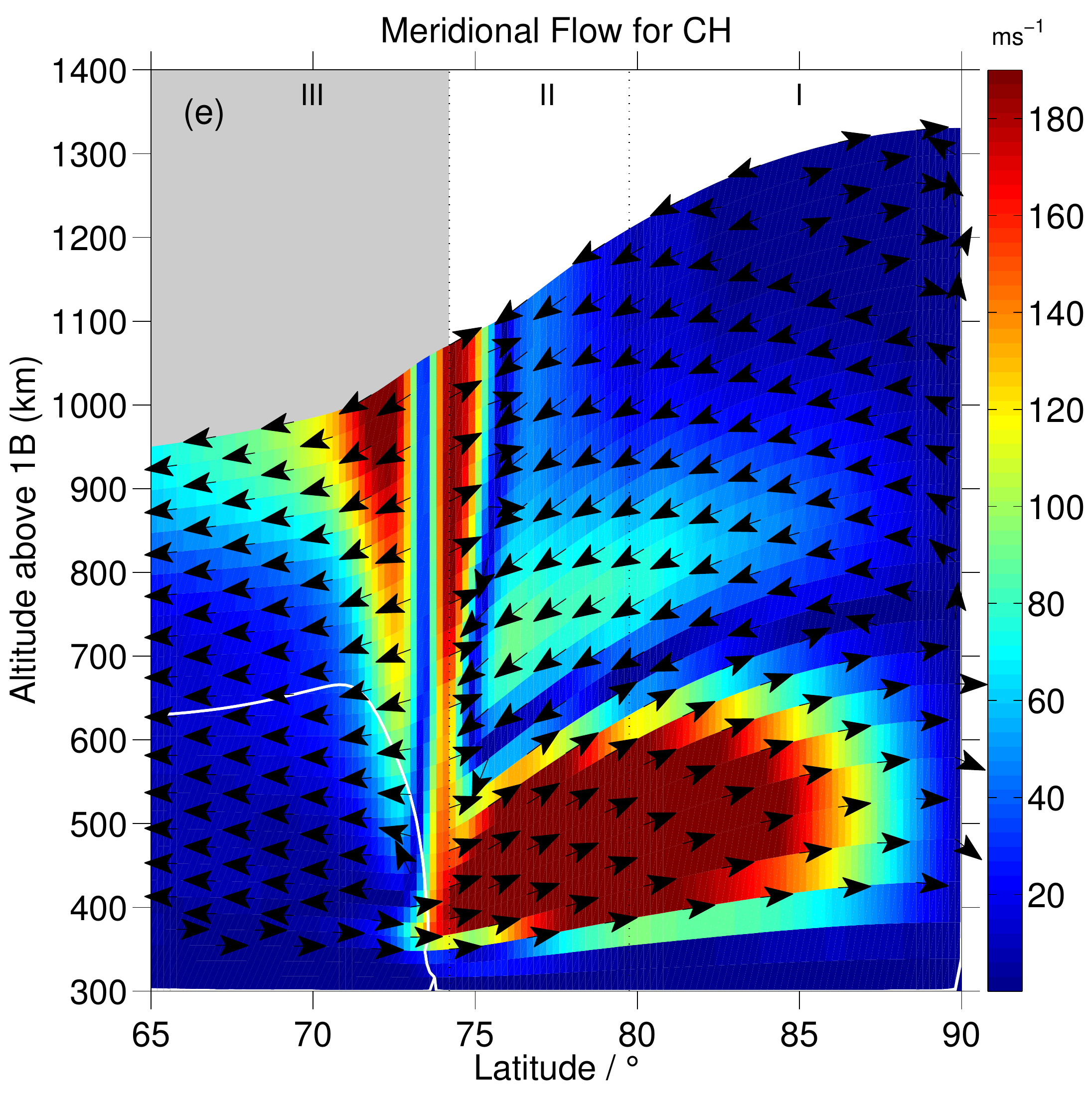}
      \includegraphics[width= 0.65\figwidth]{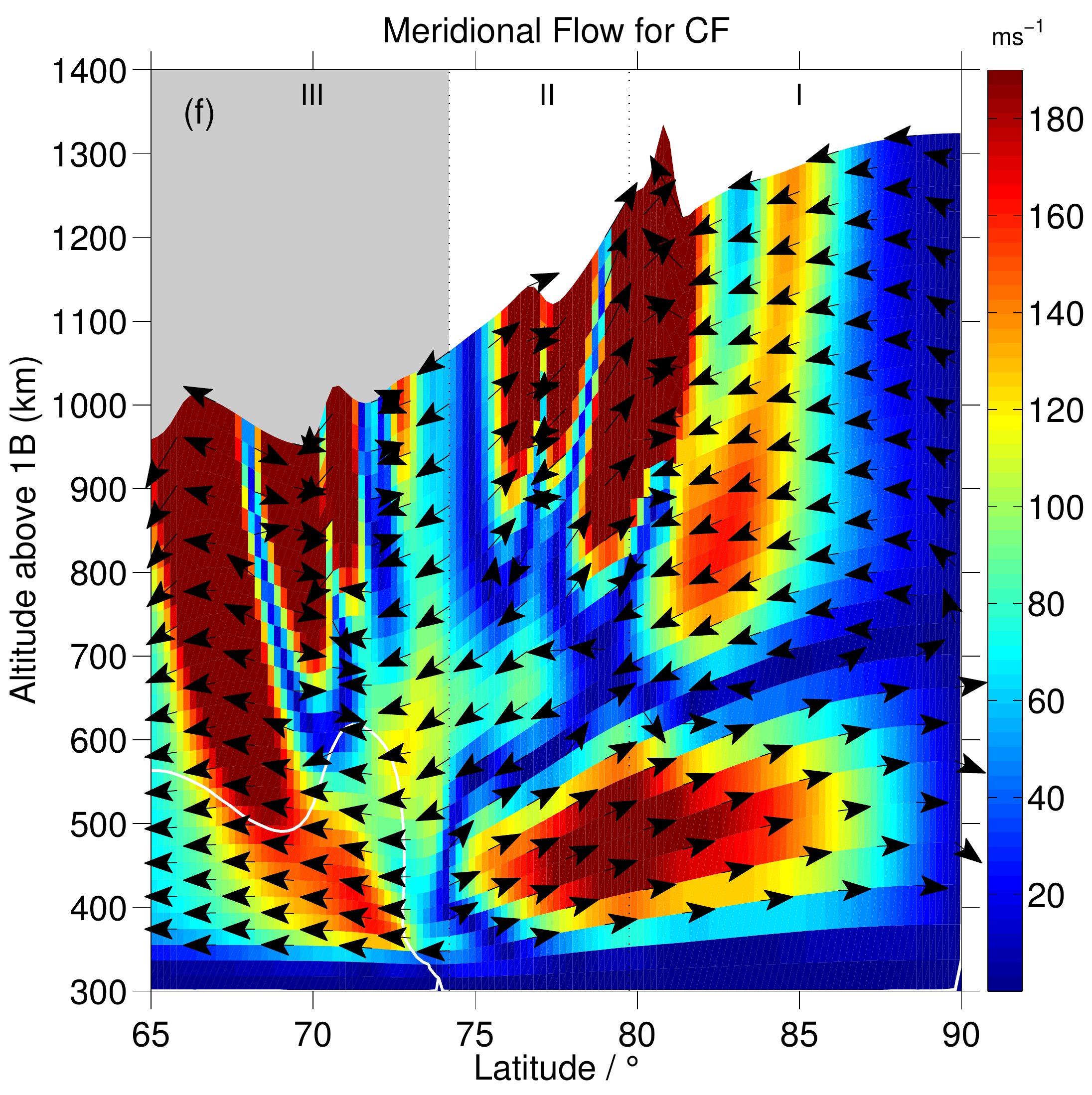} \\
      
      \includegraphics[width= 0.65\figwidth]{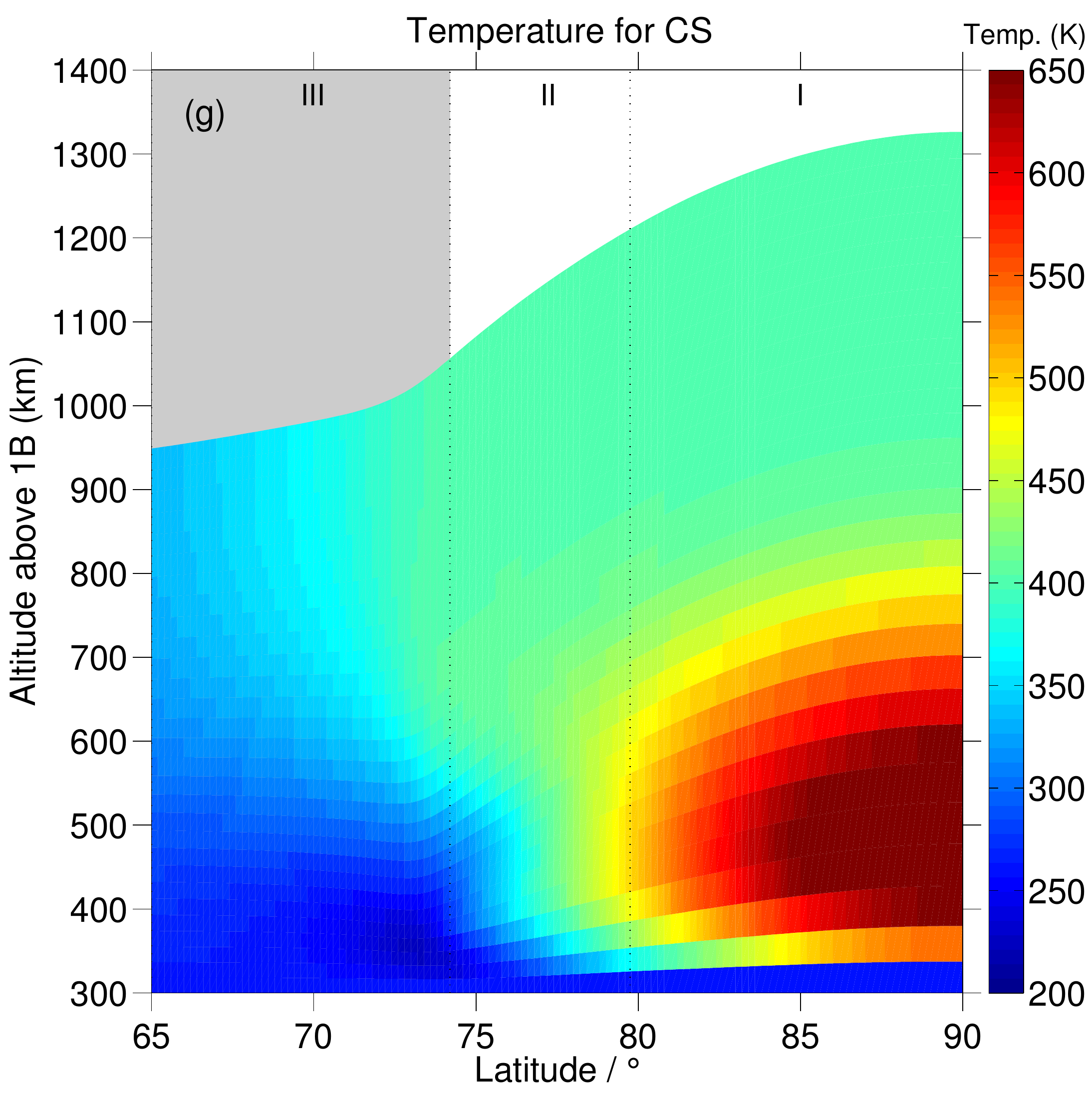}
      \includegraphics[width= 0.65\figwidth]{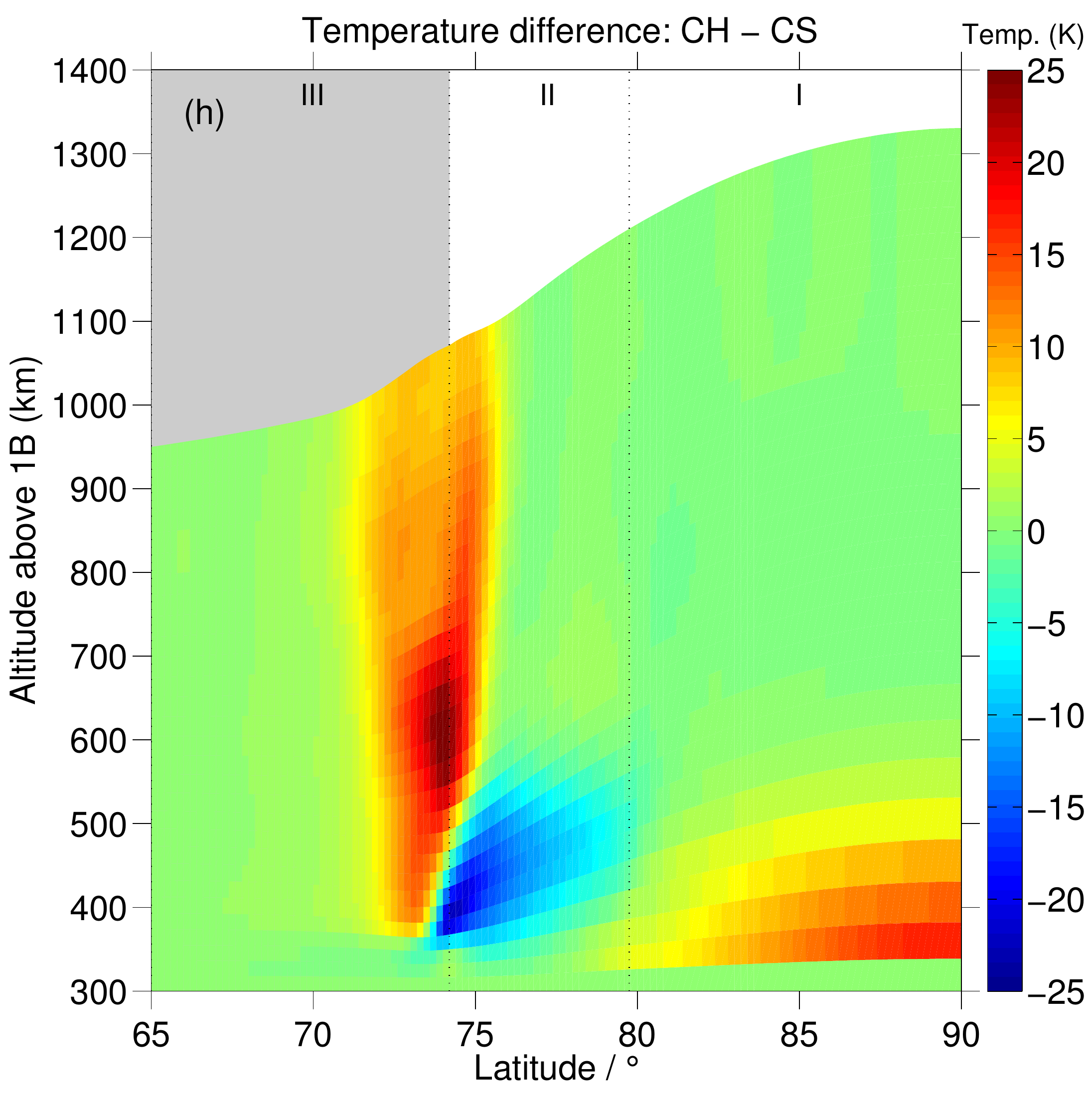}
      \includegraphics[width= 0.65\figwidth]{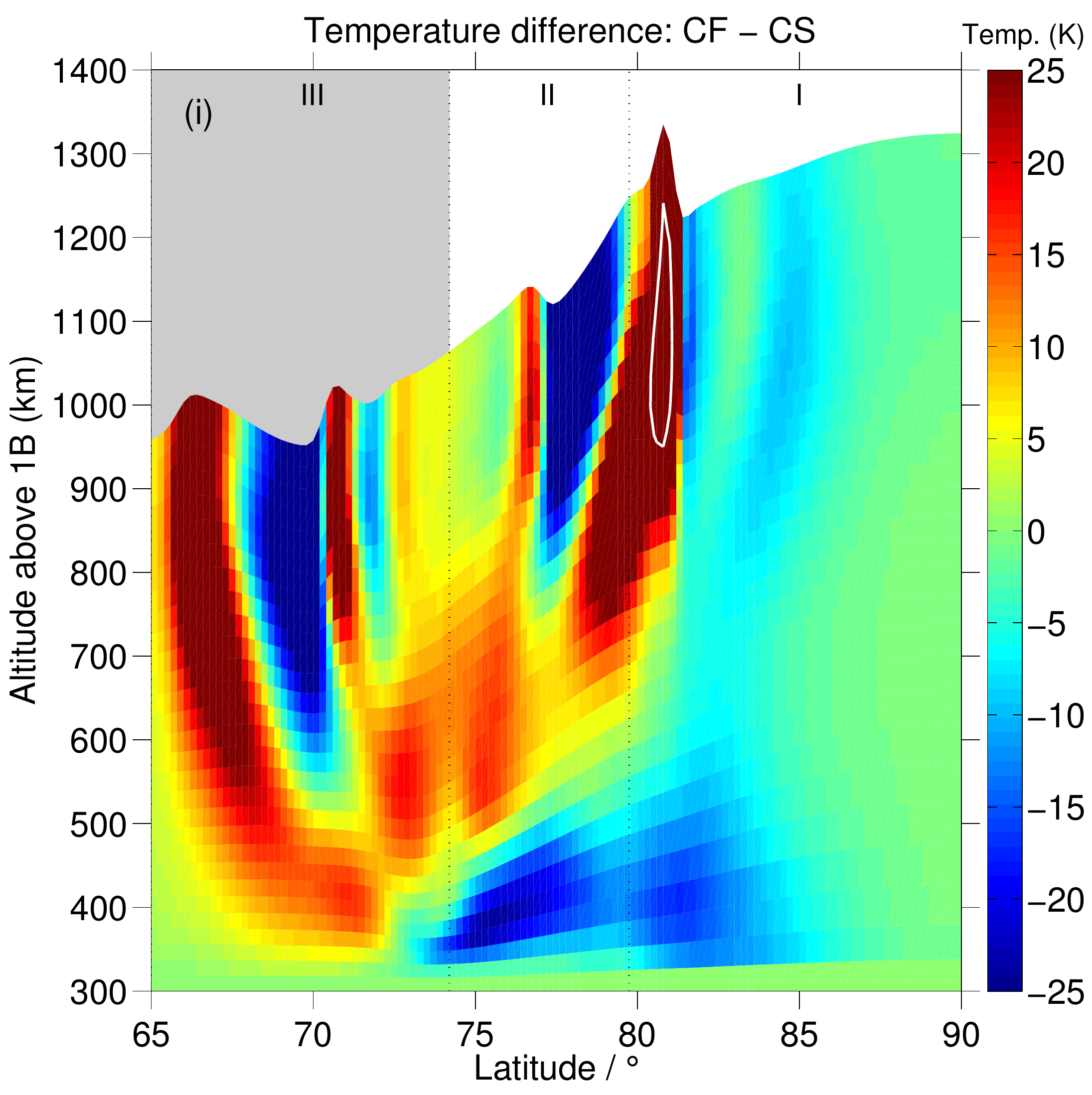} 
      
      \caption{ (a)-(c) show the variation of thermospheric azimuthal velocity (colour scale) in the 
      			corotating reference frame for cases CS-CF respectively (left to right). Positive values (dark red) 
      			indicate super-corotation, whilst negative values (light red to blue) indicate 
      			sub-corotation. The arrows show the direction of meridional flow and the solid white 
      			line indicates the locus of rigid corotation. The solid black encloses regions of super-corotation 
      			($>\unitSI[25]{m\,s^{-1}}$) and the dashed white line encloses regions that are sub-corotating at a rate 
      			$<\unitSI[-2500]{m\,s^{-1}}$. The magnetospheric regions 
      			(region III is shaded) are separated by dotted black lines and labelled. (d)-(f) show the 
      			meridional velocity in the thermosphere for cases CS-CF. The colour scale indicates 
      			the speed of flows. All other labels and are as for (a)-(c). (g) shows the thermospheric 
      			temperature distributions for case CS whilst (h)-(i) show the temperature difference 
      			of cases CH and CF with case CS. \Rev{The white contour encloses temperature differences 
      			$>\unitSI[100]{K}$.} All labels are as in (a)-(c).
      		}

      \label{fig:thvel_cmp}
 \end{figure}

 \Figs{\ref{fig:thvel_cmp}a-c} show the variation of thermospheric azimuthal velocity (in the corotating 
 reference frame) in the high latitude thermosphere for cases CS-CF respectively. Positive (resp. negative) values of 
 azimuthal velocity indicate super (resp. sub)-corotating regions. The direction of meridional flow is 
 indicated by the black arrows, the locus of rigid corotation is indicated by the solid white line, strong super-corotation 
 ($<\unitSI[25]{m\,s^{-1}}$) is indicated by the black contour, strong sub-corotation ($>\unitSI[-2500]{m\,s^{-1}}$) 
 is indicated by the dashed white contour. Magnetospheric regions are labelled and separated by black dotted 
 lines. Zonally, there are two prominent features in our transient compression cases: 
 
 \Rev{\emph{i)} a low altitude small super-corotating jet, centred at ${\sim}\unitSI[72]{^{\circ}}$. In 
 case CS, this jet is created by a small excess in the zonal Coriolis and advection momentum terms compared to 
 the ion drag term. At low altitude, the Coriolis force is primarily directed eastwards and unopposed can promote 
 super-corotation in the neutrals \citep{smith09,yates2012}.}
 
 \Rev{\emph{ii)} a large sub-corotating jet, from region III - I (blue region in \Figs{\ref{fig:thvel_cmp}a-c}). 
 This sub-corotational jet is caused by the drag of the sub-corotating magnetosphere on the thermosphere. Zonal 
 flows in this region are generally sub-corotational and acceleration terms are balanced in case CS, as the 
 thermosphere is in steady-state.} \\
 
 \Figs{\ref{fig:thvel_cmp}d-f} shows the variation of meridional flows in the high latitude thermosphere for our 
 transient compression cases. Magnetospheric labels, locus of corotation and arrows are the same as in 
 \Figs{\ref{fig:thvel_cmp}a-c}. These figures show the meridional flow patterns in the thermosphere, as well as 
 localised accelerated regions (red/brown hues). \Rev{In steady state, flow patterns are as described by 
 \citet{smith2007Natur}, \citet{smith09} and \citet{yates2012} - where at:}

 \emph{i) Low-altitude (${<}\unitSI[600]{km}$)}, ion drag acceleration becomes strong due to the Pedersen 
 conductivity layer (maximum value of \unitSI[0.1163]{mho\,m^{-1}} at ${\sim}\unitSI[370]{km}$). An imbalance is 
 created between ion drag, Coriolis and pressure gradient terms; thus, giving rise to advection of momentum, which 
 restores equilibrium in this low altitude region. This results in mostly sub-corotational, poleward accelerated flow 
 as shown \Rev{by the black arrows and brown hues} in \Rev{\Fig{\ref{fig:thvel_cmp}d}}.
 
 \emph{ii) High-altitude (${>}\unitSI[600]{km}$)}, conditions are quite different, meridional 
 Coriolis and pressure gradient accelerations are essentially balanced, whilst terms such as ion drag, advection 
 and zonal Coriolis are small and insignificant. This creates a `jovistrophic' condition, whereby flow 
 is directed very slightly equatorwards and is sub-corotational \Rev{(see black arrows in \Fig{\ref{fig:thvel_cmp}d})}.\\
 
 \Rev{The above descriptions of zonal and meridional flow patterns pertain to steady state conditions 
 (\Figs{\ref{fig:thvel_cmp}a and d respectively}). Zonal flows for case CH (\Fig{\ref{fig:thvel_cmp}b}) show 
 little change from the steady state zonal flow patterns described above. The main differences lie in the magnitude 
 of the velocities; velocity in the super-corotational jet doubles and the magnitude of azimuthal velocity has decreased by 
 ${\sim}\unit[4]{\%}$ in the sub-corotational jet. Meridional flows in \Fig{\ref{fig:thvel_cmp}e} show two additional 
 local acceleration regions either side of ${\sim}\unitSI[73]{^{\circ}}$ latitude and from altitudes ${>}\unitSI[500]{km}$.} \\
 
 \Rev{In addition, low altitude flow in region III is now directed purely equatorward. This is in marked contrast to the 
 steady state flow patterns. All the changes in flows discussed for 
 case CH result from the super-corotation of the magnetosphere which causes a reversal in the coupling current, 
 subsequently leading to a change in the sign of ion drag momentum terms in region III (see \Fig{\ref{fig:merid_cmp}} 
 and corresponding discussion).}\\
 
 \Rev{Zonal and meridional flows of case CF are respectively show in \Fig{\ref{fig:thvel_cmp}c and f}. The overall 
 flow patterns are as described above for case CH: i) a large sub-corotational jet combined with low altitude poleward 
 flows and high altitude equatorward flows in regions II and I, and ii) a low small low altitude super-corotational jet 
 combined with equatorward flows in region III. However, the degree and spatial extent of super-corotation has decreased 
 and a number of local accelerated regions exist where the direction of meridional flow changes on relatively small 
 spatial scales. These complex flow patterns result from the highly perturbed nature of the case CF thermosphere and 
 the imbalance between ion drag, Coriolis, pressure gradients and advection of momentum terms. }
  
\subsection{Thermospheric heating}
\label{sec:heat_cmp}

\begin{figure}
      \centering
      \includegraphics[width= 0.65\figwidth]{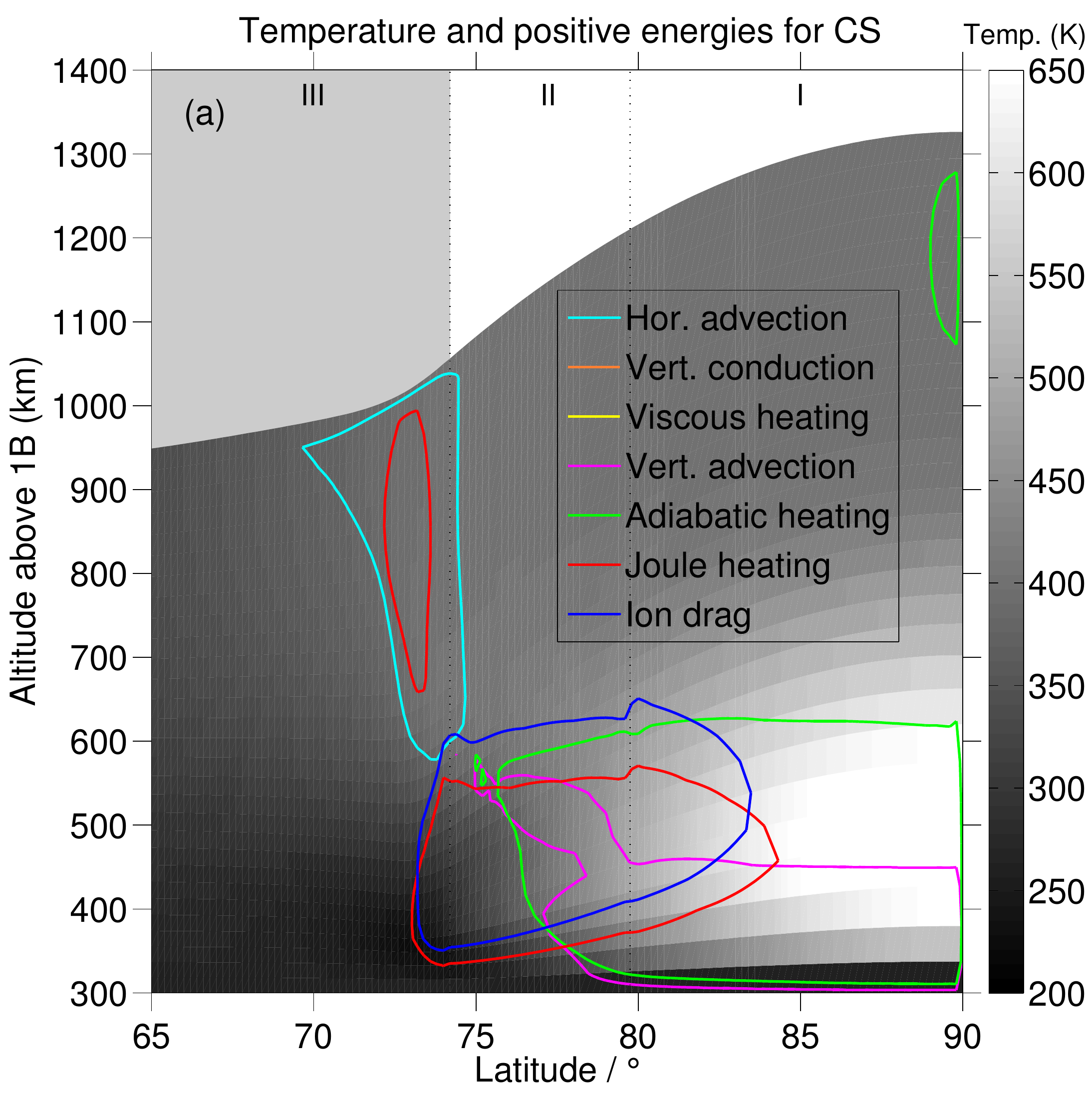} %width=99\figwidth
      \includegraphics[width= 0.65\figwidth]{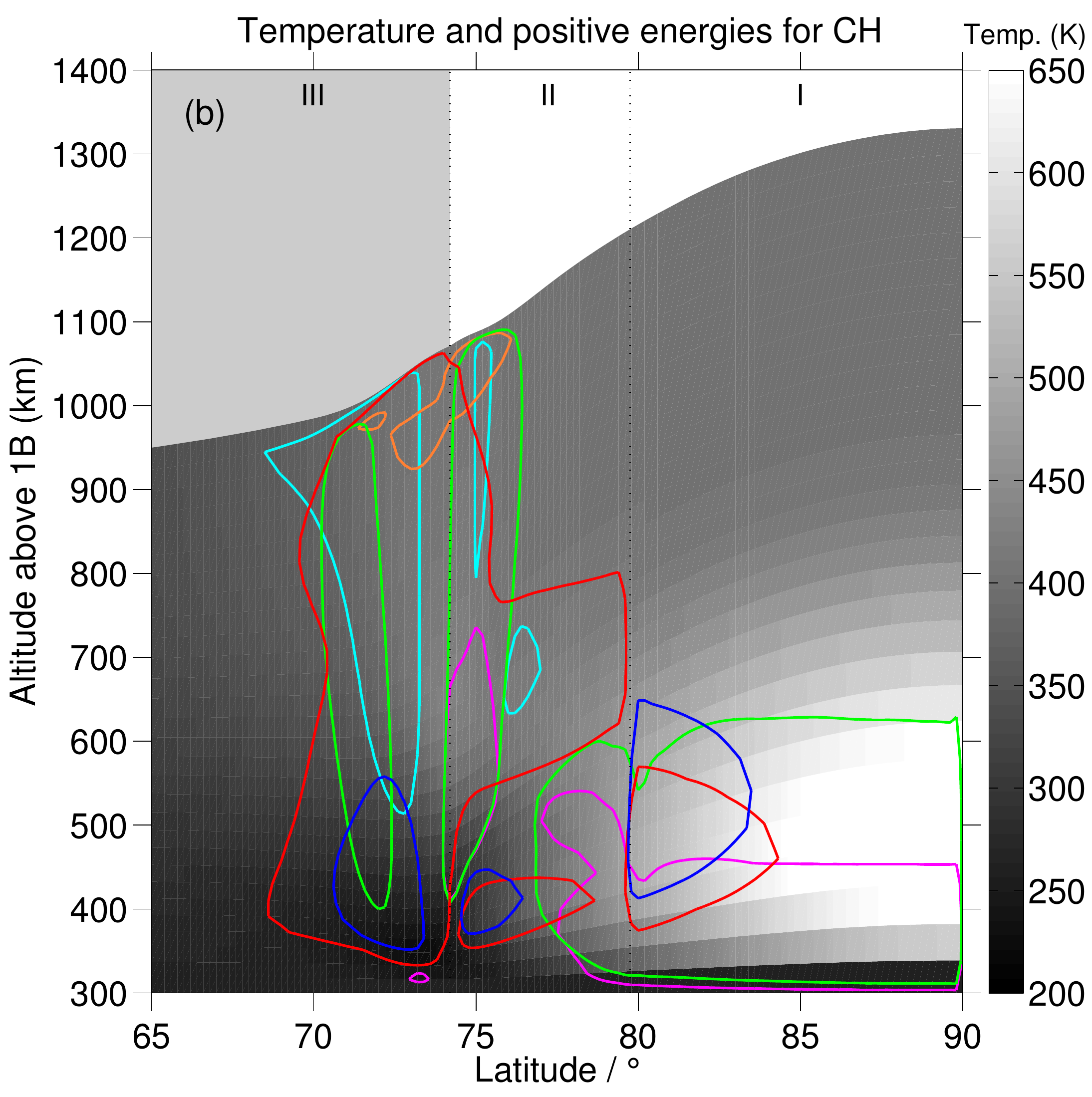}
      \includegraphics[width= 0.65\figwidth]{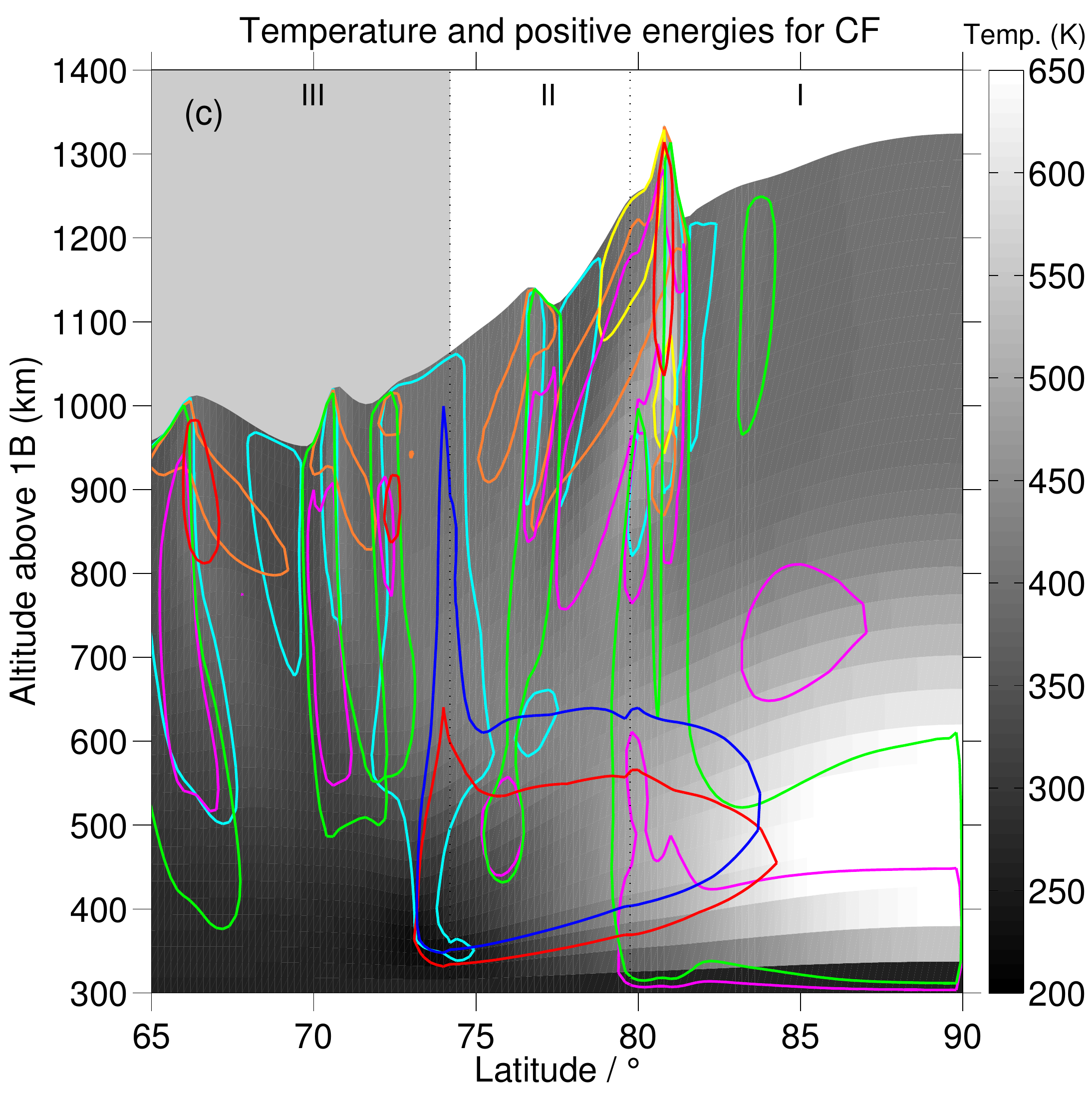} \\
      
      \includegraphics[width= 0.65\figwidth]{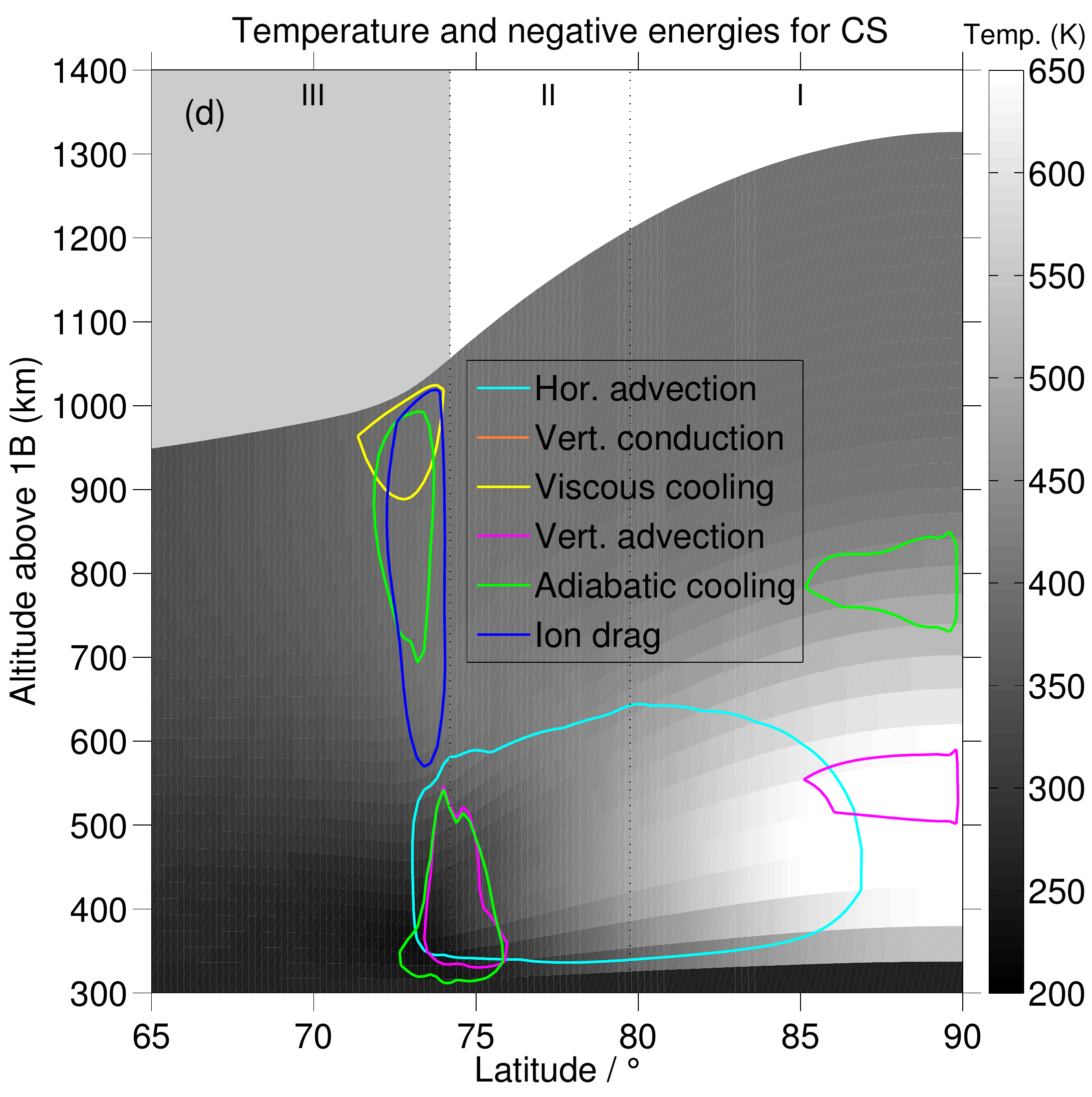}
      \includegraphics[width= 0.65\figwidth]{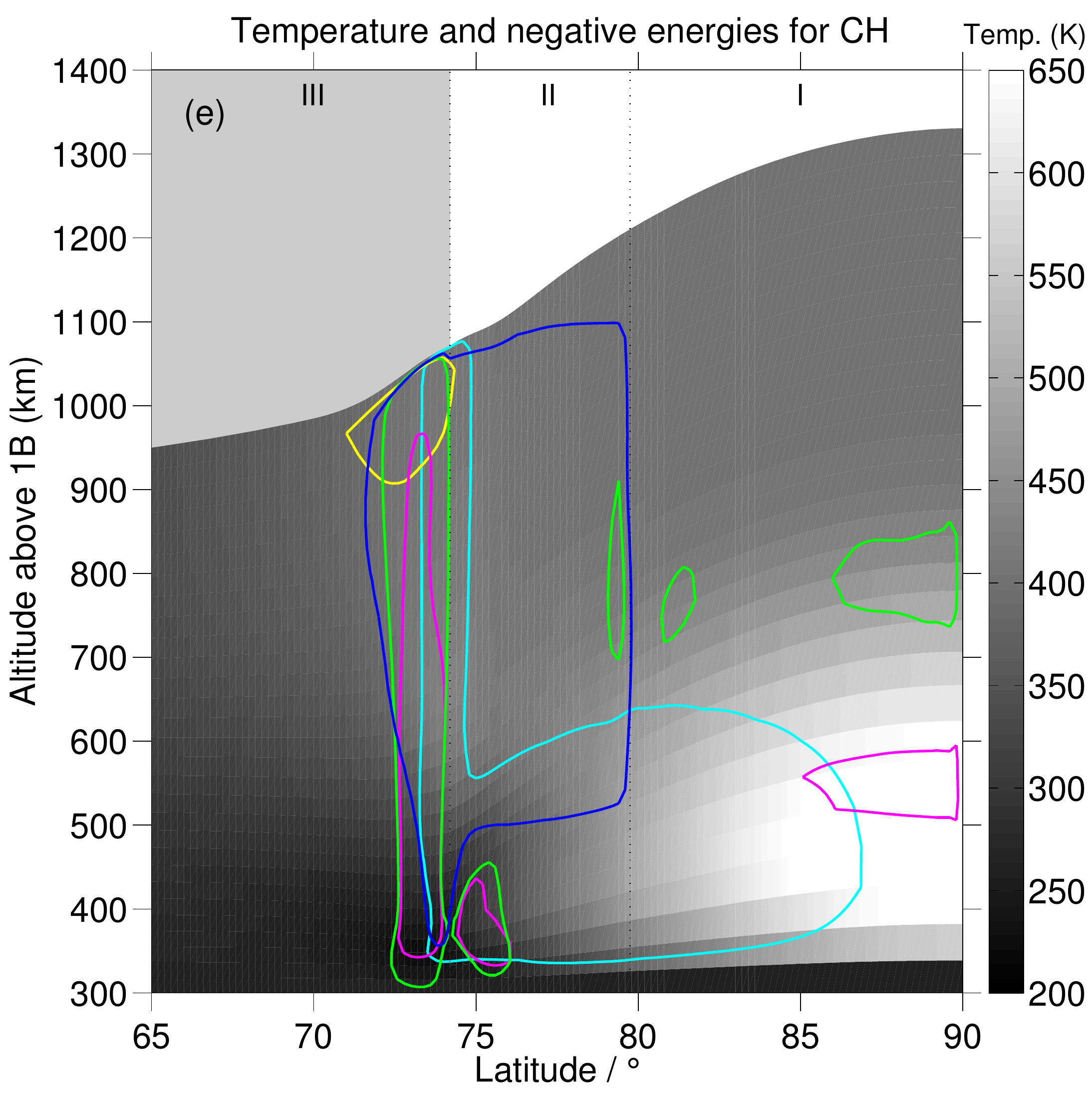}
      \includegraphics[width= 0.65\figwidth]{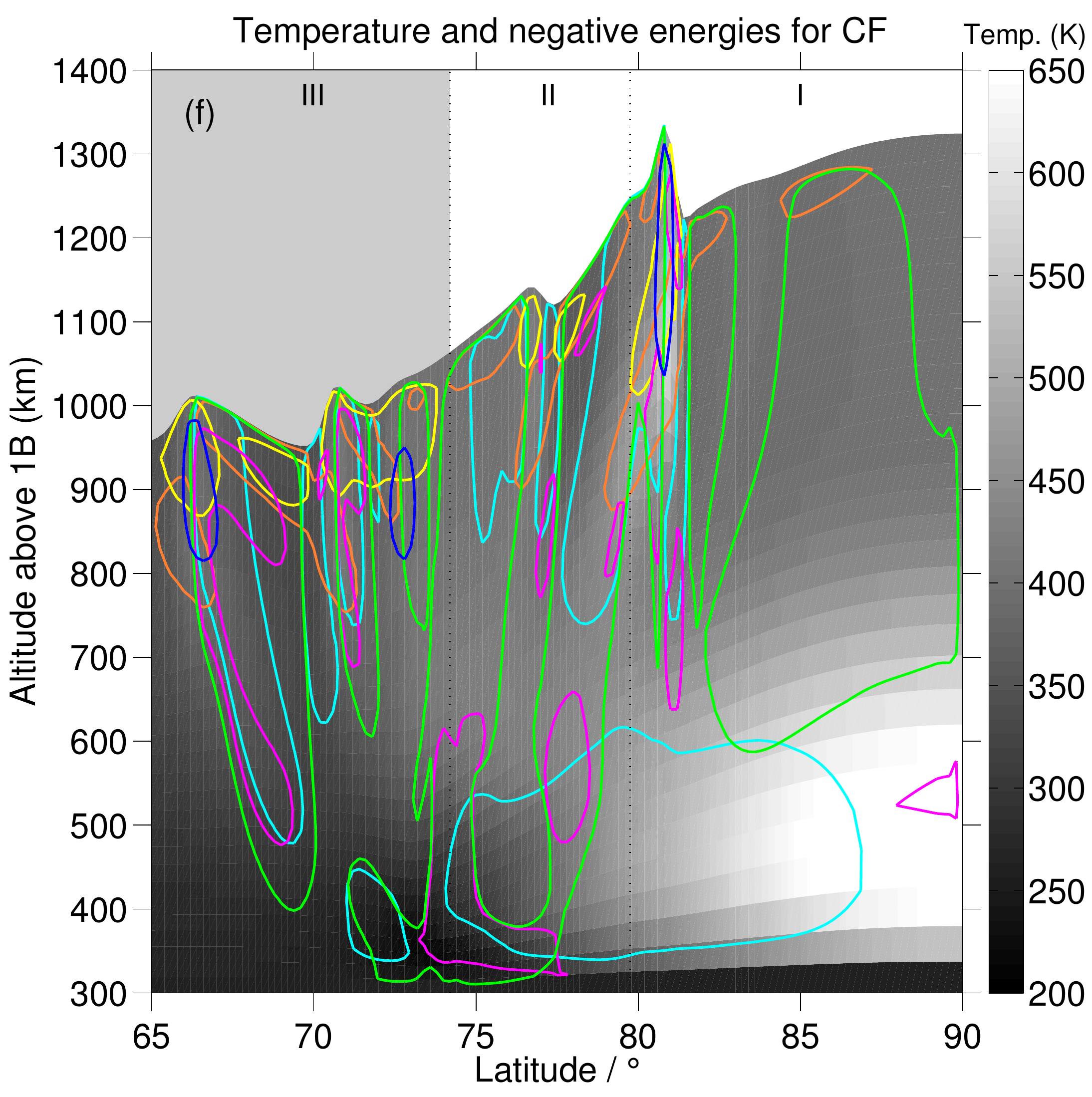} \\

      \includegraphics[width= 0.65\figwidth]{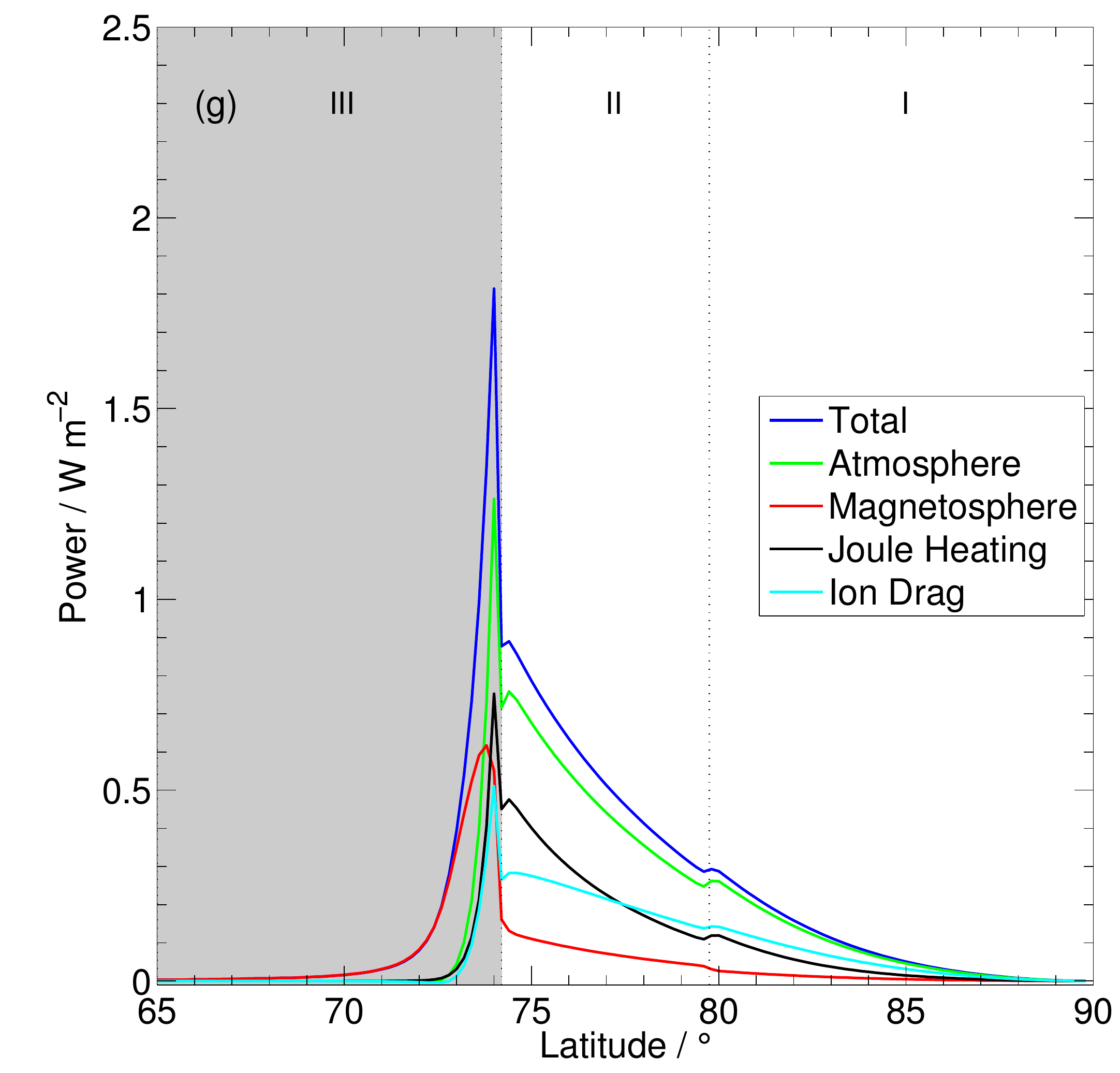}
      \includegraphics[width= 0.65\figwidth]{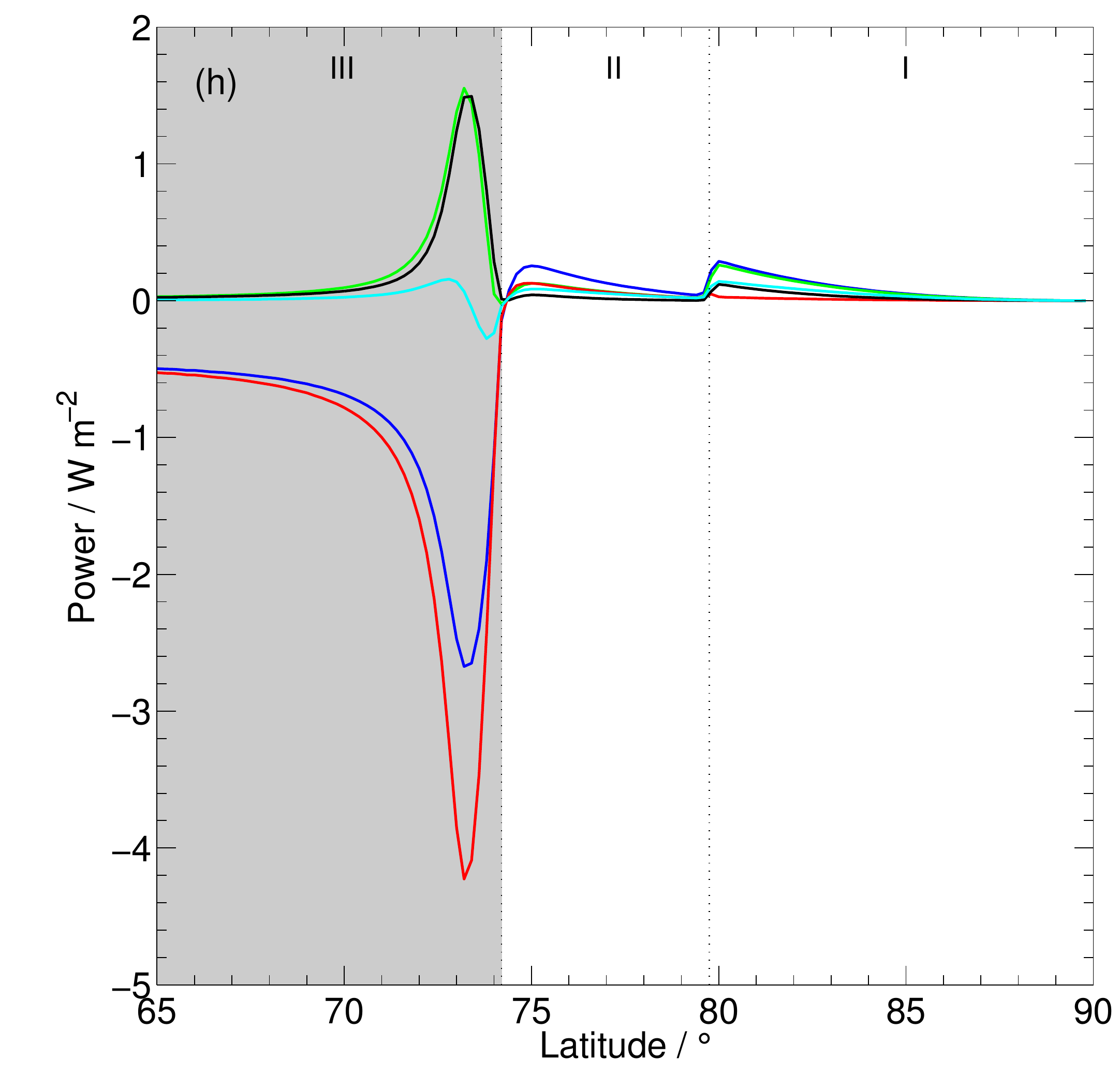}
      \includegraphics[width= 0.65\figwidth]{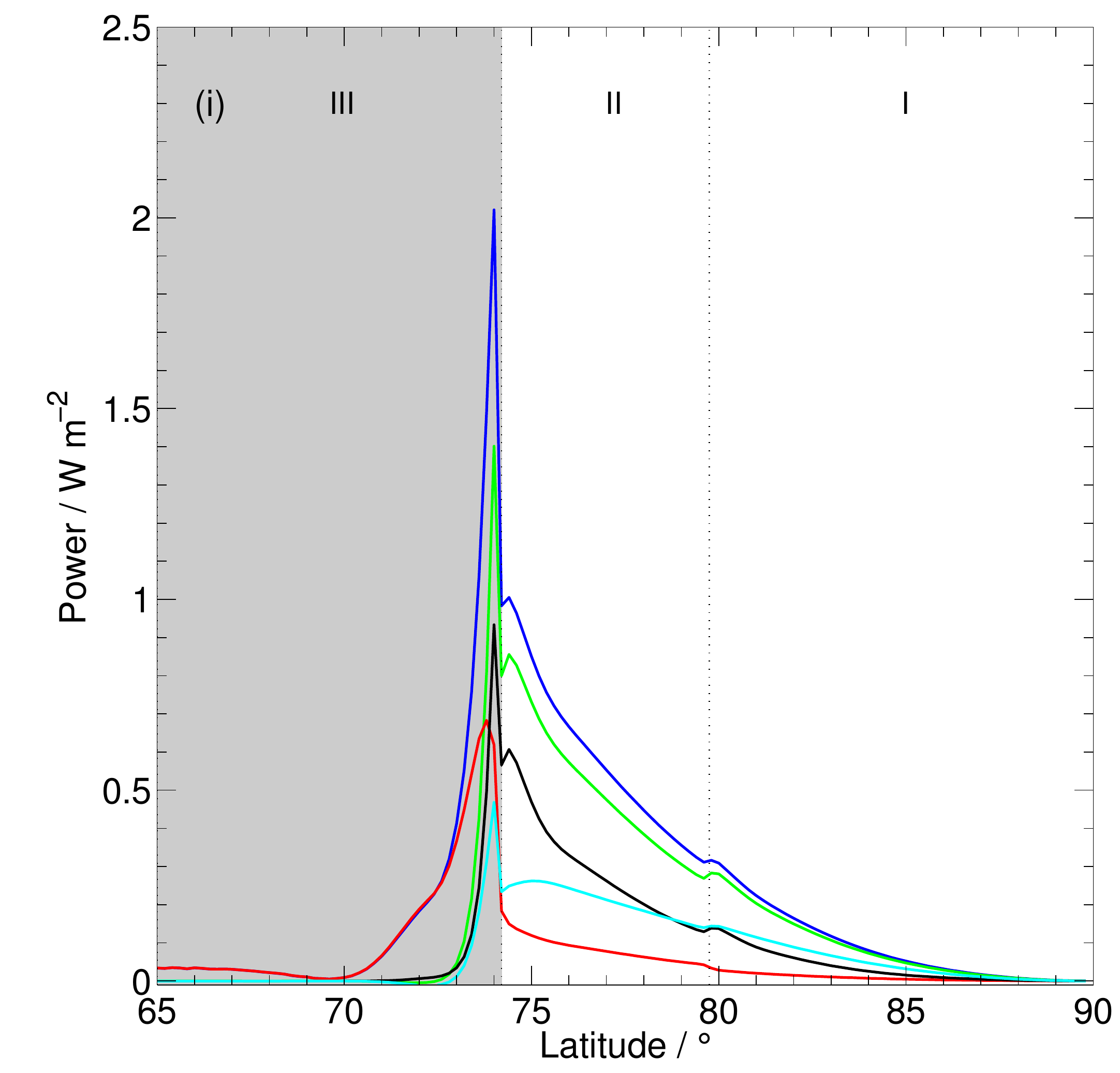} 
      
      \caption{ (a)-(c) shows the variation of atmospheric heating terms with altitude, 
      			latitude and temperature (colour bar) for cases CS, CH and CF (left to right). 
      			The contours enclose regions where heating/kinetic energy rates exceed 
      			$\unitSI[20]{W\,kg^{-1}}$. \Revnew{Ion drag energy}, Joule heating, vertical and horizontal 
      			advection of energy, adiabatic heating/cooling, viscous heating and heat conduction 
      			(vertical and turbulent) are represented by blue, red, yellow and magenta, green, 
      			cyan and orange lines. The magnetospheric regions are separated and labelled. 
      			(d)-(f) show the variation of atmospheric cooling terms where the contours enclose 
      			regions where heating/kinetic energy are decreasing (cooling) with rates exceeding 
      			$\unitSI[20]{W\,kg^{-1}}$. All colours and labels are 
      			as in (a)-(c). 
      			(g)-(i) show how the power per unit area varies for our transient compression cases. 
      			The blue line represents total power which is the sum of magnetospheric power 
      			(red line) and atmospheric power (green line); atmospheric power is the sum of 
      			both Joule heating (black solid line) and \Revnew{ion drag energy} (cyan solid line). Other 
      			labels are as for (a)-(c).
              }
      \label{fig:heating_cmp}
 \end{figure}

 \Fig{\ref{fig:thvel_cmp}g} shows thermospheric temperature as a function of altitude and latitude for case CS. 
 \Figs{\ref{fig:thvel_cmp}h-i} show the difference in thermospheric temperature between cases CH and CS, and 
 cases CF and CS. We will use \Figs{\ref{fig:heating_cmp}a-f}, showing contour plots for various thermospheric 
 heating (\Figs{\ref{fig:heating_cmp}a-c}) and cooling (\Figs{\ref{fig:heating_cmp}d-f}) terms (see plot legends 
 for details) to interpret the temperature response. \\
 
 In \Fig{\ref{fig:thvel_cmp}g} we see a clear temperature difference between upper (${>}\unitSI[75]{^{\circ}}$) and 
 lower (${<}\unitSI[75]{^{\circ}}$) latitudes; lower latitudes are cooled whilst upper latitudes are significantly 
 heated \citep{smith2007Natur,smith09}. We see a `hotspot' (in region I) with a peak temperature of 
 ${\sim}\unitSI[705]{K}$. This arises from the poleward transport of Joule heating (from regions III and II) by the 
 accelerated meridional flows shown in \Fig{\ref{fig:thvel_cmp}d} \citep{smith09,yates2012}. \\
 
 \Fig{\ref{fig:thvel_cmp}h} shows the temperature difference between cases CH and CS. There are three prominent 
 features in \Fig{\ref{fig:thvel_cmp}h}: \\
 
 \emph{i)} Temperature increase up to ${\sim}\unitSI[26]{K}$ across the region III/II boundary 
 ($\unit{z}{\geq}\unitSI[400]{km}$) resulting from a large (\unit[\times 2]{}) increase in Joule heating and the 
 addition of other heat sources, such as adiabatic heating (see \Fig{\ref{fig:heating_cmp}b}). The large increase in 
 Joule heating is caused by the increase in the rest-frame electric field, and the corresponding Pedersen current density.
 
 \emph{ii)} Temperature decrease down to ${\sim}\unitSI[{-}22]{K}$, at low altitudes of region II. \Fig{\ref{fig:heating_cmp}b} 
 shows that at low altitudes (${\leq}\unitSI[500]{km}$) of region II there is, on average, a \unit[20]{\%} decrease in 
 energy \Rev{dissipated} by Joule heating and \Revnew{ion drag energy}. This, coupled with the presence of energy lost 
 by \Revnew{ion drag energy} (\Fig{\ref{fig:heating_cmp}e}) in this region causes the significant decrease in temperature 
 shown above. All the factors discussed above result from the reversal and decrease (in magnitude) of the flow shear 
 between the magnetosphere and thermosphere in case CH.
 
 \emph{iii)} A maximum of ${\sim}\unitSI[17]{K}$ increase at low altitudes in region I. The meridional velocity of case 
 CH increases slightly (${\sim}\unitSI[2]{\%}$) in this region and, as such, can transport heat from Joule heating 
 and \Revnew{ion drag energy} polewards somewhat more efficiently than in case CS.\\
 
 \Fig{\ref{fig:thvel_cmp}i} shows the temperature difference between cases CF and CS. Immediately, we can see that there 
 are changes in the distribution of temperature in the upper thermosphere of case CF. There are four 
 `finger-like' regions with local temperature increases ${\geq}\unitSI[50]{K}$ (maximum of \unitSI[175]{K}\Rev{; white 
 contour encircles regions where temperature difference is ${\geq}\unitSI[100]{K}$}) and three regions 
 with temperature decreases ${\leq}\unitSI[40]{K}$. These alternating temperature deviations increase with altitude and 
 are collocated with accelerated meridional flow regions. Considering \Figs{\ref{fig:heating_cmp}c and f}, we see that 
 the heating and cooling terms are now quite complex, with advective and adiabatic terms dominating 
 (${\geq}\unitSI[10]{\times}$ Joule heating and \Revnew{ion drag energy} terms). The CF thermosphere appears to be 
 transporting heat, 
 both equatorward and poleward from the region III/II boundary (see \Fig{\ref{fig:thvel_cmp}f}). \citet{achilleos98} also 
 shows a similar phenomenon (see top left of Fig. 9 in \citet{achilleos98}), whereby \Rev{perturbations} of high temperature 
 are transported away from the auroral region \Rev{by meridional winds}. The energy deposited in the auroral regions heats 
 the local thermosphere which increases local pressure gradients. Advection then attempts to redistribute this heat which 
 momentarily cools the local area until enough heat is deposited again and the process restarts. \\
  
 % FIG 5 ere
 
 \Figs{\ref{fig:heating_cmp}g-i} show powers per unit area as functions of ionospheric latitude for cases CS, CH and CF, 
 respectively \Rev{(calculated using \Eqs{\ref{eq:totpower} - \ref{eq:iondrag}})}. \Rev{Blue lines represent total 
 power transferred to the ionosphere from planetary rotation which is divided into the power used to accelerate the 
 magnetospheric plasma (magnetospheric power; red lines) and power dissipated in the thermosphere (for atmospheric 
 heating and changing kinetic energy; green lines). Atmospheric power is subdivided into Joule heating (black 
 lines) and \Revnew{ion drag energy} (cyan lines).} In case CS, magnetospheric power is dominant up to 
 ${\sim}\unitSI[73]{^{\circ}}$, where atmospheric power quickly 
 dominates for all poleward latitudes (see \Fig{\ref{fig:heating_cmp}g}). This indicates that a relatively expanded \MI 
 system (in steady-state) generally dissipates more heat in the atmosphere than in acceleration of outward-moving 
 plasma \citep{yates2012}. In case CF, powers per unit area closely resemble those for case CS. There are 
 increases in peak magnetospheric power (${\sim}\unitSI[10]{\%}$) and Joule heating (${\sim}\unitSI[25]{\%}$) leading to 
 an overall maximum increase in available power of ${\sim}\unitSI[10]{\%}$. \Rev{This increase in total power is ultimately 
 due to the lag in response of the thermosphere.} For case CH, \Fig{\ref{fig:heating_cmp}h}, we see the effects of plasma 
 super-corotation in region III, where magnetospheric power reverses (now negative) \Rev{and energy is now 
 transferred from magnetosphere to thermosphere. As a consequence, heat 
 dissipated as Joule heating doubles, positive \Revnew{ion drag energy} decreases by ${\sim}\unitSI[70]{\%}$ and negative 
 \Revnew{ion drag energy} increases by two orders of magnitude. These effects lead to the local temperature variations 
 seen above. Powers decrease in region II due to the decrease in azimuthal flow shear between the magnetosphere and 
 thermosphere (see \Fig{\ref{fig:angvel}a}). } 
 
%%%%%%%%%%%%%%%%%%%%%%%%%%%%%%%%%%%%%%%%%%%%%%%%%%%%%%%%%%%%%%%%%%%%%%
\section{Magnetospheric Expansions}
\label{sec:exp}

  This section presents our findings for a transient magnetospheric expansion event with a three hour duration.
  
 \subsection{Auroral currents}
 \label{sec:current_exp}
 
% FIG 6 here
 
\begin{figure}
      \centering
      \includegraphics[width= 0.99\figwidth]{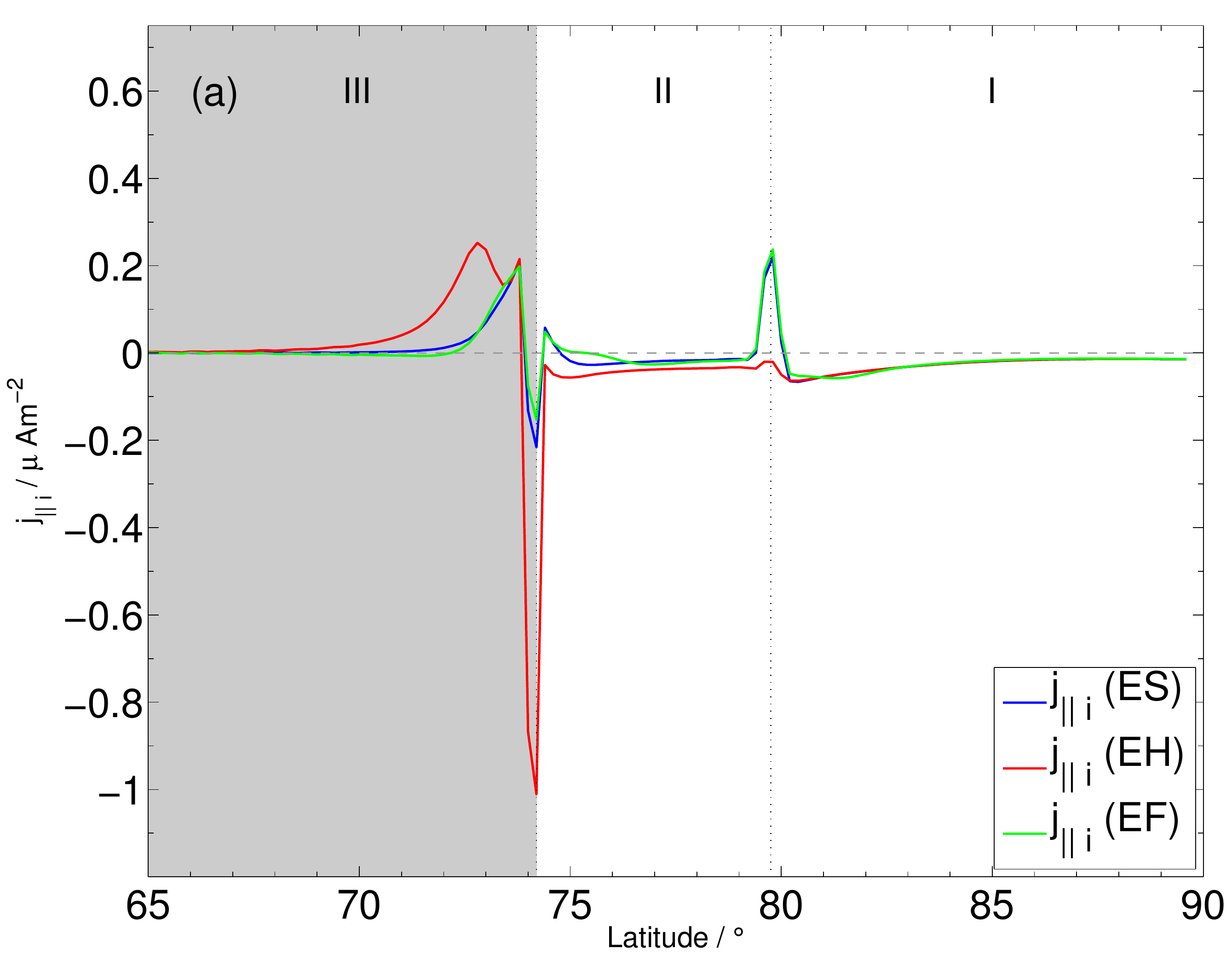}
      \includegraphics[width= 0.99\figwidth]{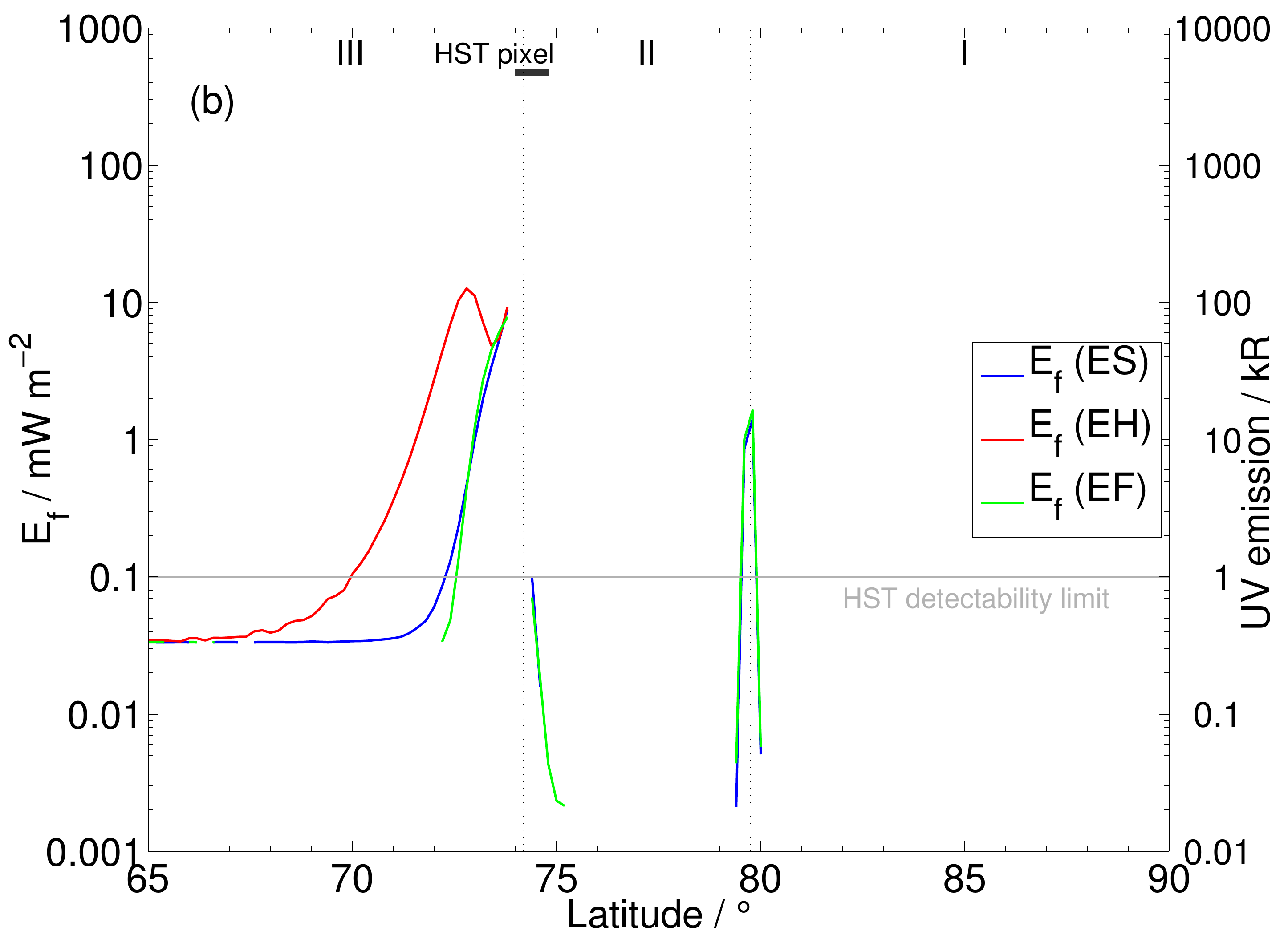}
      
      \caption{ (a) \FAC densities in the high latitude ionospheric region for our transient expansion cases. 
      			The \Rev{blue} line represents case ES whilst the \Rev{red} and \Rev{green} lines indicate cases EH and 
      			EF respectively. The conjugate magnetospheric regions (region III is shaded) are separated by 
      			dotted black lines and labelled. 
                (b) Shows the latitudinal variation of the precipitating electron flux \Rev{(on the left axis) and 
                the corresponding UV auroral emission (on the right axis)} for the transient expansion cases. 
                The colour codes and in plot labels are the same as (a). The latitudinal size of an ACS-SBC HST pixel 
                located near the main auroral emission is represented by the dark grey box. The solid grey line indicates 
                the limit of detectability of the HST \citep{cowley07}. 
      			}
      \label{fig:jpar_exp}
 \end{figure}

 \FAC densities in the high latitude region are plotted for cases ES \Rev{(blue line)}, EH \Rev{(red line)} 
 and EF \Rev{(green line)} in \Fig{\ref{fig:jpar_exp}}. Comparing cases ES with EH we see three main differences: 
 i) EH has two upward \FAC{}s peaks in region III 
 (of similar magnitude to the peak in case ES) creating a large area of upward-directed \FAC, ii) the magnitude of 
 downward \FAC near the region III boundary has increased by a factor of four (from ES to EH) and iii) \FAC densities 
 at the region II/I boundary are entirely downward-directed, unlike case ES. As the magnetosphere expands, its magnetic 
 field strength and plasma angular velocity decrease. This change in \unit{\Omega_M} (see \Fig{\ref{fig:angvel}b}) 
 increases the flow shear between the magnetosphere and thermosphere and thus increases the \FAC density in region III by 
 ${\sim}\unitSI[15]{\%}$. The strong downward \FAC results from the large gradients in \unit{\Omega_M} through 
 the poleward boundary of region III, \Rev{where magnetodisc plasma moves from a region with angular velocity of 
 ${~}\unitSI[0.9]{\Omega_J}$ to a region moving at ${~}\unitSI[0.2]{\Omega_J}$}. The lack of a peak at the region II/I 
 boundary is due to the small change in 
 \unit{\Omega_M} as the model traverses these two regions. Case EF shows only small differences with case ES due to 
 the lag in response time of the thermosphere to transient magnetospheric changes on this time scale. \\
 
 Looking now at case EH, and comparing \FAC densities with the corresponding result from \citet{cowley07} (expansion from 
 \unitSI[45{-}85]{R_J}), we notice a few differences: i) the magnitude of peak upward \FAC in case EH is 
 ${\sim}\unitSI[25]{\%}$ larger than that in \citet{cowley07} and, ii) case EF has no upward \FAC at the region 
 II/I boundary, contrary to results in \citet{cowley07}. These differences emphasise the effect of using a 
 time-dependent \GCM for the thermospheric response. For example, the `double peak' structure in the upward Region III 
 FACs is due to additional modulation of current density by thermospheric flow.\\
 
 We interpret our \FAC density profiles by considering the corresponding precipitating electron energy fluxes, shown 
 in \Fig{\ref{fig:jpar_exp}b}. Fluxes are plotted as functions of latitude. The line style code and labels are the same 
 as in \Fig{\ref{fig:jpar_exp}a}, \Rev{the latitudinal size of a HST ACS-SBC pixel is indicated by the dark grey rectangle} 
 and the grey solid line indicates the limit of present HST detectability (${\sim}\unitSI[1]{kR}$; \citet{cowley07}). 
 We begin by comparing \Rev{profiles for} case ES with EF, which are almost identical and both \Rev{have maxima} 
 at ${\sim}\unitSI[74]{^{\circ}}$ 
 latitude, equivalent to the location of the `main auroral oval', and at ${\sim}\unitSI[80]{^{\circ}}$, the boundary between 
 open (region I) and closed field lines (region II). Therefore, we would expect a fairly bright auroral oval of 
 ${\sim}\unitSI[88]{kR}$ for case ES and ${\sim}\unitSI[79]{kR}$ for case EF. The electron energy flux for case EF 
 \Rev{(${\sim}\unitSI[7.85]{mW\,m^{-2}}$)} is ${\sim}\unitSI[10]{\%}$ smaller than case ES 
 \Rev{(${\sim}\unitSI[8.8]{mW\,m^{-2}}$)} due to $\unit{\Omega_T (ES)}{>}\unit{\Omega_T (EF)}$ leading to a smaller flow 
 shear. Our model also predicts the possibility of observable polar emission (region II/I boundary) of 
 ${\sim}\unitSI[15]{kR}$ for both cases ES and EF. However, this region is strongly dependent on the plasma flow 
 model used and poorly constrained by observations. \\
 
 Energy flux \unit{E_f} for case EH \Rev{is non-existent}, poleward of ${\sim}\unitSI[74]{^{\circ}}$ latitude, due to 
 the downward (negative) \FAC density in this region. In region III, there are two upward \FAC peaks, separated by 
 ${\sim}\unitSI[1]{^{\circ}}$. 
 The first one, located at ${\sim}\unitSI[73]{^{\circ}}$ is ${\sim}\unitSI[37]{\%}$ larger than the second, at 
 ${\sim}\unitSI[74]{^{\circ}}$. These peaks result from the large degree of magnetospheric 
 sub-corotation \Rev{and the modulation of the thermospheric angular velocity} in region III (evident in \Fig{\ref{fig:angvel}b}). 
 Comparing case EH with the equivalent expansion case in \citet{cowley07}; case EH, in region III, has a maximum value of 
 \unit{E_{f}} \Rev{(${\sim}\unitSI[12.6]{mW\,m^{-2}}$)} that is twice that in \citet{cowley07}. This study represents the 
 thermosphere with a \GCM which responds \Rev{self-consistently} to time-dependent changes in \unit{\Omega_M} profiles. 
 Our results indicate that 
 this response is not as strong as that in \citet{cowley07}, who use a simple `slippage' relation to model the thermospheric 
 angular velocity. At the open-closed field line (region II/I) boundary, \citet{cowley07} obtain larger energy fluxes due 
 to their large change in \unit{\Omega_M} across these regions; in our study, we obtain negligible changes in 
 \unit{E_f} due to our smaller change in imposed \unit{\Omega_M} across this boundary.  

 \subsection{Thermospheric dynamics}
 \label{sec:dyn_exp}
  
% FIG 7 here

 The altitude-latitude variation of azimuthal and meridional thermospheric velocities and temperature 
 are shown in \Fig{\ref{fig:thvel_exp}}. The first column in \Fig{\ref{fig:thvel_exp}} shows thermospheric 
 outputs for case ES; cases EH and EF are represented in columns two and three respectively.\\
 
 For case ES, the zonal \Rev{(\Fig{\ref{fig:thvel_exp}a})} and meridional \Rev{(\Fig{\ref{fig:thvel_exp}d})} 
 flows are very similar to those discussed in \citet{yates2012} as the 
 only difference between both steady-state compressed cases is that here we assume a constant height-integrated 
 Pedersen conductivity whilst in \citet{yates2012} the conductivity is enhanced by \FAC. \Rev{Zonal flows show 
 a low-altitude super-corotational jet in region III and two sub-corotational jets across regions II and I. The 
 meridional flows show the previously discussed flow patterns i.e. low-altitude poleward flow and high-altitude equatorward 
 flow.}\\
 
 Thermospheric flows for case EH are slightly different from those of case ES. \Rev{A magnetospheric expansion decreases 
 the degree of plasma corotation which subsequently decreases the thermospheric zonal velocities i.e. they become more 
 sub-corotational (see \Fig{\ref{fig:thvel_exp}b}). In the meridional \Rev{sense} (\Fig{\ref{fig:thvel_exp}e}), low altitude 
 flows remain poleward but with an increased magnitude (up to ${\sim}\unitSI[30]{\%}$) and all flow in region II is now 
 directed poleward. Extra heating (see section~\ref{sec:heat_exp}) near the region III/II boundary causes the forces in 
 the local thermosphere to become unbalanced leading to accelerated flows in the poleward and equatorward 
 \Rev{(high-altitude of region III)} directions.}\\
 
 \Rev{The thermospheric \Rev{velocities} at the end of the transient expansion event (case EF) are shown in the third 
 column of 
 \Fig{\ref{fig:thvel_exp}}. We see that the only change in zonal flow patterns is a slight increase in the zonal 
 velocity \Rev{(algebraic increase)}. The meridional winds show a large poleward accelerated flow originating at low 
 altitudes in region III 
 and reaching the high altitudes of region I. Two smaller regions of accelerated equatorward flow arise in the upper 
 altitudes of regions III and II. As the magnetosphere returns to its initial configuration, it weakly 
 super-corotates over most of region III; this transfers angular momentum to the sub-corotating thermosphere which acts 
 to `spin up' the thermospheric gas.}
 
 \subsection{Thermospheric heating}
 \label{sec:heat_exp}
  
% FIG 8 here

\begin{figure}
      \centering
      \includegraphics[width= 0.65\figwidth]{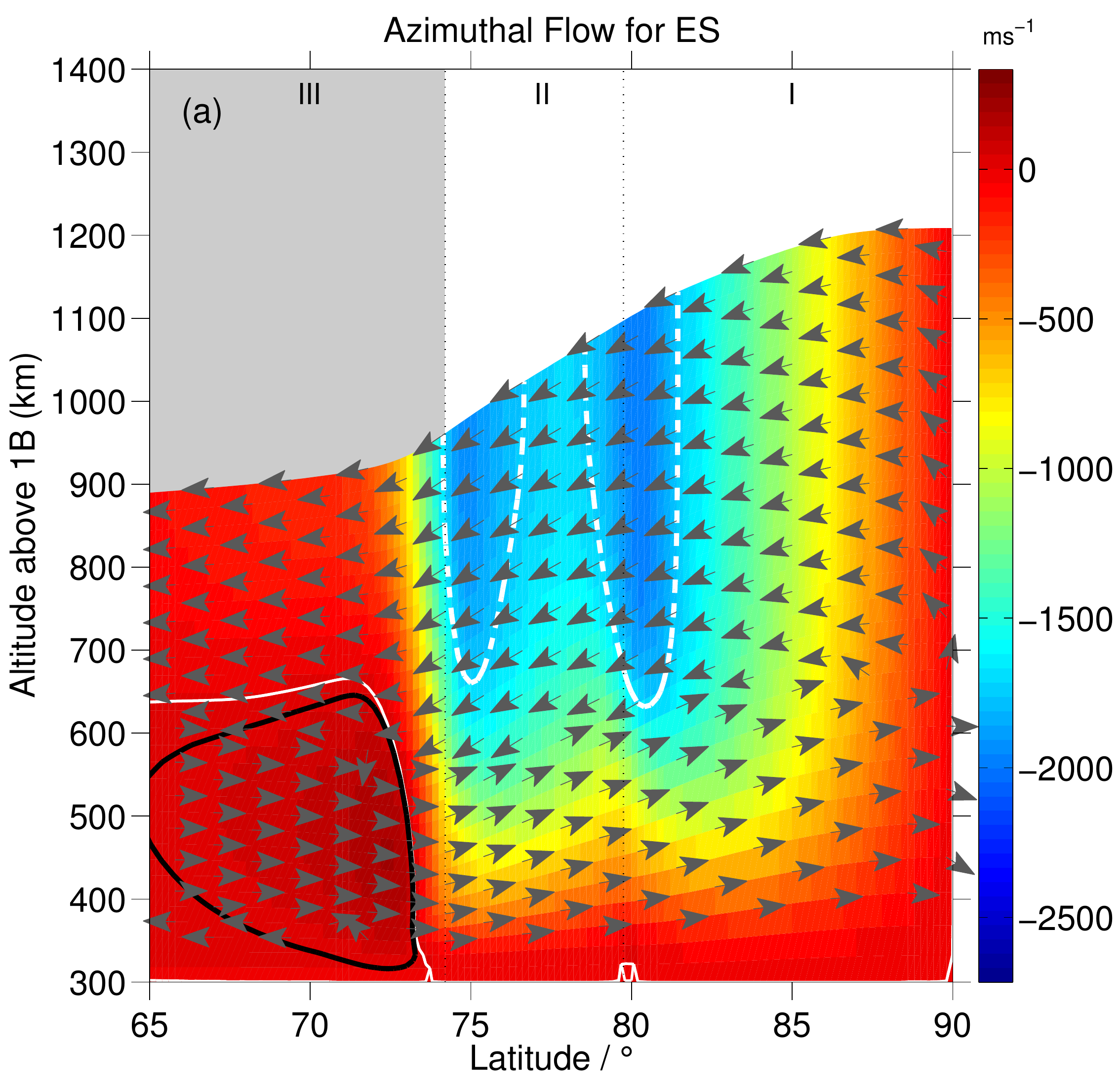}
      \includegraphics[width= 0.65\figwidth]{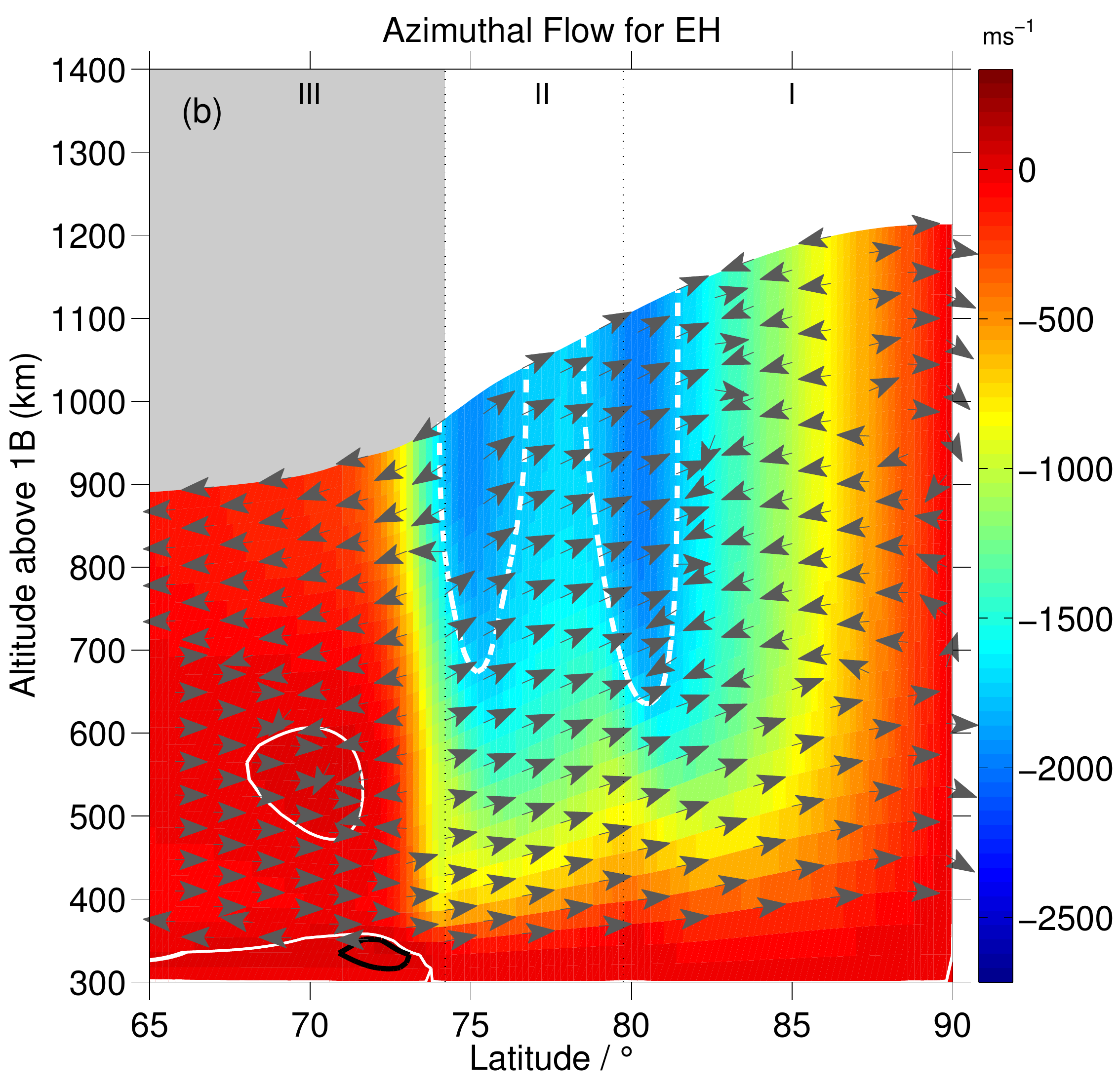}
      \includegraphics[width= 0.65\figwidth]{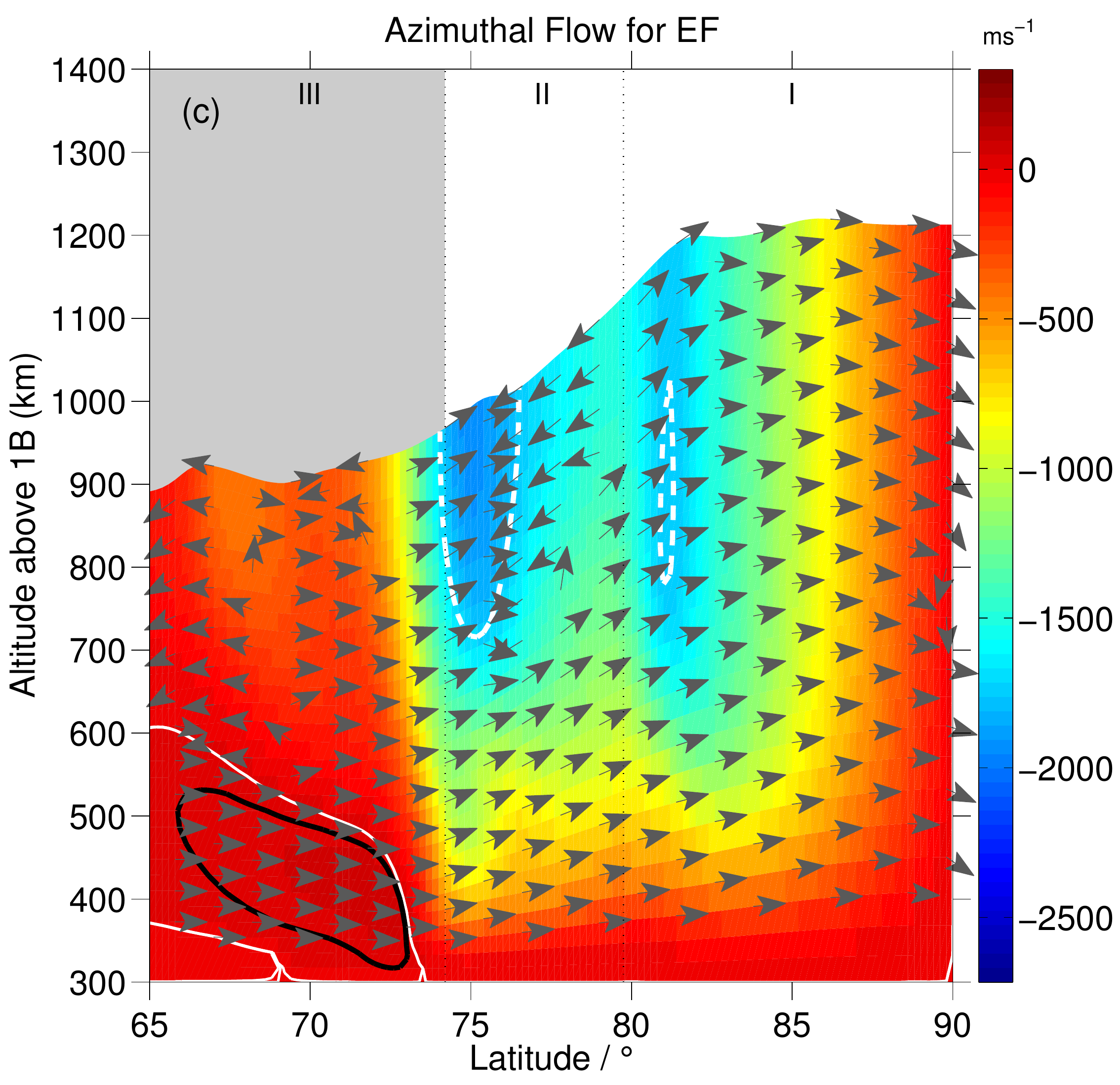} \\
      
      \includegraphics[width= 0.65\figwidth]{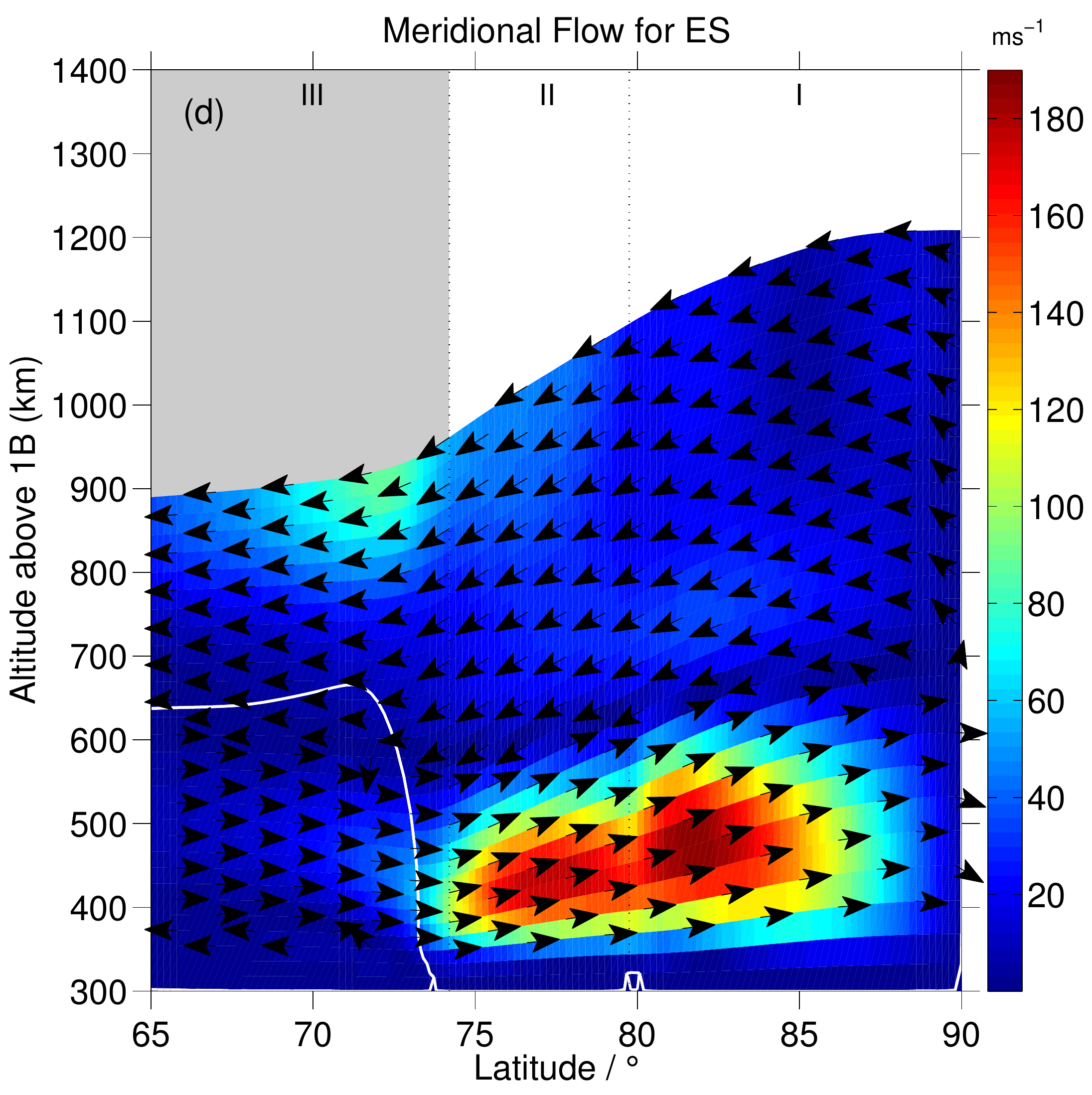}
      \includegraphics[width= 0.65\figwidth]{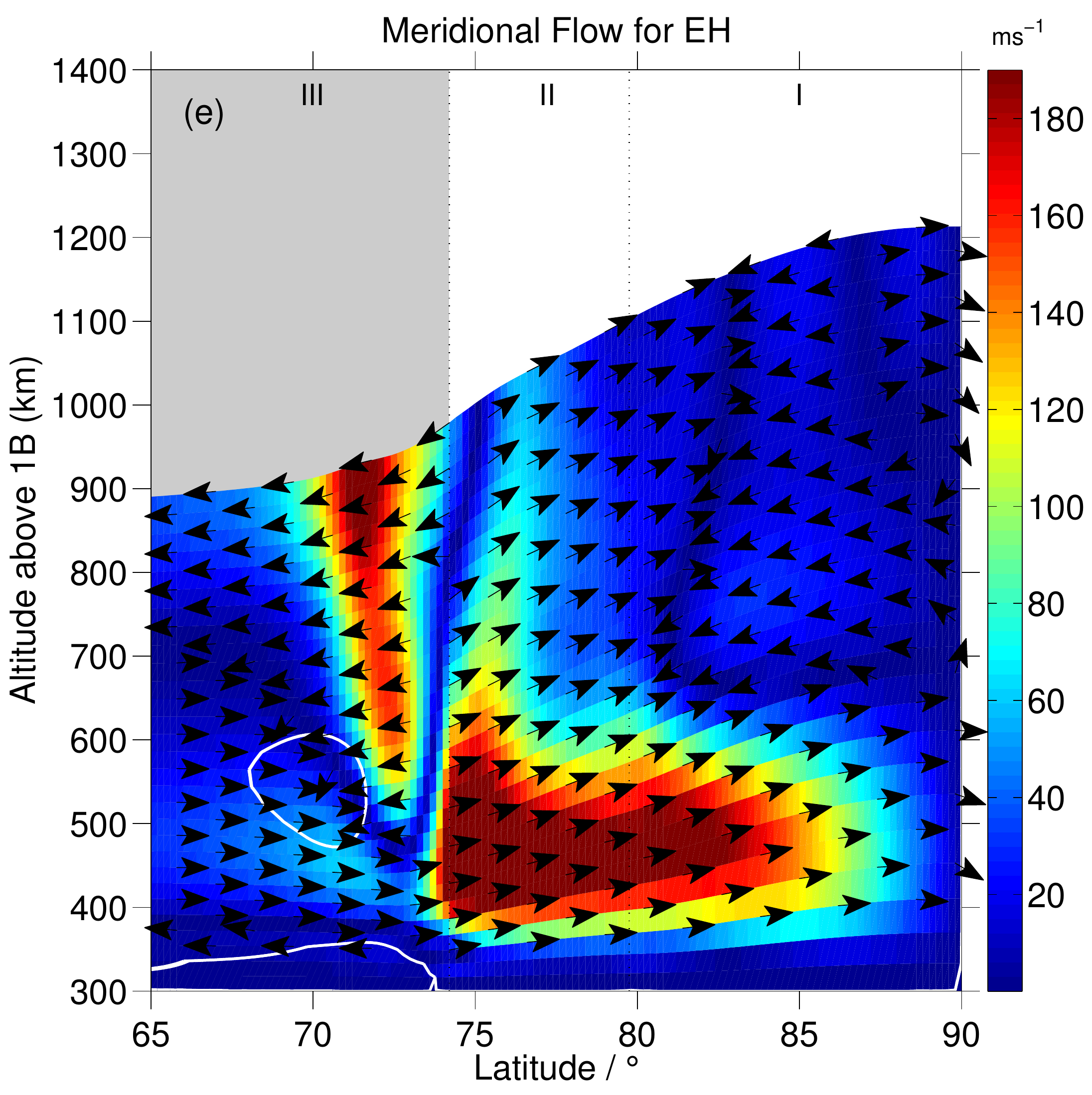}
      \includegraphics[width= 0.65\figwidth]{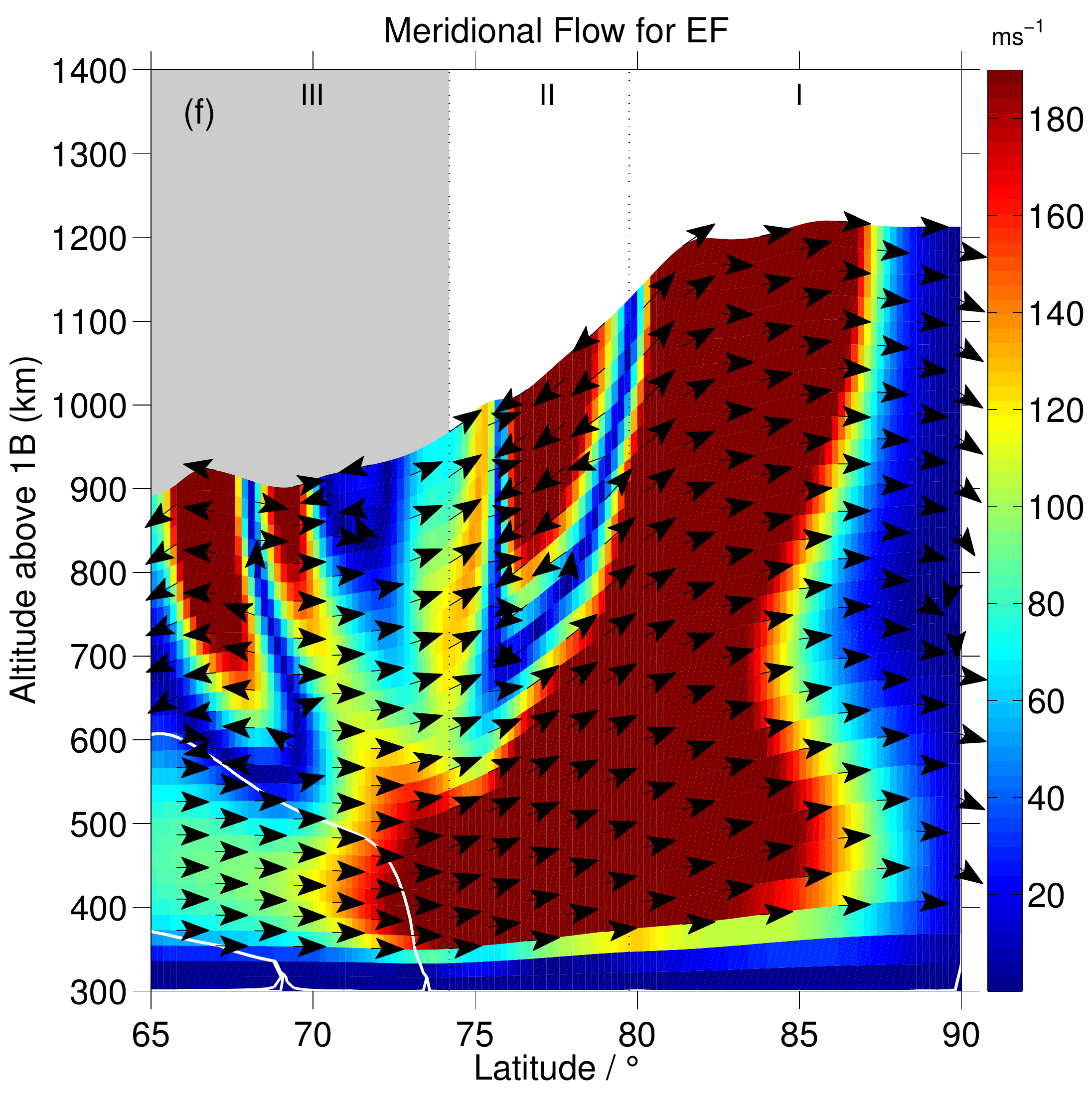} \\

      \includegraphics[width= 0.65\figwidth]{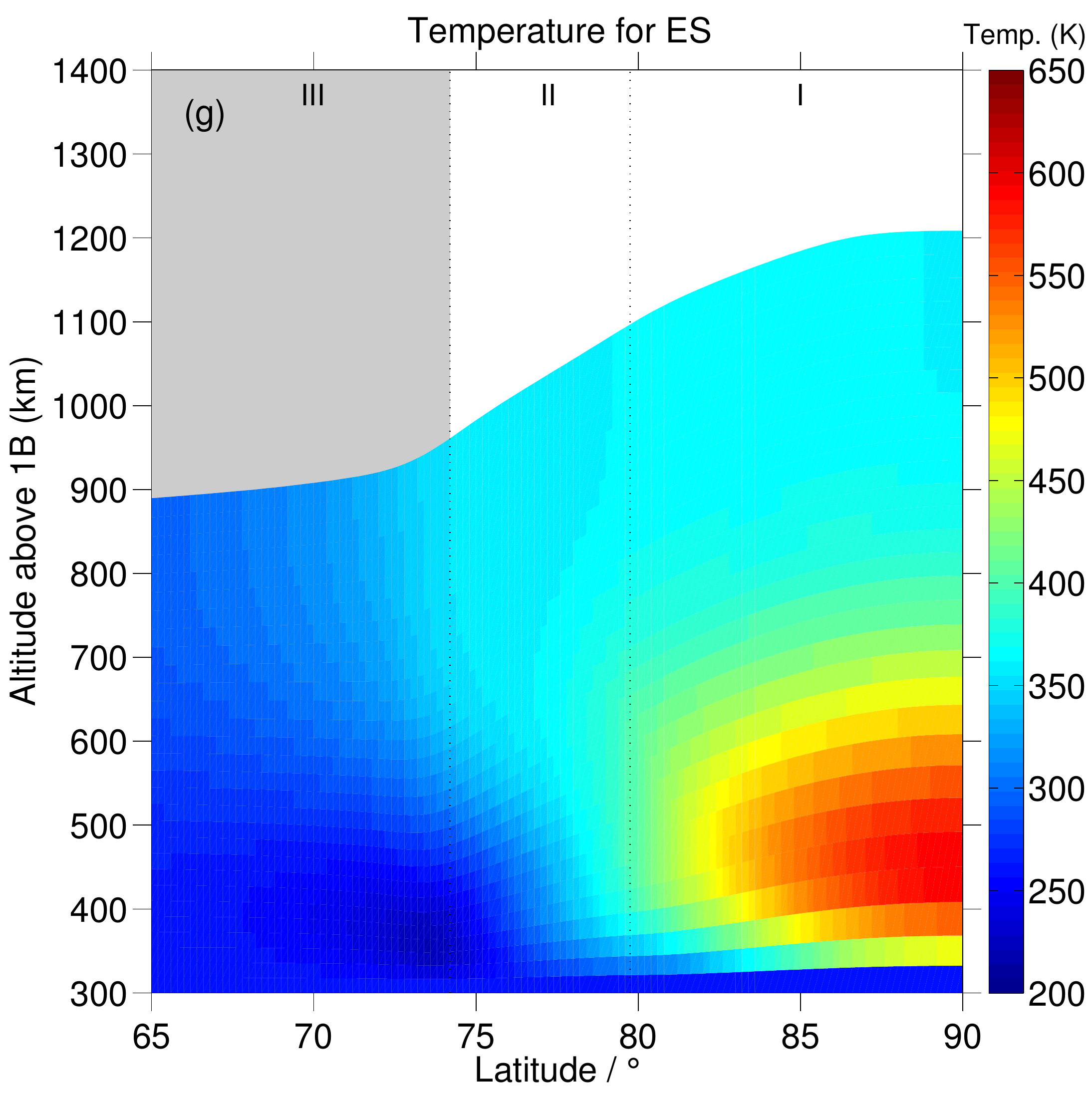}
      \includegraphics[width= 0.65\figwidth]{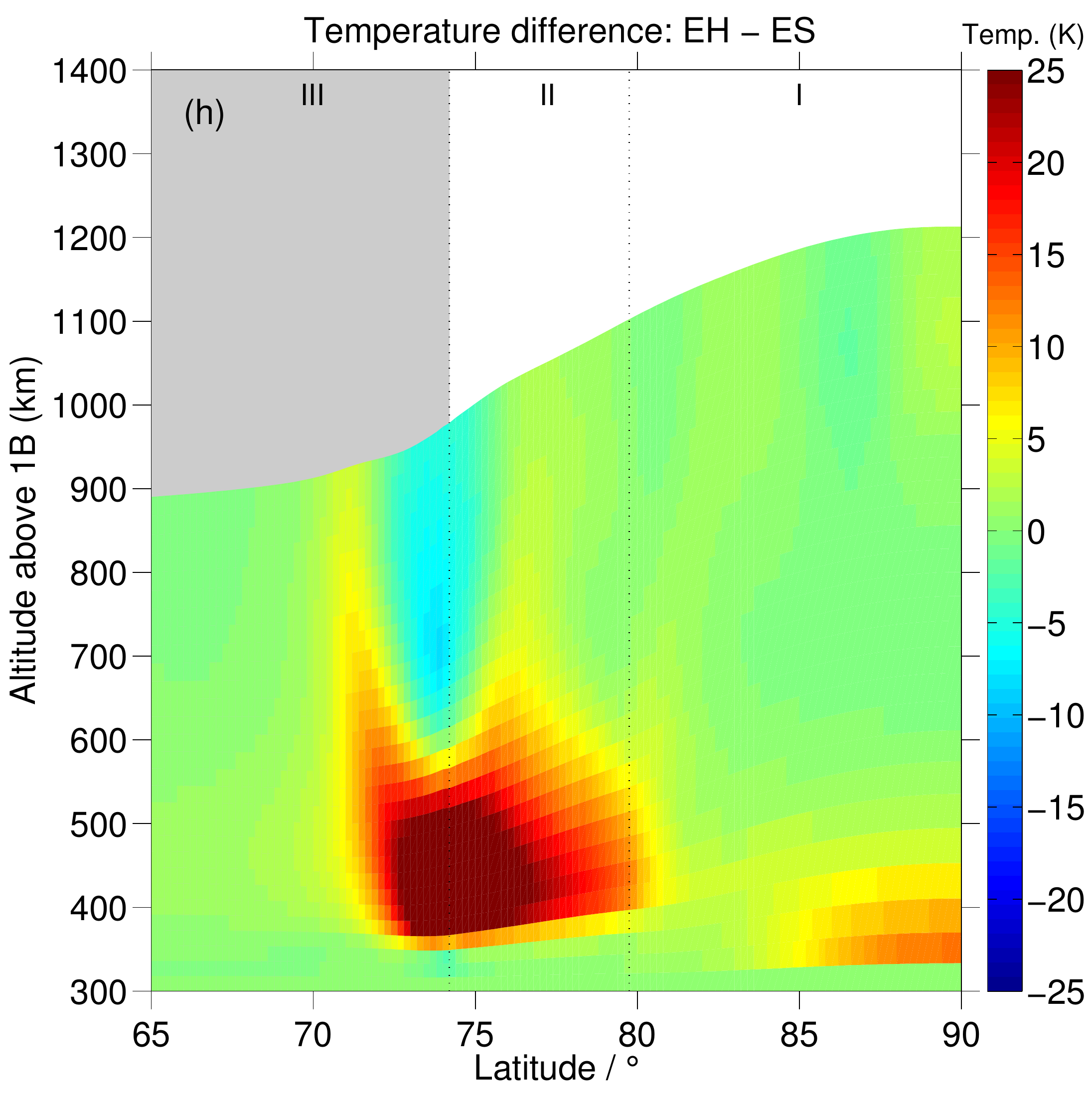}
      \includegraphics[width= 0.65\figwidth]{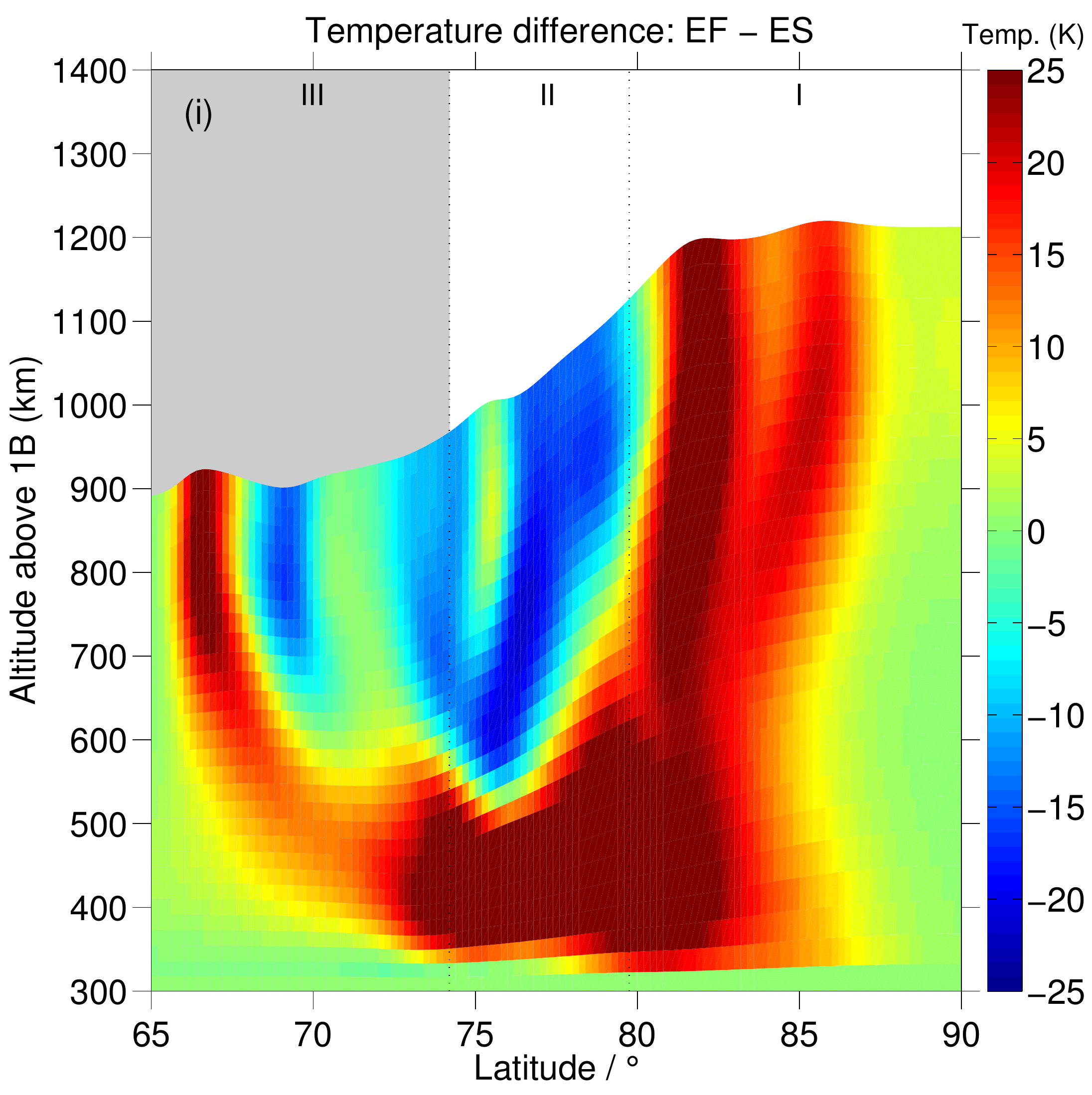} 
      
      \caption{ (a)-(c) show the variation of thermospheric azimuthal velocity (colour scale) in the 
      			corotating reference frame for cases ES-EF respectively (left to right). Positive values (dark red) 
      			indicate super-corotation, whilst negative values (light red to blue) indicate 
      			sub-corotation. The arrows show the direction of meridional flow and the white 
      			line indicates the locus of rigid corotation. The solid black encloses regions of super-corotation 
      			($>\unitSI[25]{m\,s^{-1}}$) and the dashed white line encloses regions that are sub-corotating at a rate 
      			$<\unitSI[-1750]{m\,s^{-1}}$. The magnetospheric regions 
      			(region III is shaded) are separated by dotted black lines and labelled. (d)-(f) show the 
      			meridional velocity in the thermosphere for cases ES-EF. The colour scale indicates 
      			the speed of flows. All other labels and are as for (a)-(c). (g) shows the thermospheric 
      			temperature distributions for case ES whilst (h)-(i) show the temperature difference 
      			between cases EH and EF and case ES. All labels are as in (a)-(c). 
      			}

      \label{fig:thvel_exp}
 \end{figure}

\begin{figure}
      \centering
      \includegraphics[width= 0.65\figwidth]{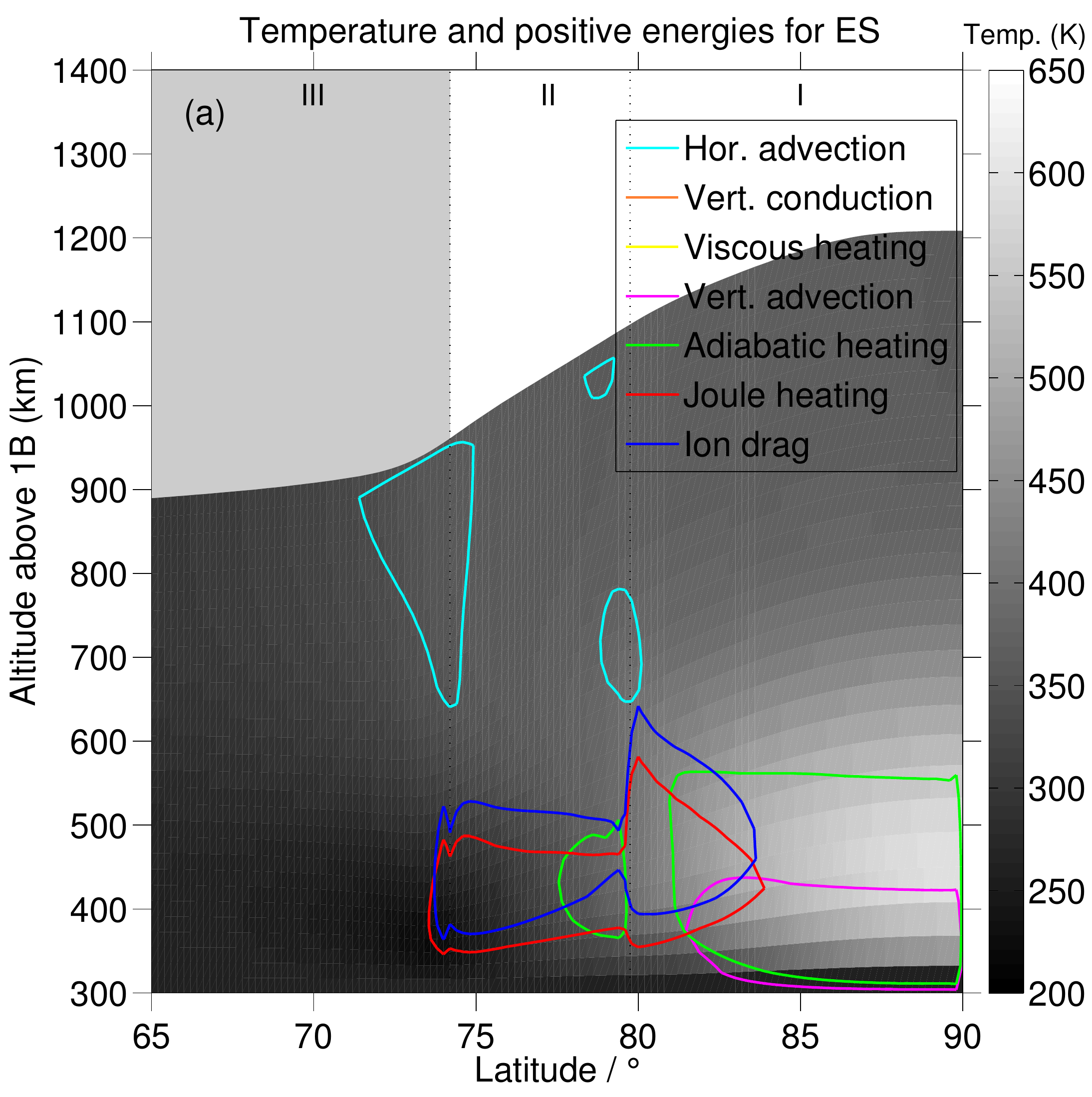} %width=99\figwidth
      \includegraphics[width= 0.65\figwidth]{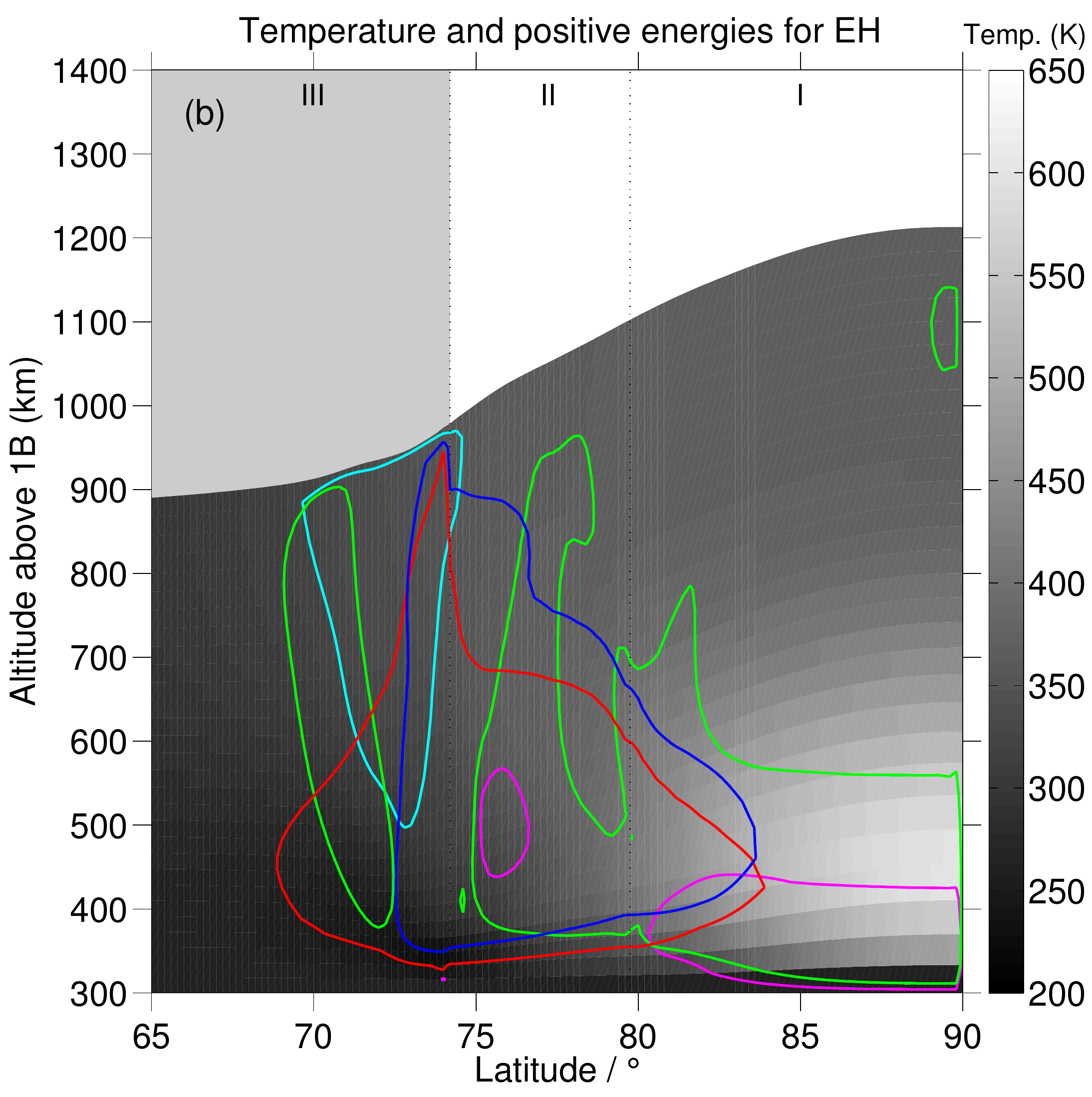}
      \includegraphics[width= 0.65\figwidth]{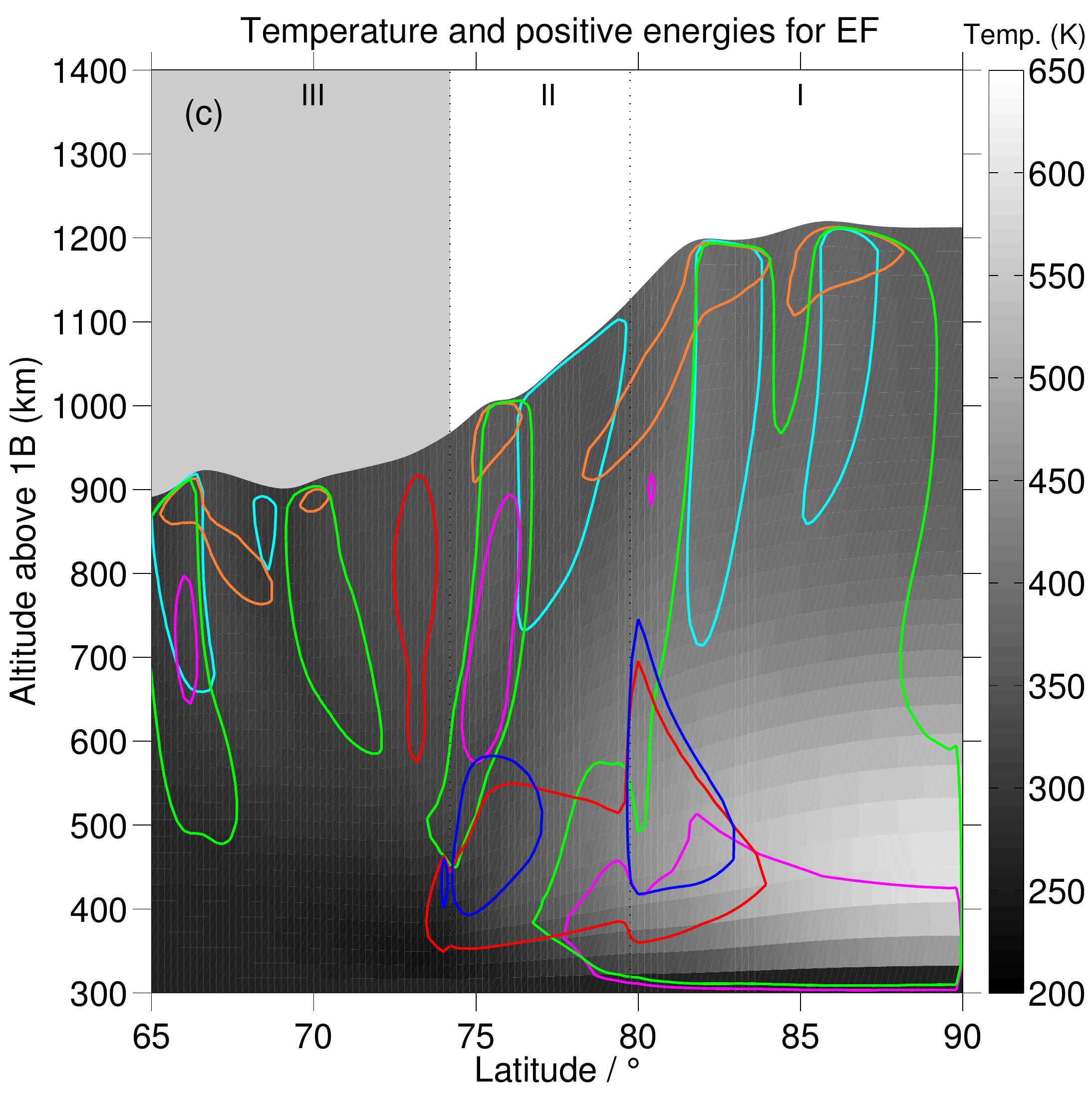} \\
      
      \includegraphics[width= 0.65\figwidth]{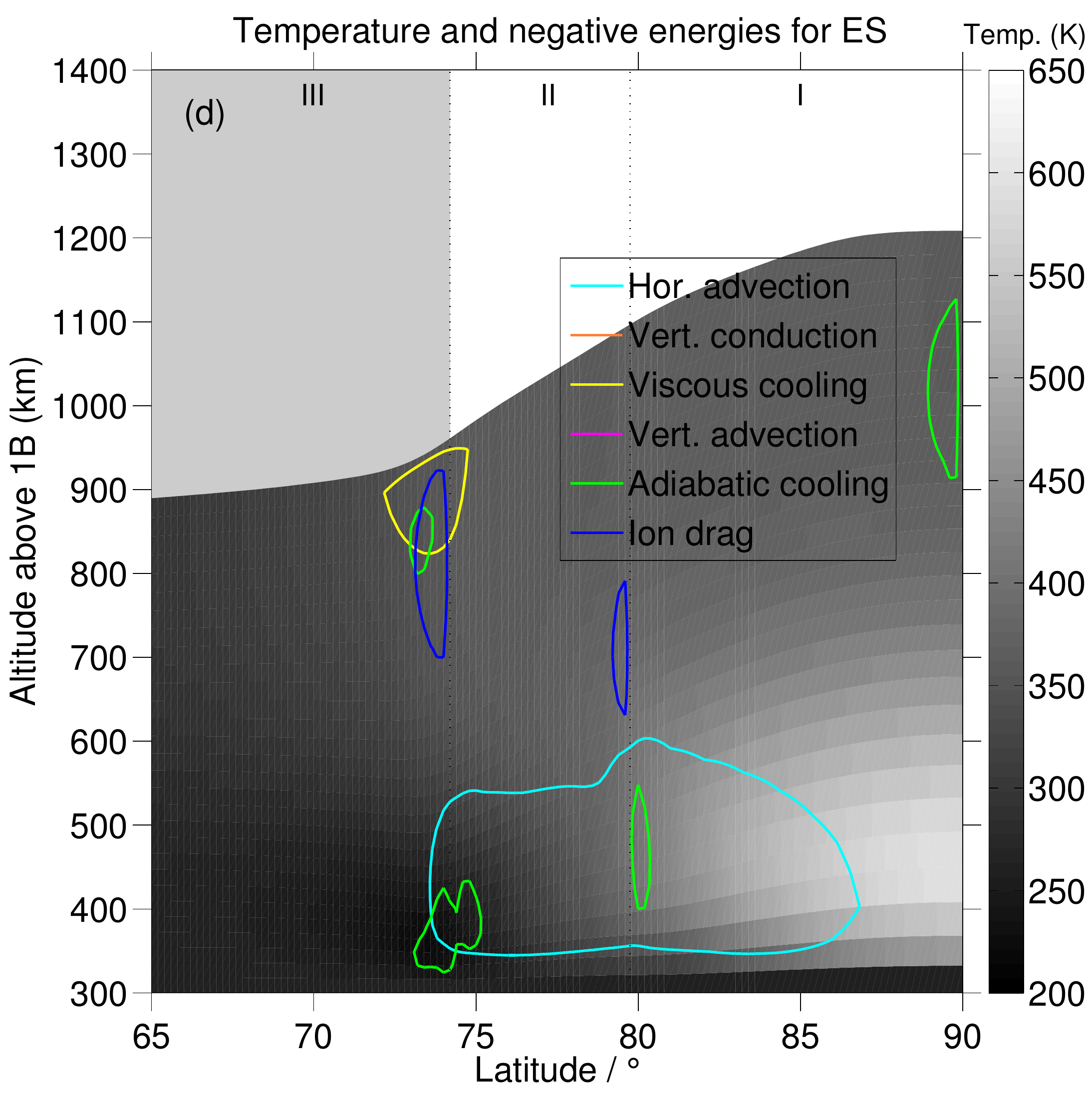}
      \includegraphics[width= 0.65\figwidth]{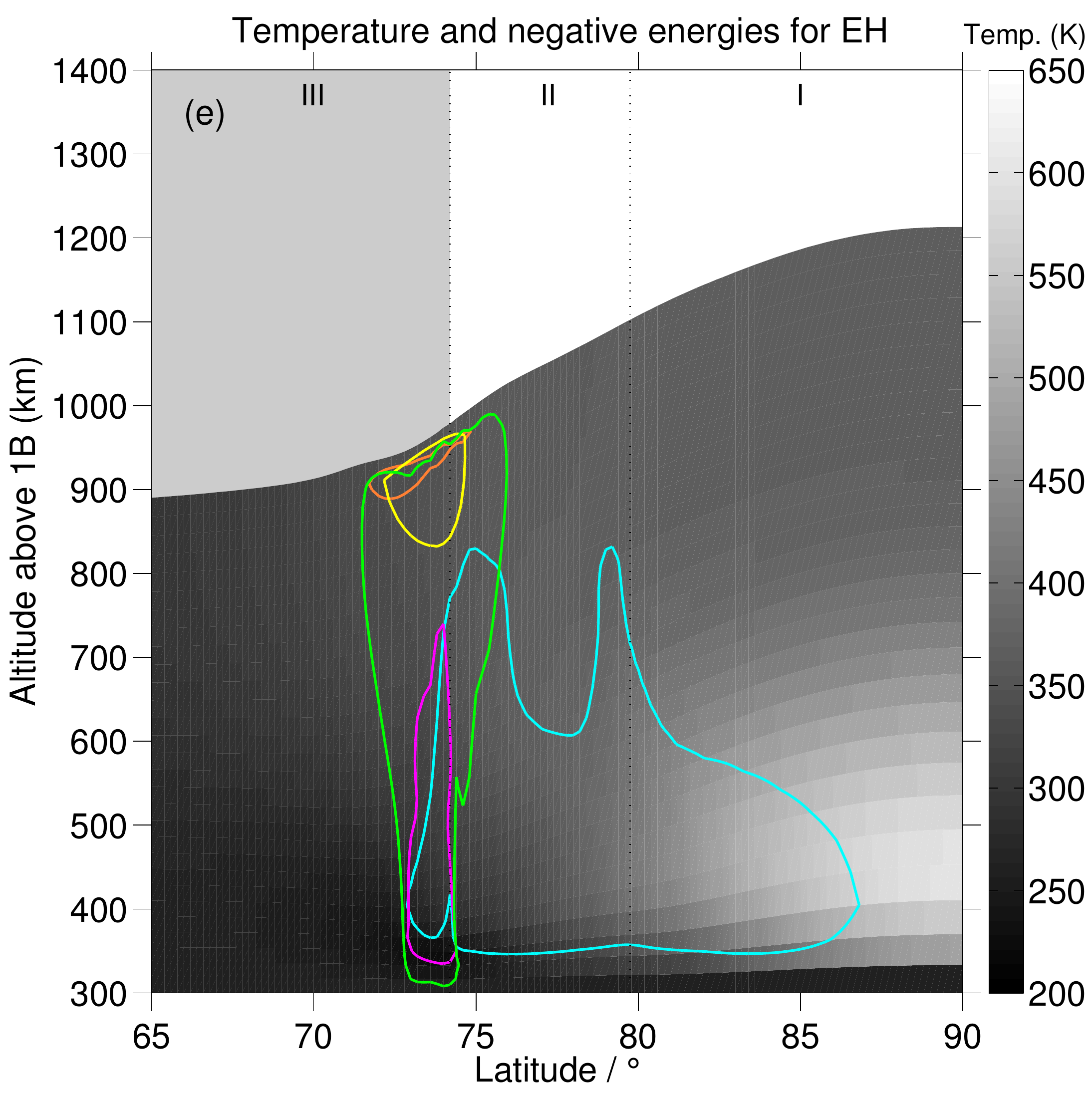}
      \includegraphics[width= 0.65\figwidth]{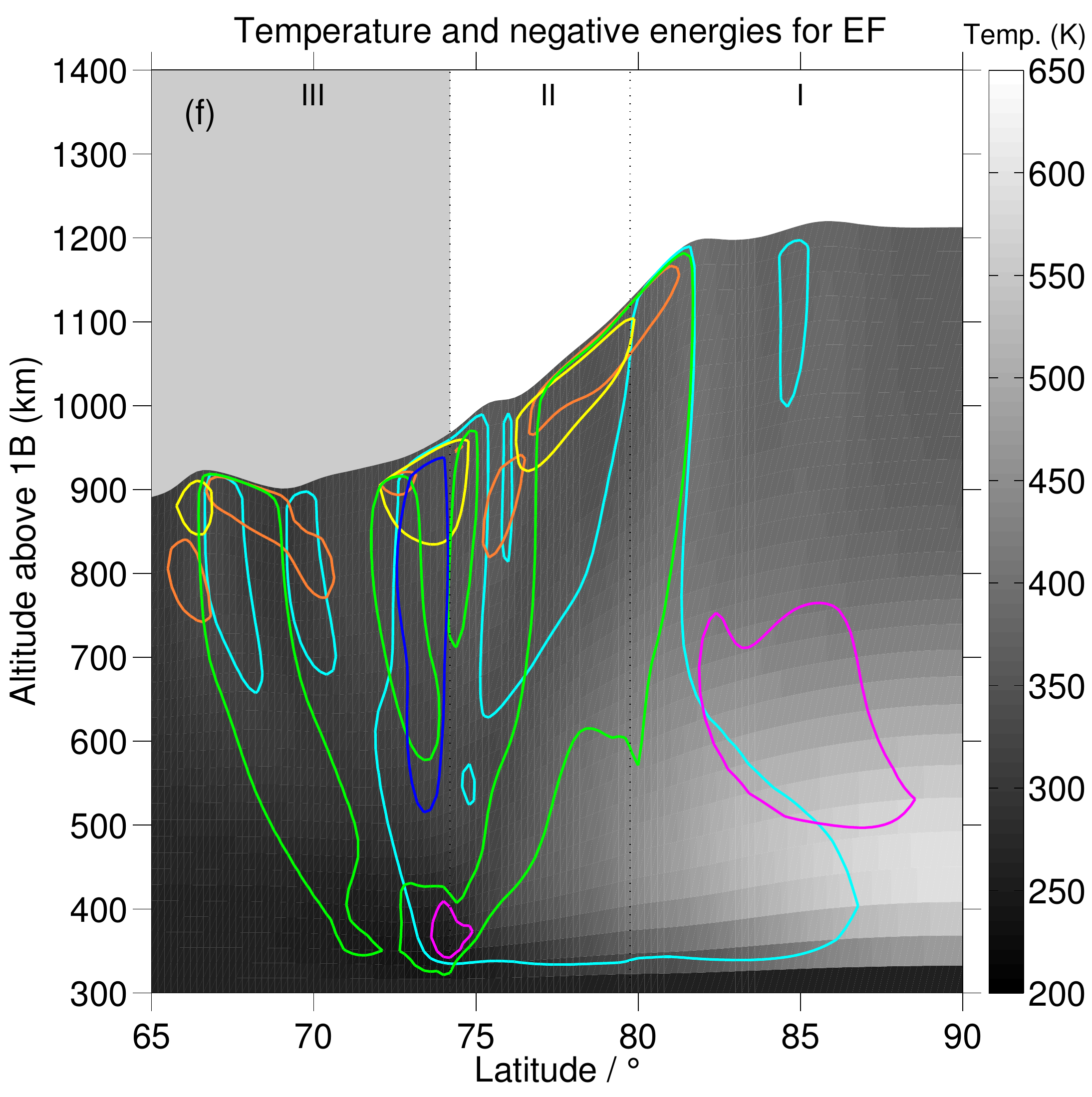} \\

      \includegraphics[width= 0.65\figwidth]{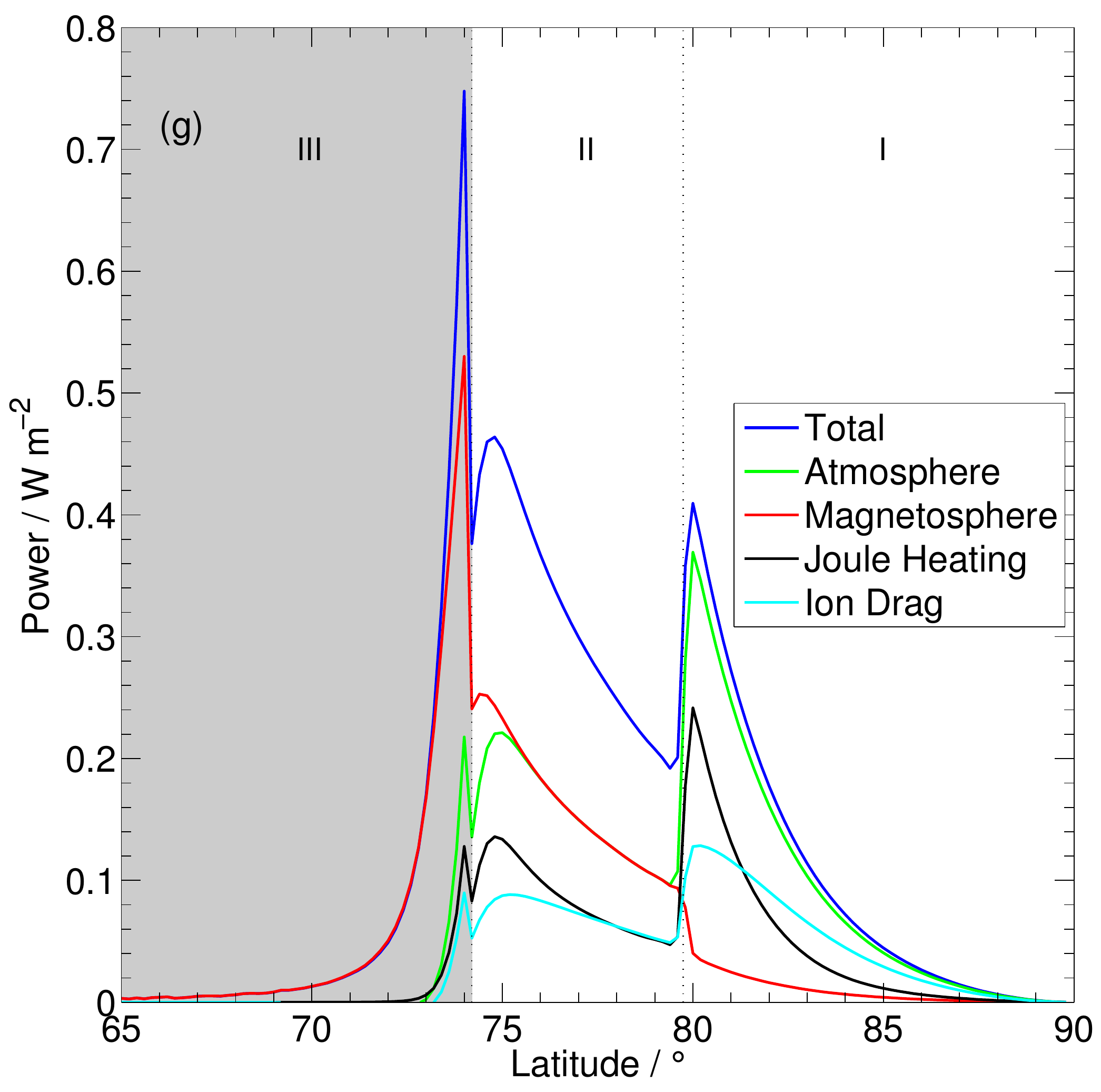}
      \includegraphics[width= 0.65\figwidth]{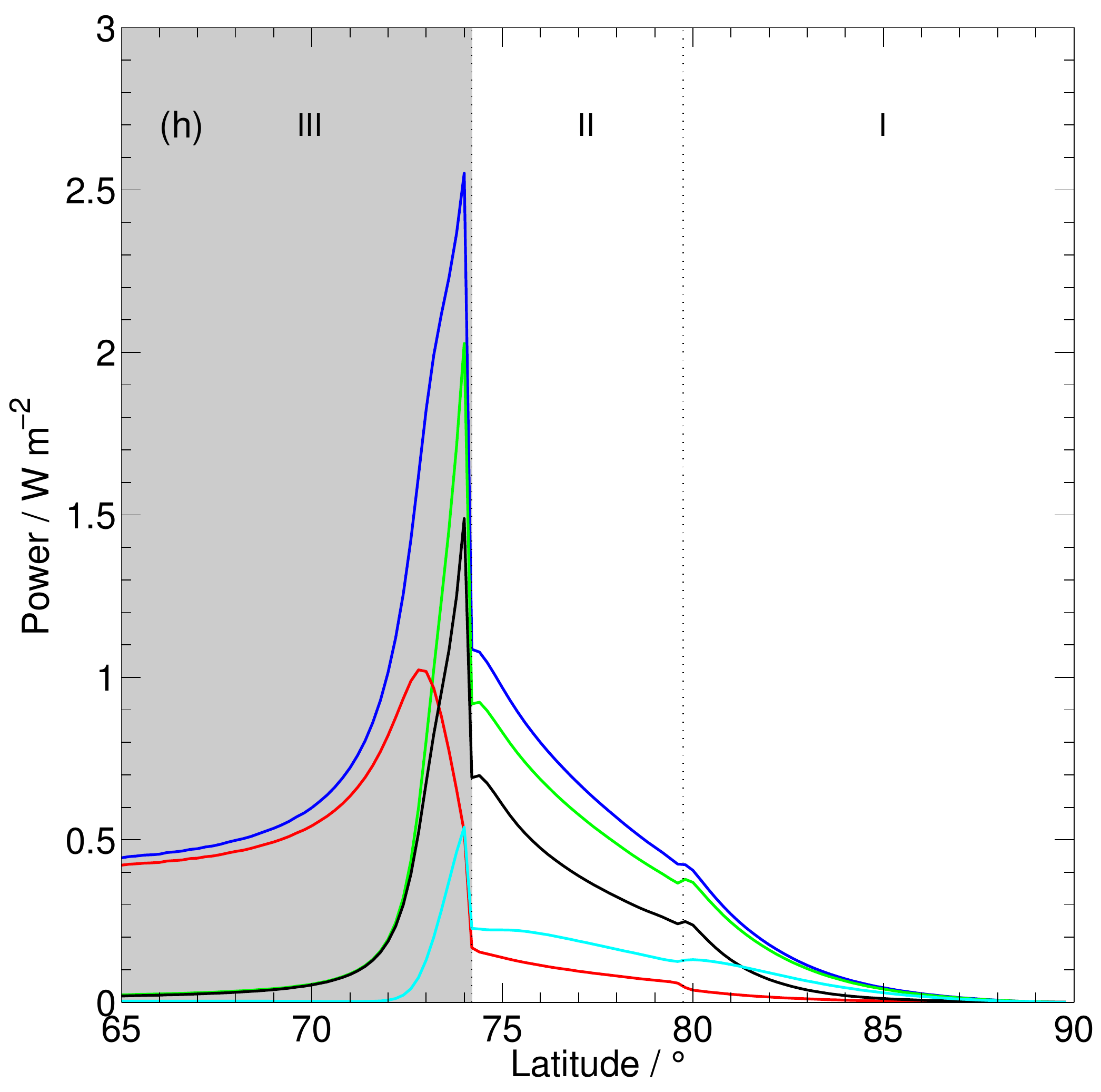}
      \includegraphics[width= 0.65\figwidth]{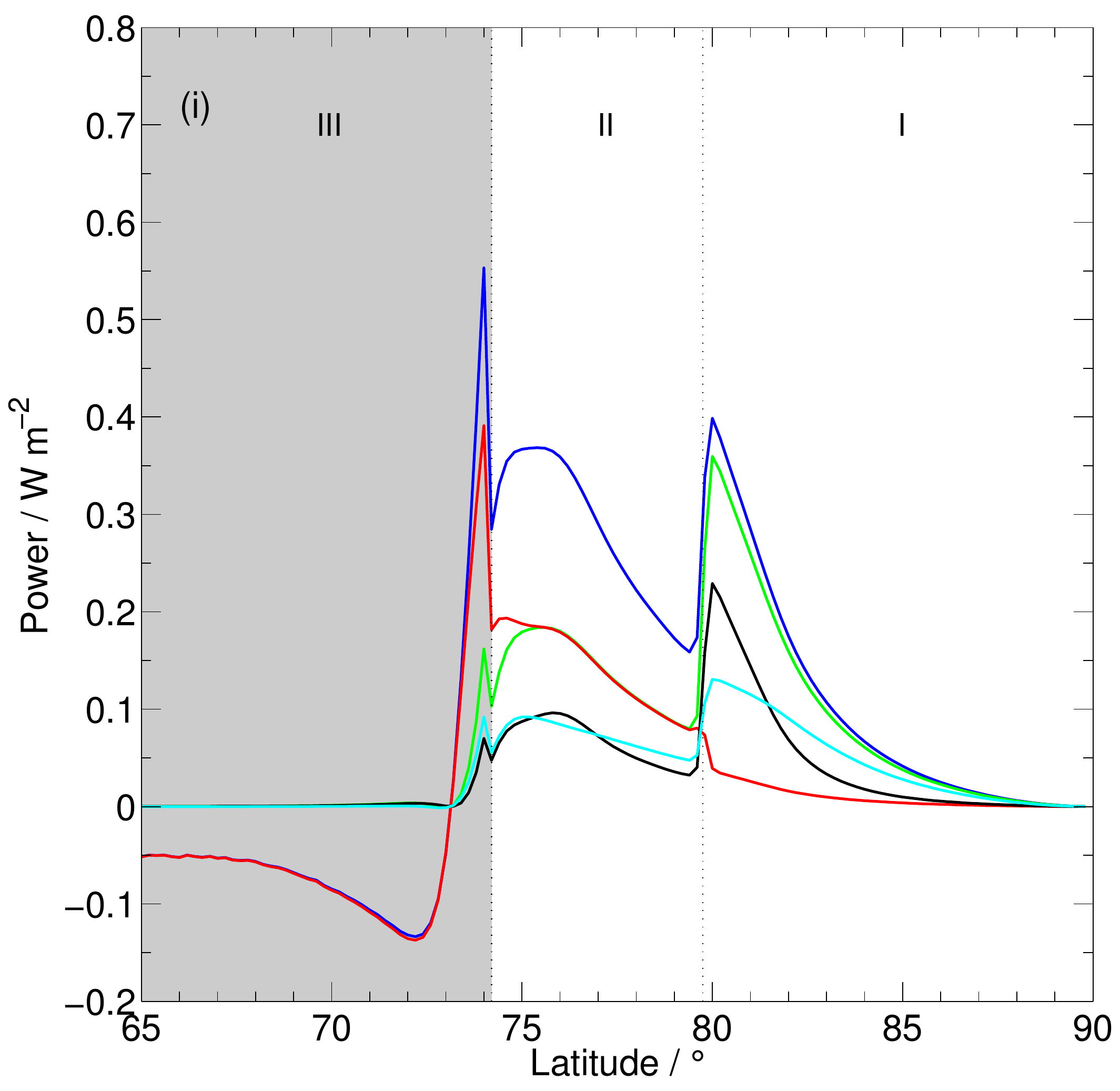}
      
      \caption{ (a)-(c) shows the variation of atmospheric heating terms with altitude, 
      			latitude and temperature (colour bar) for cases ES, EH and EF (left to right). 
      			The contours enclose regions where heating/kinetic energy rates exceed 
      			$\unitSI[20]{W\,kg^{-1}}$. \Revnew{Ion drag energy}, Joule heating, vertical and horizontal 
      			advection of energy, adiabatic heating/cooling, viscous heating and heat conduction 
      			(vertical and turbulent) are represented by blue, red, yellow and magenta, green, 
      			cyan and orange lines. The magnetospheric regions are separated and labelled. 
      			(d)-(f) show the variation of atmospheric cooling terms where the contours enclose 
      			regions where heating/kinetic energy are decreasing (cooling) with rates exceeding 
      			$\unitSI[20]{W\,kg^{-1}}$. All colours and labels are 
      			as in (a)-(c). 
      			(g)-(i) show how the power per unit area varies for our transient expansion cases. 
      			The blue line represents total power which is the sum of magnetospheric power 
      			(red line) and atmospheric power (green line); atmospheric power is the sum of 
      			both Joule heating (black solid line) and \Revnew{ion drag energy} (cyan solid line). All other 
      			labels are as for (a)-(c).
      		   }
      \label{fig:heating_exp}
 \end{figure}

\begin{figure} 
      \centering
      \includegraphics[width= 0.99\figwidth]{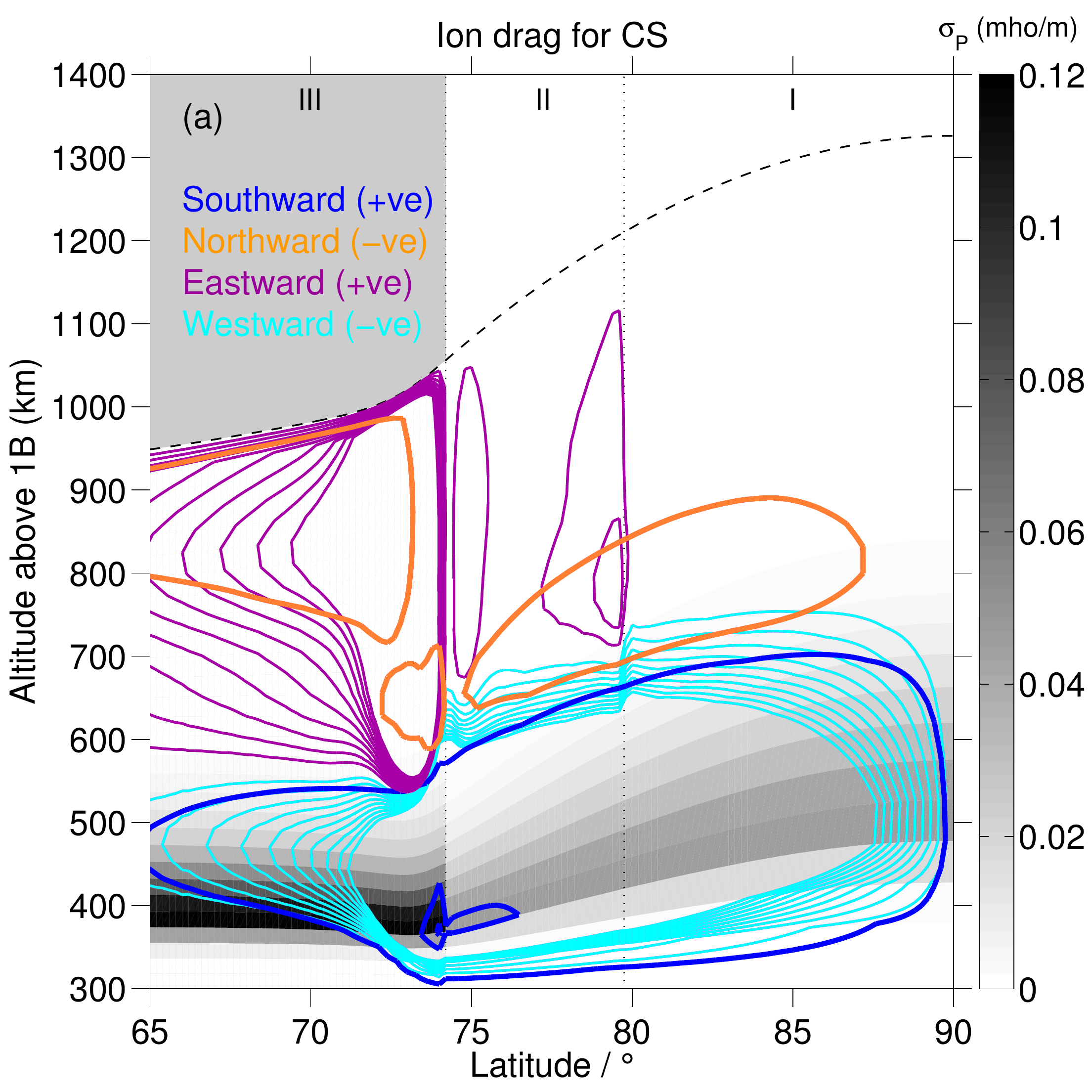}
      \includegraphics[width= 0.99\figwidth]{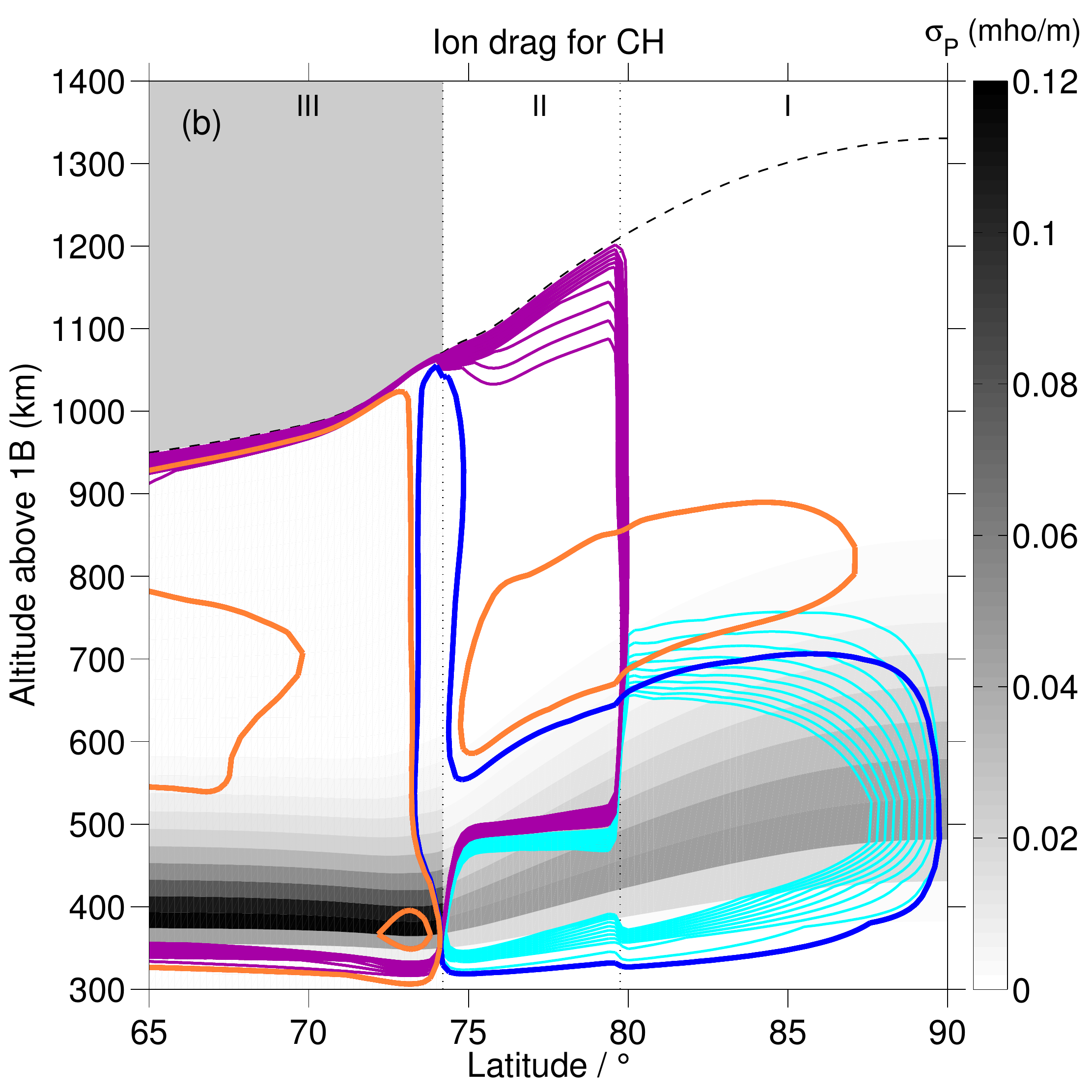}
      
      \caption{ (a) Shows the variation of zonal and meridional ion drag momentum with altitude, latitude and 
      			Pedersen conductivity (colour bar) for case CS. The meridional contours (blue and orange) 
      			range from $\unitSI[1-500]{mm\,s^{-2}}$ with an interval of $\unitSI[50]{mm\,s^{-2}}$ and with blue 
      			being positive (southward) and orange being negative (northward). The zonal contours (purple and cyan) 
      			range from $\unitSI[1-10]{mm\,s^{-2}}$ with an interval of $\unitSI[1]{mm\,s^{-2}}$ and with purple 
      			being positive (eastwards) and cyan being negative (westwards). The magnetospheric regions are 
      			separated and labelled. 
      			(b) Variation of zonal and meridional ion drag momentum with altitude, latitude and 
      			Pedersen conductivity (colour bar) for case CH. All line styles and labels are as in (a).
      		  }
      \label{fig:merid_cmp}
 \end{figure}

 \Fig{\ref{fig:thvel_exp}g} shows temperature as a function of altitude and latitude for case ES. 
 \Figs{\ref{fig:thvel_exp}h-i} show the difference in thermospheric temperature between cases EH and ES, and 
 cases EF and ES, as functions of altitude and latitude. Magnetospheric regions are labelled and separated by 
 black dotted lines and temperatures are indicated by the colour bar. Interpretation of the response of thermospheric 
 temperature is aided by \Figs{\ref{fig:heating_exp}a-f}, showing contour plots for various thermospheric 
 heating (\Figs{\ref{fig:heating_exp}a-c}) and cooling (\Figs{\ref{fig:heating_exp}d-f}) terms (see plot legends for 
 details). \\
 
 \Fig{\ref{fig:thvel_exp}g} shows similar results to those described in section~\ref{sec:heat_cmp}. The main difference 
 is related to the polar `hotspot' which is considerably cooler (peak temperature of ${\sim}\unitSI[590]{K}$) than that 
 for case CS (peak temperature ${\sim}\unitSI[705]{K}$). As previously discussed, the `hotspot' results from the 
 meridional transport (via poleward accelerated flows) of Joule heating from lower latitudes 
 (${\sim}\unitSI[73{-}84]{^{\circ}}$; see \Figs{\ref{fig:thvel_exp}d and \ref{fig:heating_exp}a}) 
 \citep{smith09,yates2012}. \\
 
 \Fig{\ref{fig:thvel_exp}h} exhibits the temperature difference between cases EH and ES. The figure shows a 
 \Rev{maximum of} ${\sim}\unitSI[50]{K}$ temperature increase at low altitudes (${<}\unitSI[700]{km}$) in regions III 
 and II. Also evident are two minor temperature variations: i) ${\sim}\unitSI[10]{K}$ decrease at high altitude, centred 
 on the region III/II boundary and, ii) ${\sim}\unitSI[10]{K}$ increase in the polar `hotspot' region. 
 \Rev{\Fig{\ref{fig:heating_exp}b} shows a large (${\geq}\unitSI[4{\times}]{}$) increase in \Revnew{ion drag energy} and 
 Joule heating rates which accounts for the temperature increase across regions III and II. This low-altitude increase 
 in temperature causes a local increase in pressure gradients leading to accelerated meridional flows being able to 
 efficiently transport heat away from the region III/II boundary and towards the pole. The high altitude cool region 
 ensues from local equatorward and poleward meridional flows combined with factor-of-three increase in adiabatic cooling
 (\Fig{\ref{fig:heating_exp}e}). }\\
 
 \Fig{\ref{fig:thvel_exp}i} shows the temperature difference between cases EF and ES.  The temperature profile has 
 changed significantly from that in \Fig{\ref{fig:thvel_exp}h}. There are two regions where temperatures increase 
 by up to ${\sim}\unitSI[50]{K}$: i) extending from ${\sim}\unitSI[73{-}85]{^{\circ}}$ latitude and low altitudes 
 in regions III and II, and all altitudes in region I (these map to the large poleward-accelerated region in 
 \Fig{\ref{fig:thvel_exp}f}); and ii) high-altitude (${>}\unitSI[600]{km}$) region, centred at ${\sim}\unitSI[66]{^{\circ}}$ 
 latitude. These regions are primarily heated by horizontal advection (high-altitude only) and adiabatic terms 
 (all altitudes) as shown in \Fig{\ref{fig:heating_exp}c}; these \Rev{heating rates} have increased (from case ES) by, 
 at most, 
 $\unitSI[800]{\%}$ and $\unitSI[500]{\%}$ respectively. The final feature of note in \Fig{\ref{fig:thvel_exp}i} is the 
 region cooled by up to ${\sim}\unitSI[{-}22]{K}$, lying between the two heated regions at altitudes 
 ${>}\unitSI[550]{km}$. This cooling is caused by a combination of local increases in horizontal advection and adiabatic 
 cooling, by factors of three and greater. Similar to case CF, case EF's meridional flows seem to be transporting 
 heat equatorward and poleward, although the majority of these flows act to transport thermal energy poleward.\\ 
  
 \Figs{\ref{fig:heating_exp}g-i} show powers per unit area as functions of ionospheric latitude for cases ES-EF 
 respectively. \Rev{Colour codes and labels are as in \Figs{\ref{fig:heating_cmp}g-i}}. \Fig{\ref{fig:heating_exp}g} 
 shows the energy balance in the thermosphere for case ES. As discussed in \citet{yates2012}, most of the energy in 
 region III is expended in accelerating magnetospheric plasma; in region II we have a situation where magnetospheric 
 power and atmospheric power (the sum of Joule heating and \Revnew{ion drag energy}) are equal, due to 
 $\unit{\Omega_M}{=}\unitSI[0.5]{\Omega_J}$. Atmospheric power is dominant in region I. For case EH 
 (\Fig{\ref{fig:heating_exp}h}), the magnetodisc \Rev{plasma sub-corotates to a large degree which causes the majority 
 of available power to be used in accelerating the sub-corotaing plasma. Poleward of ${\sim}\unitSI[73]{^{\circ}}$ latitude, 
 the large flow shear ($\unit{\Omega_T} - \unit{\Omega_M}$) leads to an increase in energy dissipated within the thermosphere,
 primarily through Joule heating. The magnetosphere of case EF super-corotates, compared to the thermosphere, at latitudes 
 ${\leq}\unitSI[73]{^{\circ}}$. This causes a reversal in energy transfer, which now flows from magnetosphere to atmosphere 
 and acts to spin up the sub-corotaing neutral thermosphere (see \Fig{\ref{fig:heating_exp}i}). Polewards of 
 \unitSI[73]{^{\circ}}, the energy balance is similar to that of case ES. }
  
%%%%%%%%%%%%%%%%%%%%%%%%%%%%%%%%%%%%%%%%%%%%%%%%%%%%%%%%%%%%%%%%%%%%%%
\section{Discussion}
\label{sec:discussion}

\subsection{Effect of a non-responsive thermosphere on \MI coupling currents}
\label{sec:nonresp_therm}

 \Rev{Our work makes no a priori assumptions regarding the response of the thermosphere \Revnew{to magnetospheric forcing. 
 The \GCM responds self-consistently by solving the Navier-Stokes equations for momentum, energy and continuity.} 
 For completeness, we calculated \MI coupling currents for the case of a non-responsive thermosphere \citep{gong2005phd}. 
 To do this, we assume that 
 $\unit{\Omega_T}{=}\unit{\Omega_T}(CS)$ throughout the entire transient event. In this non-responsive thermosphere 
 scenario, there is an average increase in \MI currents of ${\sim}\unitSI[20]{\%}$ midway through the pulse compared 
 to case CH (obtained using \GCM). At full-pulse, however, the non-responsive case has \MI currents that are on 
 average ${\sim}\unitSI[12]{\%}$ smaller than currents in case CF. These differences between a non-responsive 
 thermosphere and a responsive one (\GCM), are related to the flow shear between thermosphere and magnetosphere; 
 which, is maximal (resp. minimal) at half-pulse (resp. full-pulse) when using a non-responsive thermosphere. A 
 similar analysis for the expansion scenario results in an average of ${\sim}\unitSI[20]{\%}$ increase in the 
 maximum magnitude of \MI currents in a non-responsive thermosphere, compared to the \GCM thermosphere. Here, 
 $\unit{\Omega_T}(ES)$ 
 is uniformly larger than \unit{\Omega_T} for cases EH and EF (see \Fig{\ref{fig:angvel}}) so the flow shear in 
 the non-responsive scenario will always be greater than the flow shear obtained with the \GCM thermosphere. }
 
\subsection{The auroral response: predictions and comparisons with observations}
\label{sec:aurresp}
% fac and ef for comp 
 
 \Rev{ \Figs{\ref{fig:jpar_cmp}b and \ref{fig:jpar_exp}b} respectively show the change in precipitating electron energy 
 flux in response to transient magnetospheric compression and expansion events. We also indicate (on the right axis of 
 these Figures) the corresponding UV emission associated with such energy fluxes (assuming that \unitSI[1]{mW\,m^{-2}} 
 of precipitation creates ${\sim}\unitSI[10]{kR}$ of UV output). Considering the compression scenario, our results 
 suggest that the arrival of a solar wind shock would increase the UV emission of the main oval from 
 ${\sim}\unitSI[600]{kR}$ to ${\sim}\unitSI[2800]{kR}$ \Rev{(factor of \unitSI[4.7]{})} and constrict the width of 
 the oval by ${\sim}\unitSI[0.2]{^{\circ}}$. The HST detectability limit and the size of an HST pixel (dark grey box in 
 \Fig{\ref{fig:jpar_cmp}b}) suggest that such an increase in auroral emission would be detectable but the 
 constriction of the main oval may be too small to be observed. \citet{clarke2009} and \citet{nichols2009} observed 
 that the brightness of UV auroral emission increased by a factor of two, in response to transient (almost 
 instantaneous) increases in solar wind dynamic pressure (${\sim}\unitSI[0.01{-}0.3]{nPa}$ or equivalently 
 ${\sim}\unitSI[109{-}72]{R_J}$). The increase in UV emission was also found to persist for a few days following the 
 solar wind shock. \citet{nichols2009} also observed poleward shifts (constrictions) in main oval 
 emission on the order of ${\sim}\unitSI[1]{^{\circ}}$ corresponding to the arrival of solar wind shocks.}\\
 
 \Rev{Total emitted 
 UV power may \Rev{also} be used to describe auroral activity, assuming that this quantity is ${\sim}\unitSI[10]{\%}$ 
 of the integrated electron energy flux per hemisphere \citep{cowley07}. Case CH has a total UV power of 
 ${\sim}\unitSI[1.58]{TW}$ (compared to ${\sim}\unitSI[420]{GW}$ for case CS), which is a factor of two to three times 
 larger than UV powers observed by both \citet{clarke2009} and \citet{nichols2009}. The profile of case CH also 
 indicates the possibility of observable polar emission at region II/I (open-closed) boundary. This conclusion is, 
 however, sensitive to our model assumptions (see section~\ref{sec:mod_lim}). }\\
 
 %fac and ef for exp
 \Rev{Our model results predict very different behaviour for the expansion scenario (\Fig{\ref{fig:jpar_exp}b}). 
 At maximum expansion we would expect a small increase (${\sim}\unitSI[40]{kR}$) in peak main oval brightness 
 along with a ${\sim}\unitSI[1]{^{\circ}}$ equatorward shift (expansion) of the oval. We also note the possible 
 observation of a somewhat bifurcated main oval (see HST pixel in Figure); with emission peaking at 
 ${\sim}\unitSI[73]{^{\circ}}$ and ${\sim}\unitSI[74]{^{\circ}}$ latitude. The main oval would, either way, appear 
 considerably broader (${\sim}\unitSI[2{-}3]{^{\circ}}$) as a result of the large increase in the spatial region 
 of magnetospheric sub-corotation. \citet{clarke2009} observed little change in auroral brightness near the arrival 
 of a solar wind rarefaction region, however \citet{nichols2009} have seen changes in main oval location. The total UV 
 power in case EH is ${\sim}\unitSI[270]{GW}$ (compared to ${\sim}\unitSI[78]{GW}$ in case ES). While this power is 
 considerably smaller than that in case CH, it is comparable to UV powers calculated in \citet{clarke2009} and 
 \citet{nichols2009}, following solar wind rarefactions (${\sim}\unitSI[200-400]{GW}$).}

\subsection{Global thermospheric response} 
\label{sec:therm_resp}

% flows
 \Rev{ The arrival of solar wind shocks or rarefactions has, for the most part, a similar effect on thermospheric flows. 
 Our modelling shows a general increase (resp. decrease) in the degree of corotation with solar wind dynamic pressure 
 increases (resp. decreases). Zonal flow patterns remain essentially unchanged with a large sub-corotational jet and a 
 small super-corotational jet. Meridional flow cells however, respond to transient magnetospheric reconfigurations 
 somewhat chaotically, with numerous poleward and equatorward accelerated flow regions developing (at altitudes 
 $>\unitSI[600]{km}$) with time throughout the event. The overall low-altitude poleward flow remains fixed with solar wind 
 rarefactions but reverses in response to a solar wind shock (see cases CH and CF in \Figs{\ref{fig:thvel_cmp}e and f}). 
 This flow becomes equatorward due to a reversal in the direction of ion drag acceleration in the region III, as 
 shown in \Fig{\ref{fig:merid_cmp}}. This reversal, in turn, arises from the super-corotation of magnetospheric 
 plasma. }\\
 
% fig 9 here 

% stability
 Compared to the transient compression case CF, the EF thermosphere seems fairly stable i.e. there are no sharp peaks 
 and troughs in the upper boundary. Our interpretation is that for the compression scenario the magnetosphere transfers 
 a large amount of angular momentum to the thermosphere due to its large degree of super-corotation. This surge in 
 momentum and energy input to the thermosphere over a short time scale causes significant strain on the thermosphere 
 and thus requires a drastic reconfiguration in order to attempt to re-establish dynamic equilibrium. On the other hand, 
 in our 
 expansion scenario the magnetosphere significantly sub-corotates for most of the event and only super-corotates compared 
 to the planet and thermosphere (slightly) nearing the end of the event. Thus, for the majority of the expansion event the 
 thermosphere is losing angular momentum to the magnetosphere. This implies that its dynamics and energy input are 
 generally smaller than the transient compression scenario, which leads to a less `drastic' response. \\
 
 % fig 10 here
 
\begin{figure} 
      \centering
      \includegraphics[width= 0.99\figwidth]{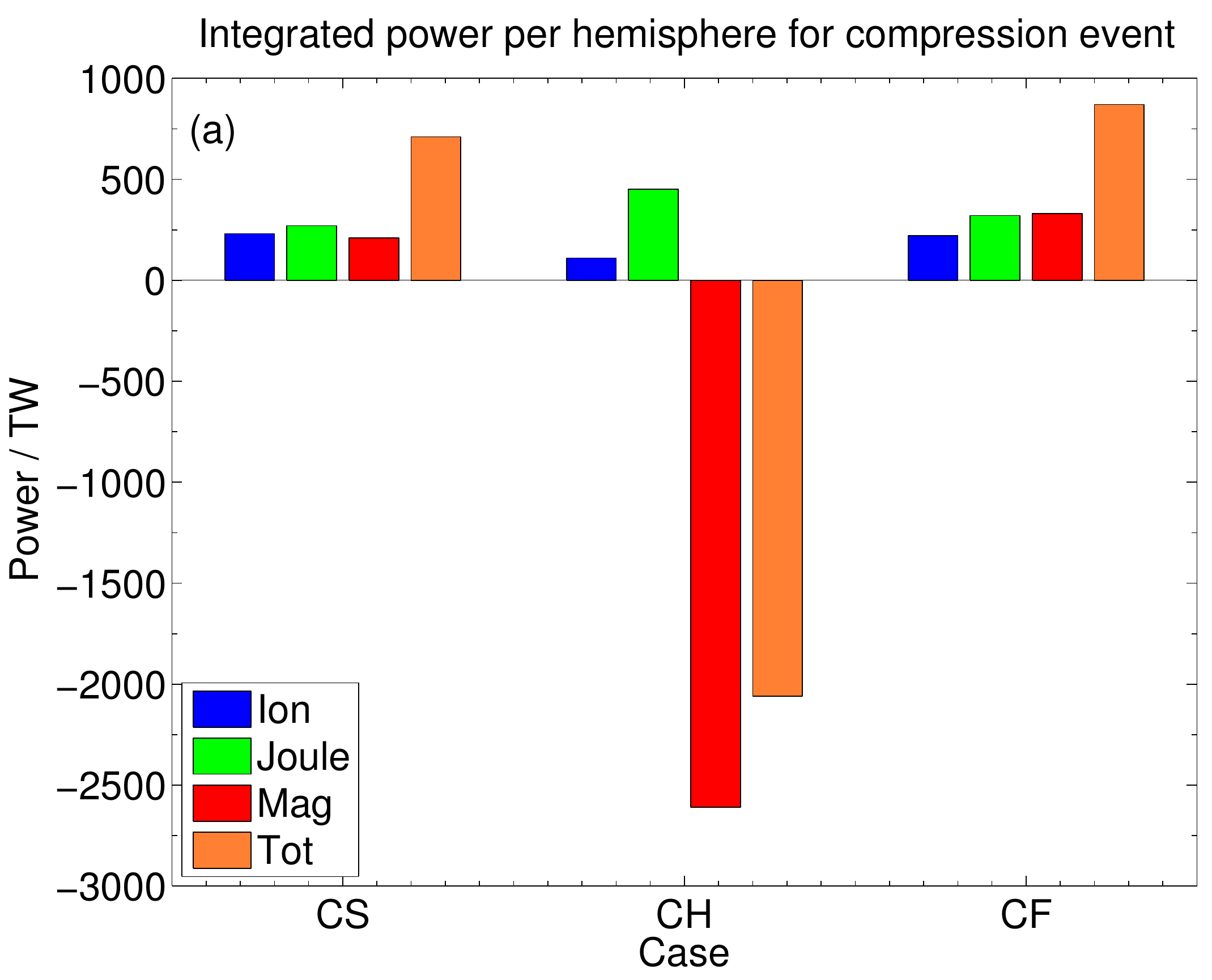}
      \includegraphics[width= 0.99\figwidth]{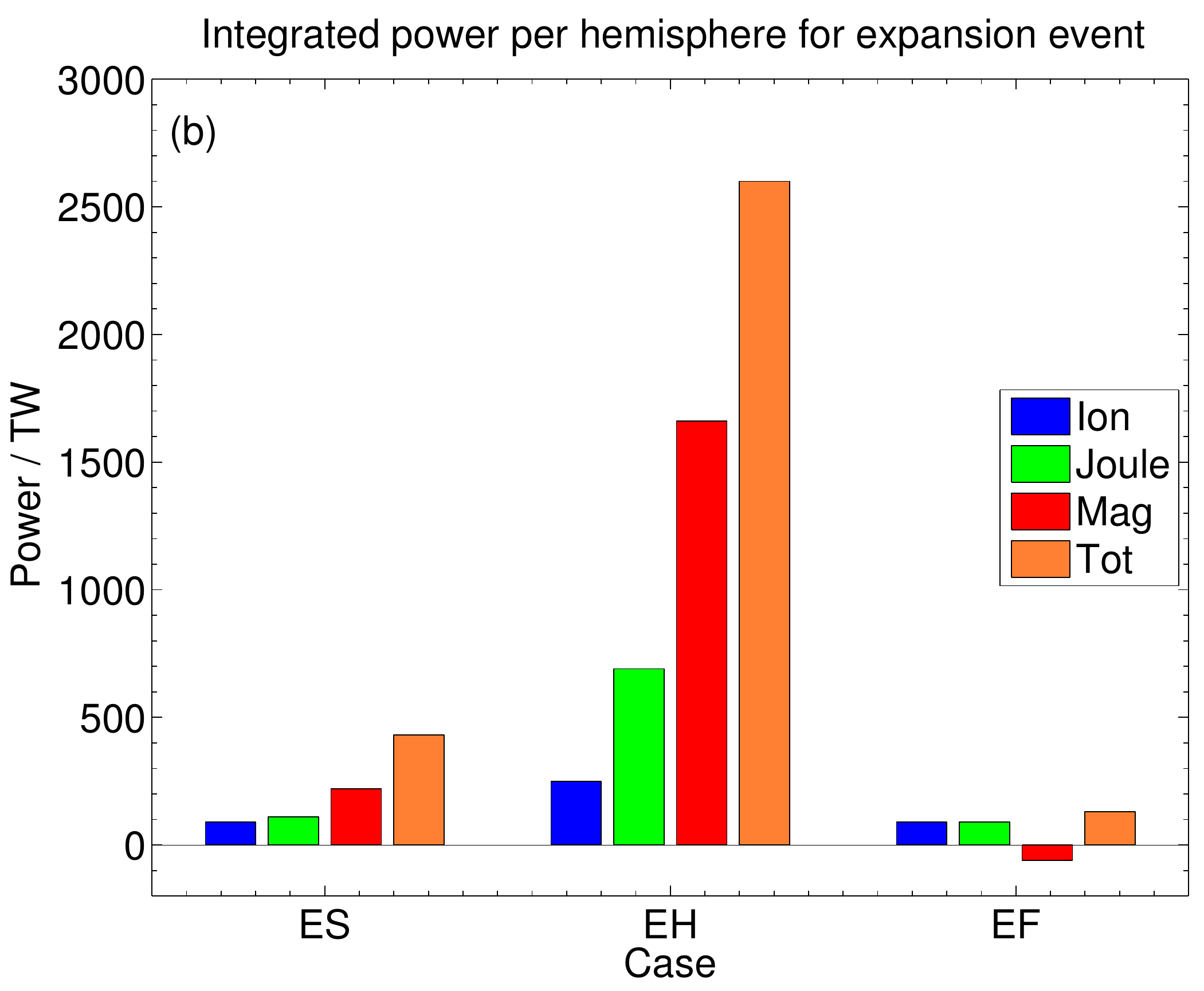}
      
      \caption{ (a) Integrated ionospheric powers per hemisphere for cases CS-CF are represented in this figure. 
      			\Revnew{Ion drag energy} is represented by blue bars, Joule heating by green bars, magnetospheric power 
      			by red bars and total (sum of all above) is represented by orange bars. See text for further detail.
      			(b) Integrated ionospheric powers per hemisphere for cases ES-EF are represented in this figure. 
      		  }
      \label{fig:powerbar}
 \end{figure}

 \Rev{ The magnetospheric reconfigurations discussed above have been shown to have a significant impact on the dynamics 
 and energy balance of the thermosphere. We now attempt to globally quantify such changes in energy by calculating the 
 integrated power per hemisphere obtained from the power densities in \Figs{\ref{fig:heating_cmp}g-i and 
 \ref{fig:heating_exp}g-i}. These integrated powers are presented in \Figs{\ref{fig:powerbar}a and b} for the 
 compression and expansion scenarios respectively. Blue bars represent the kinetic energy dissipated by ion drag, green bars 
 indicate Joule heating, red bars represent the power used to accelerate magnetospheric plasma and orange bars 
 simply represent the sum of all the above terms. Positive powers indicate energy dissipated in/by the thermosphere whilst 
 negative powers indicate energy deposited into the thermosphere.}\\
 
 Midway through the compression event (case CH), magnetospheric plasma super-corotates compared to the thermosphere and deep 
 atmosphere. This reverses the direction of momentum and energy transfer so that energy is now being transferred from 
 the magnetosphere to the thermosphere. Our results indicate that ${\sim}\unitSI[2000]{TW}$ of total power (magnetospheric, 
 Joule heating and \Revnew{ion drag energy}) is gained by the coupled system as a result of plasma super-corotation. Note 
 that this is considerably larger than the ${\sim}\unitSI[325]{TW}$ (closed and open field regions) calculated in 
 \citet{cowley07} for a responsive thermosphere scenario. This energy transfer from the magnetosphere would act to, 
 essentially `spin up' the planet \citep{cowley07} and increase the thermospheric temperature. In case CF, plasma 
 is not super-corotating; thus the picture is fairly similar to case CS. The main difference is that there is a 
 ${\sim}\unitSI[20]{\%}$ increase in total power dissipated in the atmosphere and in acceleration of the magnetosphere. 
 This arises from increases in flow shear due to the `lagging' thermosphere (see \Fig{\ref{fig:angvel}}) and inevitably 
 leads to the local temperature increases seen in \Fig{\ref{fig:thvel_cmp}i} and discussed above. The finite time 
 required for thermospheric response results in the described `residual' perturbations to the initial system (CS) even 
 after the pulse has subsided (CF).\\
 
 A transient magnetospheric expansion event creates a significant increase in both power dissipated in the 
 atmosphere due to Joule heating (${\sim}\unitSI[6{\times}]{}$ that of ES) and \Revnew{ion drag energy} 
 (${\sim}\unitSI[3{\times}]{}$ that of ES). Moreover, the power used to accelerate the magnetosphere towards 
 corotation is ${\sim}\unitSI[7{\times}]{}$ 
 that of ES, and is shown in \Fig{\ref{fig:powerbar}b}. These increases lead to a total power per hemisphere of 
 ${\sim}\unitSI[2600]{TW}$ which is three times larger than the responsive thermosphere case 
 in \citet{cowley07}. These changes in heating and cooling create the local temperature increases discussed above. 
 For case EF, where we now have the magnetosphere rotating faster than the thermosphere, there is a 
 ${\sim}\unitSI[75]{\%}$ decrease in the magnitude of `magnetospheric' power. The magnetosphere is thus transferring 
 power to the thermosphere in this case, albeit a relatively small amount. This effectively `pulls' the 
 thermosphere along, increasing its angular velocity in order to return to the steady-state situation where 
 $\unit{\Omega_T {>} \Omega_M}$. We note that energy dissipated via Joule heating also decreases slightly 
 due to the small decrease in flow shear. Overall, then, the total power per hemisphere in case EF is only \unitSI[30]{\%} 
 that of case ES. \\
 
 \Rev{Results for cases CH and EH show large (approximately three orders of magnitude larger than solar heating) 
 increases in energy either being deposited or dissipated in the thermosphere. 
 Observations by \citet{stallard2001,stallard2002} of an auroral heating event at Jupiter were ananlysed by \citet{melin2006}. 
 These authors found that during this auroral heating event, which they attribute to being caused by a decrease in solar 
 wind dynamic pressure, the combined \Revnew{ion drag energy} and Joule heating rates increase from 
 \unitSI[67]{mW\,m^{-2}} to 
 \unitSI[277]{mW\,m^{-2}} over three days. They proposed that this extra heat must then be transported equatorward 
 from the auroral regions by an increase in equatorward meridional winds \citep{waite1983}. If we assume that their auroral 
 region ranges from \unitSI[65]{^{\circ}} to \unitSI[85]{^{\circ}} latitude and that these heating rates are constant across 
 such a region, the total energy dissipated by Joule heating and \Revnew{ion drag energy} increases from 
 ${\sim}\unitSI[193]{TW}$ to ${\sim}\unitSI[800]{TW}$. This increase is comparable to the increase of Joule heating 
 and \Revnew{ion drag energy} in our expansion scenario, going from case ES (${\sim}\unitSI[201]{TW}$) to EH 
 (${\sim}\unitSI[942]{TW}$). Increase in Joule heating and \Revnew{ion drag energy} from case CS to CH is more 
 modest (${\sim}\unitSI[499]{TW}$ to ${\sim}\unitSI[555]{TW}$) due to the 
 reversal of kinetic energy exchange between atmospheric neutrals and ions and despite an increase in Joule heating. 
 Our modelling supports the work of \citet{melin2006} in terms of i) the magnitude of energy dissipated in the 
 thermosphere and ii) the type of magnetospheric reconfiguration required. We do not however,
 see a significant increase in equatorward flows in our expansion scenario. Our compression scenario, however, shows a 
 large change in meridional flow patterns with a large portion of the thermosphere flowing equatorwards. }

\subsection{Model limitations} 
\label{sec:mod_lim}
 
 \Rev{The main limitation to our transient model is the use of a fixed model for relative changes in conductivity with 
 altitude, and a uniform Pedersen conductance \unit{\Sigma_P} for the ionosphere (see section~\ref{sec:ion_model}).  
 Whilst not ideal, we feel it is a suitable first step to developing a fully self-consistent, time-dependent model of 
 the Jovian magnetosphere-ionosphere-thermosphere system. Use of an enhanced conductivity model would concentrate 
 all but background levels of conductance just equatorward of the main auroral oval location 
 (${\sim}\unitSI[74]{^{\circ}}$) \citep{yates2012}; effectively increasing the coupling between the atmosphere and 
 magnetosphere in this region. We would thus expect the magnitude of current densities to increase in the region 
 near the main oval (region III/II boundary in our model), along with an increase in the Joule heating rate. }\\
 
 \Rev{ The 
 high conductivity at latitudes between \unitSI[60]{^{\circ}} and \unitSI[70]{^{\circ}} in the present model, combined 
 with the super-corotation of the thermosphere, allows for the plasma magnetically mapped to these ionospheric latitudes 
 to super-corotate slightly in steady state. With an enhanced conductivity model, this region would have a 
 super-corotating thermosphere but low, background-level conductances (e.g. \citet{smith09,yates2012}). Therefore, 
 even though the super-corotating 
 thermosphere acts to accelerate the magnetodisc plasma, the low conductances inhibit how efficiently the plasma is 
 accelerated. It is worth noting that despite the fact that, in this study, both the neutral thermosphere and magnetodisc 
 plasma super-corotate compared to the deep atmosphere, as long as the plasma sub-corotates compared to the thermosphere, 
 angular momentum and energy will be transferred from the upper atmosphere to the magnetosphere as is expected in steady 
 state. We plan to incorporate enhancements in Pedersen conductance due to auroral precipitation of 
 electrons in a future study. }\\
 
 \Rev{ Other limitations to this model include:}\\
 
 \Rev{\emph{i) Assumption of axial symmetry}: Discussions in \citet{smith09} conclude that the assumption of axial 
 symmetry with respect to the planet's rotation axis does not significantly alter the thermospheric outputs of our model. 
 They find that axial symmetry leads to modelling errors on the order of ${\sim}\unitSI[20]{\%}$ which are 
 less than, or at least comparable to, errors derived from the various other assumptions and simplifications made 
 in this coupled model.}
 
 \Rev{\emph{ii) No development of field-aligned potentials}: Our model does not currently include the development of 
 field-aligned potentials, which accelerate electrons from the high latitude magnetosphere into the ionosphere. We simply 
 apply the linear approximation to the Knight relation (see section~\ref{sec:aur_energies}) to obtain precipitating 
 electron energy fluxes. 
 \citet{ray2009} show that significant field-aligned potentials develop at high-latitudes to supply the necessary 
 \FAC{}s, and hence angular momentum, demanded by the magnetospheric plasma. By applying the linear approximation to 
 the Knight relation, we assume that the top of the acceleration region is far enough from the planet such that the 
 ratio of the energy gained by a particle traveling through the potential drop to its thermal energy is  significantly 
 less than the mirror ratio between top and bottom of the acceleration region. Consequently, possible current saturation 
 effects are ignored, with the field-aligned current density increasing to values beyond those that would result from 
 the entire electron distribution accelerated into the loss cone. The \MI coupling modelling by \citet{ray2010} also 
 showed that including field-aligned potentials in a self-consistent treatment of the auroral current system alters 
 the electric field mapping between the ionosphere and the magnetosphere, decoupling the ionospheric and magnetospheric 
 flows. Their model did not explicitly include thermospheric flows; however, the presence of field-aligned potentials may 
 also plausibly alter the thermospheric angular velocity.}
 
 \Rev{\emph{iii) Fixed plasma angular velocity in the polar cap region (latitudes ${>}\unitSI[80]{^{\circ}}$)}: The 
 plasma angular velocity in the polar cap region \unit{\Omega_{Mpc}} 
 is fixed at a constant value of ${\sim}\unitSI[0.1]{\Omega_J}$, in accordance with the formulations in \citet{isbell1984} 
 which depend in part on the solar wind velocity \unit{v_{sw}}. A change in solar wind dynamic pressure \unit{p_{sw}} 
 would generally be accompanied by a corresponding change in \unit{v_{sw}}, so when we change the magnetospheric 
 configuration of our model, \unit{\Omega_{Mpc}} should also change depending on the new value of \unit{v_{sw}}. If we 
 assume that the solar wind density \unit{\rho_{sw}} remains constant and that $\unit{p_{sw} \approx \rho_{sw} v_{sw}^2}$, 
 $\unit{\Omega_{Mpc} (CS)}\approx\unitSI[0.06]{\Omega_J}$ and $\unit{\Omega_{Mpc} (CH)}\approx\unitSI[0.17]{\Omega_J}$. 
 We find the difference between the plasma angular velocities across the open-closed field line boundary with a constant 
 or variable \unit{\Omega_{Mpc}} to be negligible for both compressed and expanded magnetospheres and thus do not expect 
 this to significantly influence the results discussed above.}
  
\section{Conclusion}
\label{sec:conclusion} 

 \Rev{We investigated the effect of transient variations in solar wind dynamic pressure on the \MI coupling currents, 
 thermospheric flows, heating and cooling rates and aurora of the Jovian system. We considered two scenarios: i) a 
 transient compression event, and ii) a transient expansion event. Both of these were imposed over a 
 time scale of three hours. A transient compression event consists of an initially expanded, steady-state 
 magnetospheric configuration. The model Jovian magnetosphere then encounters a shock in the solar wind, which 
 compresses the system. As the conceptual shock propagates past the magnetosphere, a rarefaction region follows 
 and the magnetosphere subsequently expands back to its initial state. The opposite occurs for our expansion 
 event. }\\
 
 \Rev{We have made an important initial step into investigating how time-dependent phenomena affect the Jovian system. 
 In steady state, the more expanded the magnetosphere is, the hotter Jupiter's thermosphere is likely to be 
 \citep{yates2012}. The caveat to this is that only the polar 
 (high-altitude) region of the thermosphere (due to the poleward meridional winds) approaches the observable temperatures 
 of ${\sim}\unitSI[900]{K}$ \citep{seiff1998,yellemiller2004,lystrup2008}. The lower latitudes are still relatively cool with 
 temperatures of ${\sim}\unitSI[200-300]{K}$, compared to polar temperatures of up to ${\sim}\unitSI[700]{K}$. On the other 
 hand, when we consider rapid magnetospheric reconfigurations, the situation is quite different. We see a change in the 
 direction of meridional winds as well as a large (at least a factor of two) increase in Joule heating and energy being 
 dissipated in or deposited to the thermosphere. These winds 
 redistribute the extra heat, essentially sending `wave-like perturbations' of high-temperature gas (higher than ambient 
 surroundings) towards both the polar and equatorial regions \citep{waite1983,achilleos98,melin2006}. The present results 
 are not enough to increase the temperature of the equatorial thermosphere to its observed values but we stress that all 
 the results presented herein occur within a period of three hours (approximately \unit{1/3} of a Jovian day). This leads 
 to the potential of future, more realistic, time-dependent studies whereby one could vary the duration of such transient 
 events, experiment with `chains' of such events and/or more realistic solar wind dynamic pressure profiles, in order to 
 model the dynamic response of the Jovian thermosphere over more extended periods of external perturbation.}

%%%%%%%%%%%%%%%%%%%%%%%%%%%%%%%%%%%%%%%%%%%%%%%%%%%%%%%%%%%%%%%%%%%%%%
\section*{Acknowledgement} 

  JNY was supported by an STFC studentship award. NA was supported by STFC's UCL Astrophysics Consolidated Grant 
  ST/J001511/1. The authors acknowledge support of the STFC funded Miracle Consortium (part of the DiRAC facility) in 
  providing access to computational resources. The authors express their gratitude to Chris Smith who developed the \GCM 
  used herein and to Licia Ray and Fran Bagenal for our useful discussions. The authors would also 
  like to thank two anonymous referees for their useful comments and suggestions. 

%%%%%%%%%%%%%%%%%%%%%%%%%%%%%%%%%%%%%%%%%%%%%%%%%%%%%%%%%%%%%%%%%%%%%%  
  
\bibliographystyle{elsarticle-harv}
\bibliography{bibliography} \pagebreak

%%%%%%%%%%%%%%%%%%%%%%%%%%%%%%%%%%%%%%%%%%%%%%%%%%%%%%%%%%%%%%%%%%%%%%%%%%%%%%%%%%%%%%%%%%%%%%%%%%%%%%
%%%%%%%%%%%%%%%%%%%%%%%% TABLES AND FIGURES %%%%%%%%%%%%%%%%%%%%%%%%%%%%%%%%%%%%%%%%%%%%%%%%%%%%%%%%%%

  \begin{table}  
    \caption[Transient state conductivity profiles]
    	  {Transient state Pedersen conductances in our model thermosphere. General regions of the thermosphere are indicated in 
    	  the left column whilst their respective ionospheric latitude is indicated in the middle column. The right column shows 
    	  the value/profile of the height-integrated Pedersen conductance assumed in each respective region.}
    \begin{tabular}{ l  r  r r }
      \hline
      Location &  Ionospheric latitude / \unit{^{\circ}} &  \unit{\Sigma_P} / mho & Reference \\ \hline
      \hline
      Equatorial & $<\unitSI[60]{}$ & \unitSI[0.0275]{} & \citet{hill1980} \\ \hline
      Auroral &  $\unitSI[60]{}\leq\unit{\theta}\leq\unitSI[74]{}$ & \unitSI[0.5]{} & this work, see also \citet{nichols2011} \\ \hline
      Polar regions & $>\unitSI[74]{}$ & \unitSI[0.2]{} & \citet{isbell1984} \\ \hline
    \end{tabular}
    \label{tb:model:tspedcond} 
 \end{table}
 
 \begin{table}
    
    \caption
    	  {The three different stages of the transient magnetospheric reconfiguration events (compression and 
    	  expansion). The radii of the magnetodisc \unit{R_{MM}}, magnetopause \unit{R_{MP}} and corresponding 
    	  solar wind pressure \unit{P_{SW}} \citep{joy2002} are shown.}
    \begin{tabular}{ l  r  r  r  r  r  r }
      \hline
      Case &  CS & CH & CF & ES & EH & EF \\ \hline
      \hline
      \unit{R_{MM}}/ \unitSI{R_J} & \unitSI[85]{} & \unitSI[45]{} & \unitSI[85]{} & \unitSI[45]{} & \unitSI[85]{} & \unitSI[45]{}\\ \hline
      \unit{R_{MP}}/ \unitSI{R_J} & \unitSI[101]{} & \unitSI[68]{} & \unitSI[101]{} & \unitSI[68]{} & \unitSI[101]{} & \unitSI[68]{}\\ \hline
      \unit{P_{SW}}/ \unitSI{nPa} & \unitSI[0.021]{} & \unitSI[0.213]{} & \unitSI[0.021]{} & \unitSI[0.213]{} & \unitSI[0.021]{} & \unitSI[0.213]{}\\ \hline
    \end{tabular}
    
    \label{tb:trancases} 
 \end{table}
 
% Appendix figures here

\appendix 

\section{Magnetosphere-ionosphere coupling}
 \label{sec:mi_coup}
\setcounter{figure}{0}
\setcounter{table}{0}

 In this section we discuss the effect of coupling the magnetosphere and ionosphere together. The meridional 
 electric field in the rest frame of the thermosphere may be written as: 
 
 \begin{align}
	\unit{ E_{\theta} } =& \unit{ B_i \rho_{i} \left( \Omega_T - \Omega_M\right)}, \label{eq:efield}
 \end{align} 
 
 \noindent where \unit{B_i} is the magnitude of the (assumed) radial ionospheric magnetic field 
 ($\unit{B_i}{=}\unit{2 B_J}$). \\
 
 The combination of electric field, magnetic field and ion-neutral collisions causes Pedersen currents to 
 flow in the ionosphere, mainly perpendicular to the direction of the planetary 
 magnetic field. These ionospheric currents form part of a larger current circuit which includes the radial 
 current flowing in the magnetodisc and the \FAC flowing along the magnetic field lines. The height integrated 
 Pedersen current density \unit{i_P} and its azimuthally integrated form \unit{I_P (\theta_i)} are 
 \citep{cowley07,smith09} 

 \begin{align}
	\unit{ i_{P} } =& \unit{ \rho_{i} \Sigma_P \left( \Omega_T - \Omega_M\right) B_i} \label{eq:pedcur},
 \end{align}
  
 \noindent and
  
 \begin{align}
	\unit{ I_P(\theta_i) } =& \unit{ 2\pi \rho_{i}^{2} \Sigma_P \left( \Omega_T - \Omega_M\right) B_i} \label{eq:AIpedcur},
 \end{align}
 
 \noindent where \unit{\Omega_T} is a weighted average, computed over altitude, of the angular velocity of the 
 thermosphere. For a more detailed description of this the reader is referred to \citet{yates2012} and \citet{smith09}.\\
 
 The height-integrated radial current density in the magnetodisc is denoted by \unit{i_{\rho}} and can be obtained using 
 \Eq{\ref{eq:pedcur}} under the assumption of current continuity (zero divergence of current density). We 
 have \citep{nichols04,smith09}:
 
 \begin{align}
	\unit{\rho_e i_{\rho}} =& \unit{2\rho_i i_P}\label{eq:radcur},\\
	\unit{ I_{\rho} } =& \unit{ 8\pi \Sigma_P F_e \left( \Omega_T - \Omega_M\right) }, \label{eq:AIradcur}
 \end{align}

 \noindent where \unit{I_{\rho}} is the azimuthally integrated disc current.\\
 
 The third and final component of our \MI current circuit is the \FAC density. \unit{j_{||i}(\theta_i)} represents the 
 \FAC density at the ionospheric footpoint (at co-latitude \unit{\theta_i}) of the respective field lines. This current 
 density is obtained from the horizontal divergence of the Pedersen current:
 
 \begin{align}
	\unit{j_{||i}(\theta_i)} =& \unit{-\frac{1}{2\pi R_i^2 \sin\theta_i}} \unit{\frac{d I_P}{d \theta_i}}, \label{eq:jpar}
 \end{align}
 
 \noindent where the sign of \unit{j_{||i}(\theta_i)} indicates \FAC direction (positive upward from planet). 
 \Eq{\ref{eq:jpar}} corresponds to the northern hemisphere, where the magnetic field points radially outward
 (approximately, in auroral region) \citep{cowley07}.\\

 The final aspect of \MI coupling we examine in this study is the energy transfer from planetary rotation to 
 the thermosphere and magnetosphere. The \Rev{angular momentum transfer} to the magnetosphere is used to accelerate 
 magnetospheric plasma towards corotation whilst the energy dissipated within the thermosphere is used for 
 heating and increasing kinetic energy. The total power per unit area of the ionosphere transferred from 
 planetary rotation \unit{P} is 
 the sum of atmospheric power \unit{P_A} and magnetospheric power \unit{P_M} dissipated per unit area \citep{hill2001}. 
 As shown by \citet{smith05} atmospheric power consists of two components: (i) Joule heating \unit{P_J} and 
 (ii) ion drag power \unit{P_D}, some of may be viscously dissipated as heat. These power relations are \citep{cowley05}: 
 
 \begin{align}
      \unit{P} &= \unit{ \Omega_J \tau }, \label{eq:totpower}\\
      \unit{P_M} &= \unit{ \Omega_{M} \tau }, \label{eq:magpower}\\
      \unit{P_A} &= \unit{ (\Omega_{J} - \Omega_{M} ) \tau }, \label{eq:atmpower}\\
      \unit{P_J} &= \unit{ (\Omega_{T} - \Omega_{M} ) \tau }, \label{eq:jouleheating}\\
      \unit{P_D} &= \unit{ (\Omega_{J} - \Omega_{T} ) \tau }, \label{eq:iondrag}
 \end{align}
 
 \noindent where 
 
 \begin{align}
       \unit{ \tau } &= \unit{ \rho_i i_{P} B_i } \label{eq:torque}
 \end{align}
 
 \noindent represents the torque exerted by the \textbf{\emph{J\unit{\times}B}} force per unit area of 
 the ionosphere.

%%%%%%%%%%%%%%%%%%%%%%%%%%%%%%%%%%%%%%%%%%%%%%%%%%%%%%%%%%%%%%%%%%%%%% 
 \section{Auroral energies}
 \label{sec:aur_energies}
 
 Once \FAC densities have been calculated, we can use the methods of \citet{knight1973} and \citet{lundin1978}, 
 as presented in \citet{cowley07}, to calculate the enhanced precipitating electron energy flux \unit{E_f}: 
 
  \begin{align}
      \unit{E_f} &= \unit{ \frac{E_{f0}}{2} \left( \left( \frac{j_{||i}}{j_{||i0}} \right)^2 + 1 \right) },
      \label{eq:eflux}
 \end{align}
 
 \noindent where \unit{E_{f0}} is the unaccelerated electron energy flux, \unit{j_{||i0}} is the unaccelerated 
 \FAC density (or the maximum current that can be carried by the electrons in the absence of field-aligned potential 
 drops) and \unit{j_{||i}} is the upward (positive) \FAC density calculated using \Eq{\ref{eq:jpar}}. To enable 
 a comparison with similar, earlier studies, we use the same electron population values described in \citet{cowley07}, 
 which are based on observations by \citet{scudder1981} and \citet{phillips1993a,phillips1993b}. These parameters are 
 presented in \Table{\ref{tb:aur_parameters}}.

% TABLE 3/B1 here
 \begin{table}
    
    \caption{ Magnetospheric electron source parameters. This table is adapted from Table 1, \citet{cowley07}. 
              \unit{N_e} represents the electron density, \unit{W_{th}} the electron thermal energy, \unit{j_{||i0}} 
              the unaccelerated current density and \unit{E_{f0}} the unaccelerated energy flux. 
            }
	\small       
    \begin{tabular}{ l  r  r  r }
      \hline
       Parameter &  Open field lines &  Outer magnetosphere & Middle magnetosphere \\ \hline
      \hline
      \unit{N_e} / \unitSI{cm^{-3}} & \unitSI[0.5]{} & \unitSI[0.02]{} & \unitSI[0.01]{}\\ \hline
      \unit{W_{th}} / \unitSI{keV} & \unitSI[0.05]{} & \unitSI[0.25]{} & \unitSI[2.5]{}\\ \hline
      \unit{j_{||i0}} / \unitSI{\mu A\,m^{-2}} & \unitSI[0.095]{} & \unitSI[0.0085]{} & \unitSI[0.013]{}\\ \hline
      \unit{E_{f0}} / \unitSI{mW\,m^{-2}} & \unitSI[0.0095]{} & \unitSI[0.0042]{} & \unitSI[0.067]{}\\ \hline
    \end{tabular}
    \label{tb:aur_parameters} 
 \end{table}
 \normalsize
 
\end{document}